%% file: short.tex
\documentclass[aps,english,floatfix,
tightenlines,twocolumn,nofootinbib,superscriptaddress]{revtex4-2}

\include{jm-tex-macros-public}
\include{macros}

\usepackage{ragged2e}

\usepackage{array} 
\usepackage{comment}
\usepackage{cleveref}
\usepackage{makecell}
\usepackage{xspace}

\usepackage[makeroom]{cancel}

\usepackage{lipsum}
\renewcommand{\theequation}{\arabic{equation}}

\numberwithin{equation}{section}
\renewcommand{\theequation}{\arabic{section}.\arabic{equation}}

\definecolor{XLgreen}{RGB}{34,139,34}
\definecolor{JMblue}{RGB}{25,25,125}
\definecolor{BSorange}{RGB}{140,50,0}
\definecolor{RCpurple}{RGB}{128,0,128}

\newcommand{\<}{\langle}
\renewcommand{\>}{\rangle}

\def\rec{{\text{rec}}}

\begin{document}

\title{Towards Entanglement Bootstrap for Conformal Field Theory in Any Dimension}

\author{Rolando Ramirez Camasca}
\affiliation{Department of Physics, University of California at San Diego, La Jolla, CA 92093, USA}
\author{Xiang Li}
\affiliation{Department of Physics, University of California at San Diego, La Jolla, CA 92093, USA}
\author{Ting-Chun Lin}
\affiliation{Department of Physics, University of California at San Diego, La Jolla, CA 92093, USA}
\affiliation{Hon Hai Research Institute, Taipei, Taiwan}
\author{John McGreevy}
\affiliation{Department of Physics, University of California at San Diego, La Jolla, CA 92093, USA}

\begin{abstract}
Given a quantum critical wavefunction $\ket{\psi}$ in any dimension, we propose a 
reconstructed Hamiltonian
$H_\text{rec}(\ket{\psi})$, analogous to the ones found for 1+1d CFT in \cite{Lin:2023pvl} and for 2+1d bosonic liquid topologically-ordered states in 
\cite{Kim:2024amo, Li:2025sbv}.
We test numerically that, for known regularized approximate CFT groundstates (on the icosahedron and the fuzzy sphere), 
(1) they are close to the groundstate of their $H_{\rec}$, and 
(2) the spectrum of their $H_{\rec}$ on the unit sphere has CFT properties (integer spacing of descendants) and matches known low-lying energies.
We show that this provides an automated method to improve the finite-size effects in a fixed Hilbert space.

\end{abstract}

\date{\today}

\maketitle
\let\savedaddcontentsline\addcontentsline
\renewcommand{\addcontentsline}[3]{}

\section{Introduction}

Fixed points of the renormalization group (RG) are rare and precious.
The experimental phenomenon of universality is explained precisely by how rare and precious they are.
Thus, a natural goal in theoretical physics is to find a way to systematically enumerate and understand fixed points of the RG.
This goal includes the high energy question of understanding the space of quantum field theories, which can be defined as RG flows between conformal field theories (CFTs).
It also subsumes the condensed matter or statistical mechanics question of classifying phases of matter (basins of attraction of stable fixed points) and the continuous transitions between them.

Existing methods for finding RG fixed points are not yet systematic.  Besides experiment, numerical and otherwise, our knowledge of fixed points comes from essentially only three places: 
First is perturbation theory around free fixed points, perhaps controlled by some large number of species or by proximity to an upper critical dimension \cite{Wilson:1971dc}.  
Second is special constructions using supersymmetry and string theory.  Third is the conformal bootstrap \cite{Poland:2018epd}, which, so far, provides amazingly detailed information about CFTs that are close to violating some simple conditions from unitarity, and no information at all about others, which reside safely inside the so-called `continent' of thus-far allowed behavior (though see \cite{Erramilli:2026ykj}).

The Entanglement Bootstrap offers a route toward such a systematic enumeration. 
The Entanglement Bootstrap is a program to understand the universal properties of quantum many-body states.  
By a `universal property' we mean a property of a fixed point, and therefore of its whole basin of attraction under the RG.
The idea of Entanglement Bootstrap is to extract this universal information from the local entanglement structure of a single representative wavefunction.  The key step is to identify conditions on the local entanglement structure that characterize an RG fixed point.  These conditions can then be used as axioms to prove structural properties of the universal data, or as cost functions to numerically minimize on the space of states in order to systematically find such fixed points.

The thus-far-most-well-developed aspect of Entanglement Bootstrap is its application to liquid gapped phases 
\cite{Shi:2019mlt, Shi:2020jxd, Shi:2019ngw, Shi:2018krj, Shi:2018bfb, Shi:2020rne, knots-paper, Kim:2021tse, Kim2021, Shi:2023kwr, Kim:2023ydi, Kim:2024amo, Vir, kim2024conformalgeometryentanglement, no-go, figure-eight, Kim:2024gtp, Yang:2025pke, Li:2025sbv}. 
Gapped implies that the correlation length is finite in the thermodynamic limit.  
Liquid means that the low-energy properties are not sensitive to small changes in the lattice.  
Under these assumptions, we expect the low-energy physics to be governed by a topological quantum field theory (TQFT).  
In 2+1 dimensions, much of the structure of TQFT can be derived from two simple axioms on the entanglement in the groundstate
\cite{Shi:2019mlt}.
The strategy also works for liquid gapped bosonic states in 3+1 dimensions and higher, where the topological excitations whose data provides universal labels on the state are more diverse than just particles \cite{knots-paper, Shi:2023kwr}. 

A current frontier for Entanglement Bootstrap is its extension to gapless states.  
We proposed an analogous numerically-robust condition for a 1+1 dimensional CFT groundstate called the vector fixed-point equation (VFPE) \cite{Lin:2023pvl}. 
The fact that a CFT groundstate satisfies the condition follows from known properties of the CFT groundstate modular Hamiltonian of an interval \cite{Casini:2011kv, Cardy:2016fqc}.  
But this very strong condition is satisfied by quantum critical lattice groundstates, to a better and better approximation as the system size grows.  
\cite{Li:2025czz} used this condition to perform an unbiased search for 1+1d CFTs.

We are emboldened by the fact that the logic behind the criterion proposed in \cite{Lin:2023pvl} does not depend on the infinite-dimensional nature of the 1+1d conformal algebra.
In this paper, we attempt a similar series of steps for CFTs in $D > 1+1$ spacetime dimensions.  

In the following, we first remind the reader about the known entanglement structure of a CFT groundstate. Then we use this to design a linear combination of entanglement Hamiltonians whose variance should vanish in a CFT groundstate and from which the short-distance (UV) contributions cancel.  
This combination of operators also acts as a reconstruction of the CFT Hamiltonian to extract the spectrum of scaling dimensions from the groundstate with no further input. 
We then successfully test this construction numerically in various regularizations of various CFTs.  Most interestingly, we show that this information can be used to systematically improve upon a given regulated CFT groundstate at fixed system size.

\section{Entanglement of CFT groundstates}

$S^d$ is a special spatial manifold for CFT because of radial quantization: the discrete spectrum of the Hamiltonian on $S^d$ is $ \{ \Delta_i/R \}$ where $R$ is the radius of the sphere, and $\Delta_i$ are the scaling dimensions.  Thus, we will focus on this choice of IR regulator.  

Consider a CFT groundstate $\ket{\psi}$ in any dimension $D=d+1$; for a {\it round} ball $A$ of radius $r_A$, the entanglement Hamiltonian $K_A \equiv - \log\rho_A$ from the reduced density matrix $\rho_A \equiv \Tr_{\overline{A}} \ketbra{\psi}{\psi}$\footnote{Here $\overline{A}$ denotes the complement of $A$.} takes the form \cite{Casini:2011kv, Cardy:2016fqc}
\be \label{eq:K-of-CFT} K_A^\text{CFT} = \int d^d x \beta_A(x) T_{00}(x) + \Ione f_A ~.~~~  \parfig{.2}{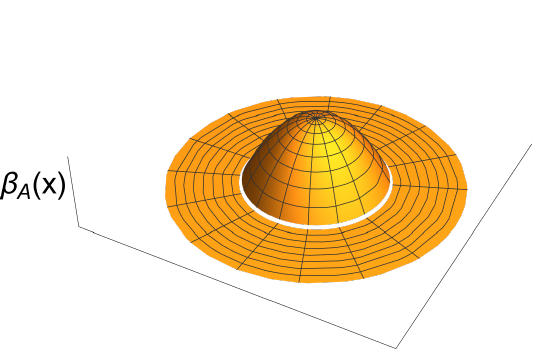} \ee
where 
for a state on $\IR^d$, $\beta_A(x) = 2\pi ( r_A^2 - r^2 ) /  (2 r_A ) \Theta(r < r_A)$ with $r = \sqrt{\vec x^2}$ is the polar coordinate centered at the center of $A$, while for a state on $S^d$ of radius $R$, 
\be \beta_A(x) = \frac{R}{\pi} \frac{\sin((\theta_A - \theta)/2) \sin((\theta_A + \theta)/2)}{\sin(\theta_A)}  \Theta(\theta < \theta_A) \ee
where $\theta$ is the polar angle from the center of the ball $A$ to its edge.  

The expectation value of $K_A^\text{CFT}$ in the CFT groundstate is the von Neumann entropy of $\rho_A$ 
\be S_A \equiv - \Tr \rho_A \log(\rho_A) = \vev{\psi | K_A^\text{CFT} | \psi} ~,\ee
which equals $f_A$ when $d$ is even or in $\IR^d$.
For $d=2$, this is 
\be \label{eq:SA}
S_A = |\partial A|/\eps - F \ee
where $|\partial A|$ is the length of the boundary of $A$, $\eps$ is a UV cutoff, and $F$ is the RG monotone \cite{Casini:2012ei,Casini:2015woa}.  

\section{Reconstructed Dilation operator and a vector fixed-point equation}
Given a CFT groundstate $\ket{\psi}$, we can look for linear combinations of its entanglement Hamiltonians $\sum_A \lambda_A K_A$ of different round balls $A$ that add up to the CFT dilation operator $D^\text{CFT} = \int d^d x T_{00}(x)$. Following the strategy of \cite{Lin:2023pvl}, we can choose $\sum_{A} \lambda_{A} \beta_{A}(x) = 1$\footnote{Recently, a very formal and non-local version of a related idea was discussed in the algebraic QFT literature \cite{Chen:2025aen}.}.

How is the linear combination designed? If the given state $\ket{\psi}$ is the ideal CFT groundstate, where $K_A^{\psi}$ takes the form of \eqref{eq:K-of-CFT}, we could simply average the $K_A$ over the whole sphere:
\begin{equation}\label{eq:ave-KA}
    \overline{K_A} = \frac{1}{\mathrm{Vol}(S^d)} \int_{S^d} d^d x K_{A(x)} ~,
\end{equation}
where $A(x)$ is a round disk of radius $r_A$ centered at $x$. The result of the averaging gives exactly $D^\text{CFT}$ up to a prefactor 
\begin{equation}\label{eq:VA}
    V_A = \frac{1}{\mathrm{Vol}(S^d)}\int d^d x \beta_A(x).
\end{equation}
In 2d unit sphere, the prefactor is $V_A = 2 \pi\sin^4(\theta_A/2)/\sin(\theta_A)$.

However, the CFT groundstate in practice can only be realized with some UV regulation scheme such as on a lattice. The UV regulation inevitably adds short-distance entanglement to the state $\ket{\psi}$ and hence we expect that the entanglement Hamiltonian is of the form $K_A = K_A^\text{CFT} + K_A^\text{UV}$.
The extra UV term, since it captures the short-distance entanglement, is localized at the boundary of $A$; we will discuss it in more detail momentarily. 
Thus, we want to further take linear combinations of multiple averaged entanglement Hamiltonians Eq.~\eqref{eq:ave-KA} to obtain the reconstructed Dilation operator 
\begin{equation}\label{eq:Hrec}
    D_{\rec} = \sum_{i=1}^k \lambda_i \overline{K_{A_i}} ~.
\end{equation}
The coefficients $\{\lambda_{A_i}\}_i$ are chosen to cancel the UV contributions $\{K_A^{UV}\}_A$. 


{\bf A model of the UV contributions.}  
Here we present a model of the contributions to $K_A$ from short-range-entangled (SRE) degrees of freedom, in the spirit of the analysis of \cite{Grover:2011fa} of the analogous problem for the entanglement entropy (EE). In that problem, the authors take a hydrodynamic approach and parametrize the SRE contribution to the EE of a region whose boundary varies smoothly on the scale of the UV cutoff as 
a derivative expansion, 
$S_A^\text{SRE} = \int_{\partial A} \left( a + b \kappa + c \kappa^2 + \cdots \right) $ (in $d=2$).
They then point out that purity of the whole state implies $S_A = S_{\bar A}$, 
while the exchange of inside and outside of $A$ reverses the sign of the extrinsic curvature $\kappa$, and thus the coefficient $b=0$ (and this is why the TEE, the term of order $|\partial A|^0$, can be universal).  

We propose an analogous operator-valued hydrodynamic expansion, again localized to the boundary of the region, for the UV contribution to the regulated entanglement Hamiltonian: 
\be\label{eq:SRE-K_A} K_A^\text{UV} = \int_{\partial A} d\ell \( \CO^{\hat n}_1(\ell) + \kappa \CO^{\hat n}_2(\ell) + \cdots \), \ee
where $\hat n$ is the direction of the normal to $\partial A$ at $\ell$.  
This model predicts, perhaps surprisingly, that $k=2$ is enough to remove the UV dependence when the operator acts on $\ket{\psi}$ and gives us a vector equation 
\begin{equation}
    D_{\rec}\ket{\psi} = \langle D_{\rec} \rangle \ket{\psi} ~.
\end{equation}
In App.~\ref{app:analysis-of-the-UV}, we will give a more detailed analysis of the cancellation of the UV operators in $D_{\rec}$ for various $k$ in Eq.~\eqref{eq:Hrec}. 
The general conclusion is that well-chosen linear combinations of entanglement Hamiltonians have a well-defined continuum limit. In particular, for $d=2$ with $k = 2$, the operator $D_{\rec}$ will converge to $D^{\text{CFT}}$ as $R \to \infty$.

In 2+1D, the choice of $\lambda_1,\lambda_2$ in Eq.~\eqref{eq:Hrec} that cancels the UV dependence in $D_{\rec}$ follows the rule 
\begin{equation}\label{eq:cancel-rule}
    \frac{\lambda_1}{\lambda_2} = - \frac{|\partial A_2|}{|\partial A_1|} ~.
\end{equation}
The overall factor is fixed by demanding 
\begin{equation}\label{eq:overall}
    \lambda_1 V_{A_1} + \lambda_2 V_{A_2} = 1,
\end{equation}
where $V_A$ is given in Eq.~\eqref{eq:VA}.

\begin{figure}
    \centering
    \includegraphics[width=0.5\linewidth]{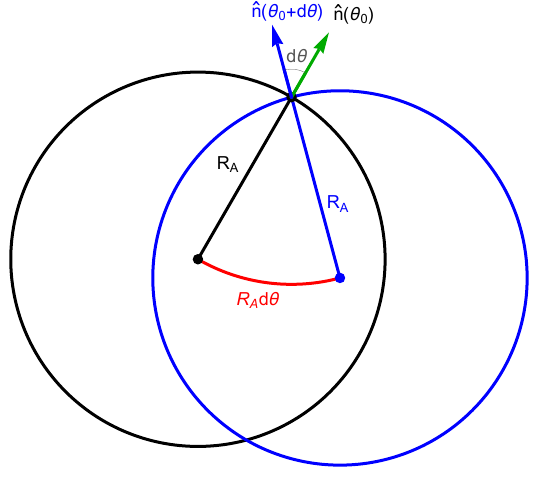}
    \caption{\justifying The geometry determining the coefficient of $\CO^{\hat n}(x)$ in $\overline{K_A}$.}
    \label{fig:coefficient-of-On}
\end{figure}

So with this choice Eq.~\eqref{eq:cancel-rule}, the area law terms cancel out in $\vev{D_\text{rec}}$.  But furthermore, the coefficient of the operator $ \CO_1^{\hat n}$ in $D_\text{rec}$ is zero.  To see this, consider the contribution to the coefficient of the operator at a point $\Omega$ with normal vector near $\hat n$ in $\overline{K_A}$.  In Fig.~\ref{fig:coefficient-of-On} we show nearby regions $A$ that contribute. The crucial point is that the measure is $ R_A d\theta $. Therefore, the coefficient of $ \CO_1^{\hat n}(x)$ after the averaging is proportional to $R_A \propto |\partial A|$. Thus, the choice Eq.~\eqref{eq:cancel-rule} also cancels the leading UV operator contribution.   

Another useful way to fix the overall factor is to demand the expectation value is $F$ in Eq.~\eqref{eq:SA}. We can define
\begin{equation} \label{eq:KF}
    K_{F}  \equiv \lambda_1^F \overline{K_{A_1}} + \lambda_2^F \overline{K_{A_2}} ~,
\end{equation}
such that $\lambda_1^F,\lambda_2^F$ not only satisfies Eq.~\eqref{eq:cancel-rule} but also makes $\langle K_F \rangle = F$. The latter requires $\lambda_1^F + \lambda_2^F = -1$. Notice the operator content in $K_F$ is proportional to $\int d^d x T_{00}(x)$, and so we have a \emph{vector fixed point equation} (VFPE)
\begin{equation}
    K_F \ket{\psi} = F \ket{\psi} ~.
\end{equation}
We will confirm this numerically by computing the error of this VFPE 
\begin{equation}
    \text{err}(\psi) \equiv \sigma(K_F) = \| K_F\ket{\psi} - \langle K_F \rangle \ket{\psi} \|. 
\end{equation}
In the following sections, we test this claim numerically in well-regularized realizations of various 2+1d CFTs.

We note that the construction described above only relies on the fact that in the CFT groundstate $K_A$ is a single integral over $A$ of some function times the Hamiltonian density, and not on the precise form of $\beta(x)$.  
As in \cite{Lin:2023pvl}, we could also look for local linear combinations $\sum_A \lambda_A K_A$ such that the operator part cancels:
$ \sum_A \lambda^F_A \beta_A(x) = 0 $. 
This strategy with a finite number of balls inside a local region has not yet been successful.

\section{Icosahedron}

We wish to study a quantum critical spin model on a discretization of the 2-sphere that preserves as much as possible of the $\gSO(3)$ symmetry.  The simplest such objects are the platonic solids, among which we focus on the icosahedron.  
The icosahedral group is a large subgroup of $\gSO(3)$ in the sense that the dimensions of its representations are large enough to keep the finite size effects quite small \cite{Lao:2023zis}.
The only shortcoming of this method is that it is difficult to systematically improve via  finite-size scaling.

\begin{figure}[!ht]
    \centering
    \includegraphics[width=0.48\linewidth]{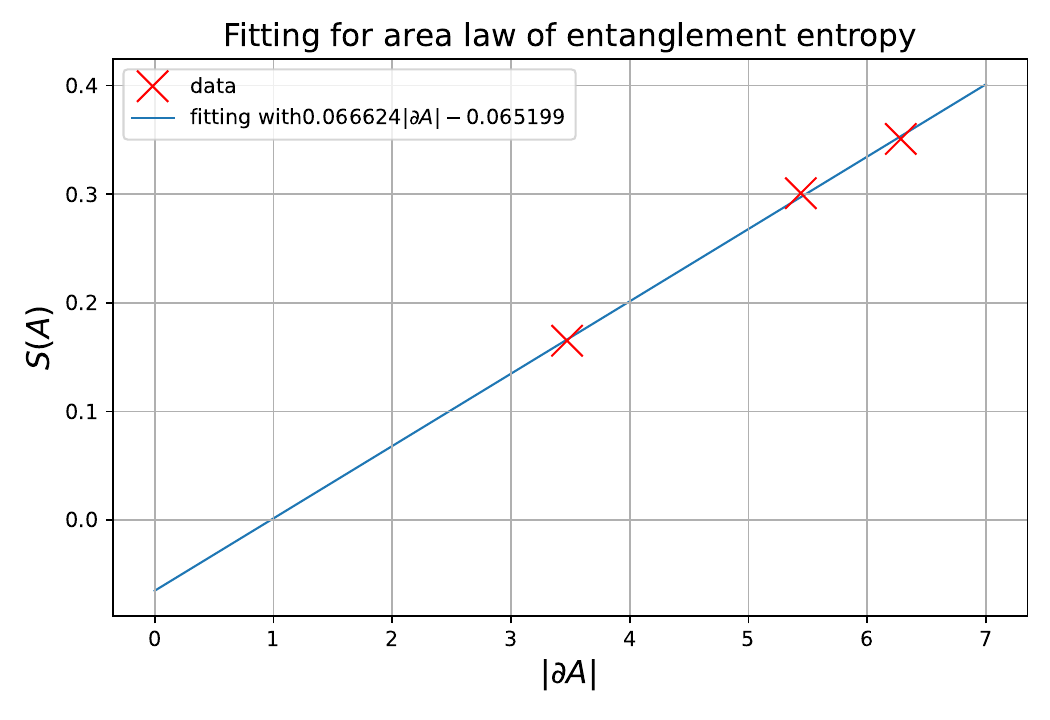}
    \includegraphics[width=0.48\linewidth]{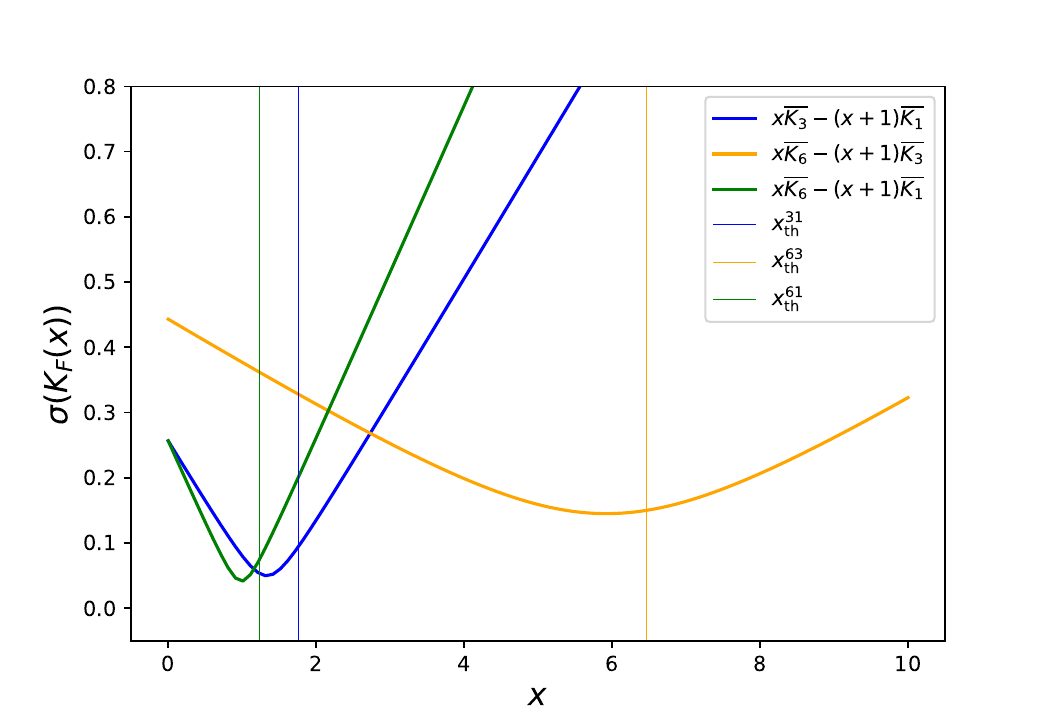}
    \caption{\justifying Left: A fit to the area law for the entanglement entropy of subsystems in the groundstate of the critical Ising model on the icosahedron ($J=1, h=4.375)$. Right: The error of the VFPE versus $x$.}
\label{fig:ising-F}
\end{figure}

For definiteness we study the transverse field Ising model (TFIM) on the icosahedron, $ H_\text{TFIM} = - \sum_{\vev{ij}} Z_i Z_j + h_x \sum_i X_i$.  which has been shown to provide a remarkably good approximation to the continuum CFT at $h_x=4.375$ \cite{Lao:2023zis}.  
First, in \Cref{fig:ising-F} we show that the groundstate of the critical Ising model on the icosahedron satisfies the area law for the entanglement entropy, $S_A = \vev{K_A} = |\partial A|^\text{eff}/\eps - F $. 
We define the boundary area 
by 
$|\partial A|^\text{eff} = 2 \pi \sin(\theta_A^\text{eff})$ with the effective angle $\theta_A^\text{eff}$ determined by the ratio of the area 
$q_A =|A|/ (4\pi)$ with $|A| = \int_A d\Omega = 2 \pi (1 - \cos(\theta_A^\text{eff}))$. The result is 
\begin{equation} \label{eq:theta-eff}
    \theta_A^\text{eff} = \cos^{-1}(1-2q_A).
\end{equation}
For icosahedron, we can use 
\begin{equation}
    q_A = \frac{\text{\# of sites in }A}{\text{total \# of sites}}.
\end{equation}
The result gives a surprisingly good value of the RG monotone $F \approx 0.0652$ [Fig.~\ref{fig:ising-F} (left)].  

Now we study $K_F$ defined in Eq.~\eqref{eq:KF}. To test the UV cancellation,
we can parametrize $\lambda_{A_1} = -(1+x)$ and $\lambda_{A_2} = x$ to express $K_F = x \overline{K_{A_2}} - (1+x) \overline{K_{A_1}}$.
We determine the remaining parameter $x$ to cancel the area law contributions, $x_\text{th} = \sin\theta_{A_1}^\text{eff}/(\sin\theta_{A_2}^\text{eff} - \sin \theta_{A_1}^\text{eff})$. 
\Cref{fig:ising-F} (right) shows that the variance of $K_F$ (whose square root we will call the error of the VFPE) is minimized by this choice $x  = x_\text{th}$,
for various choices of $A$ and $B$.

\begin{figure}[!ht]
    \centering
    \includegraphics[width=0.48\linewidth]{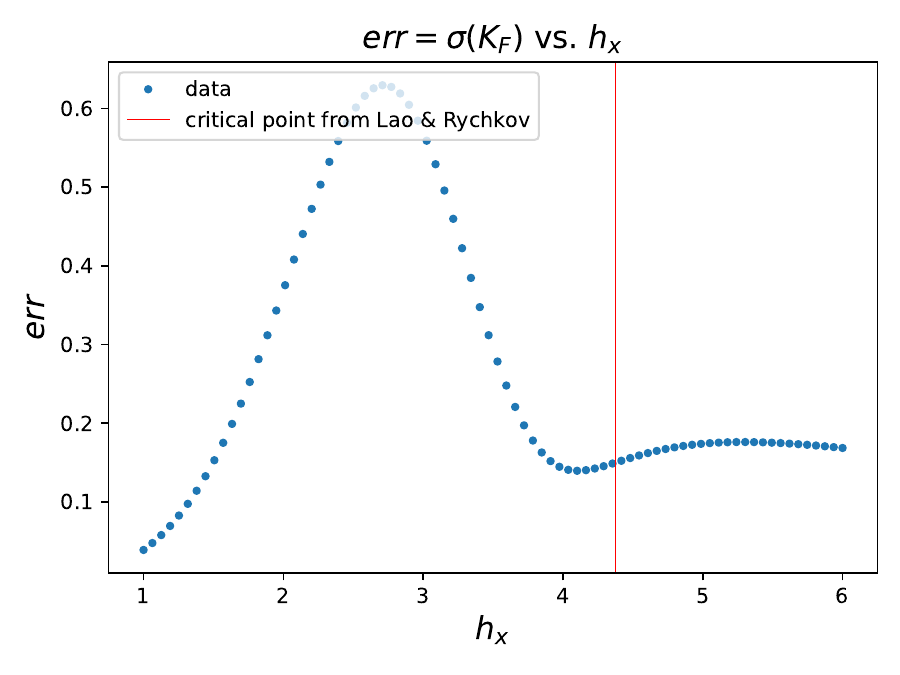}
    \includegraphics[width=0.48\linewidth]{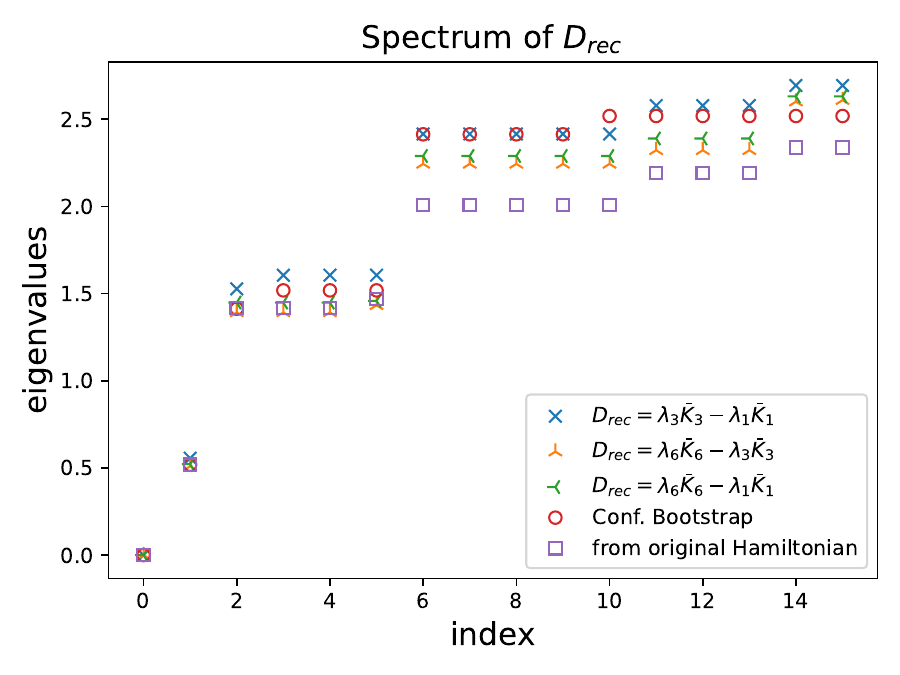}
    \caption{\justifying Left: The error of the VFPE in the groundstate of the transverse-field Ising model on the icosahedron as a function of the transverse field.  Right: The spectrum of the reconstructed Hamiltonian from the critical groundstate, for various choices of $A, B$, compared with the spectrum of the TFIM on the icosahedron at $h = 4.375$ and the conformal bootstrap spectrum \cite{El-Showk:2012cjh, El-Showk:2014dwa, Kos:2016ysd}. 
}
\label{fig:ising-err-and-spectrum}
\end{figure}

As in $D=1+1$ \cite{Li:2025czz}, one can use the error of the VFPE to detect phase transitions. 
In \Cref{fig:ising-err-and-spectrum} (left), we show the error of the VFPE computed in the groundstate of the TFIM on the icosahedron, versus the transverse field $h$. In constructing $K_F$, we are averaging entanglement Hamiltonians on 6 sites and 3 sites, and using the theoretical prediction $x_\text{th}$ for the coefficients.

In \Cref{fig:ising-err-and-spectrum} (right), we show the spectrum of the reconstructed Dilatation operator, compared with the spectrum of scaling dimensions from $H_\text{TFIM}$ and conformal boostrap. 

\section{Fuzzy sphere}

A recent breakthrough in regularizing 2+1d QFT, 
building on previous work using a fuzzy torus \cite{2018PhRvB..98w5108I, 2021PhRvL.126d5701W}, is the use of the fuzzy sphere \cite{Zhu:2022gjc,Hofmann:2023llr,Hu:2023xak,Zhou:2023qfi,Hu:2024pen,Voinea:2024ryq,Zhou:2024zud,He:2025ong,Zhou:2025rmv,Zhou:2025kng,Guo:2025odn,ArguelloCruz:2025zuq,Taylor:2025odf,Voinea:2025iun,Zhou:2025kng,Tang:2025wtj,dey2026:O3}.  
The big advantages of the fuzzy sphere are that it exactly preserves the $\gSO(3)$ spatial rotation symmetry while having small overlap between basis functions, and that the size of the single-particle Hilbert space grows linearly with the control parameter, which is the number of units of magnetic flux piercing the sphere.  
We briefly review the procedure in the Appendix \ref{app:fuzzy-review}.  

In this section, we will present the results for Ising CFT realized in \cite{Zhu:2022gjc}. The results for the other CFTs are in 
App.~\ref{app:other-CFTs}. When computing the entanglement Hamiltonian $K_A$, we will use orbital cut in this section. The results for real space cut are given in App.~\ref{app:real-space-cut}. 

First, we did a verification of the choice of the coefficients Eq.~\eqref{eq:cancel-rule} for $K_F$ in Eq.~\eqref{eq:KF}, analogous to \Cref{fig:ising-F}. To compute the theoretical prediction in terms of $|\partial A|$, we use the same strategy $|\partial A| = 2\pi R \sin(\theta_A^\text{eff})$ with $\theta_A^\text{eff}$ in Eq.~\eqref{eq:theta-eff} and $q_A = |A|/L$,
where $|A|$ stands for the number of orbitals in $A$ and $L$ is the total number of orbitals.

\begin{figure}[!ht]
    \centering
    \includegraphics[width=0.95\linewidth]{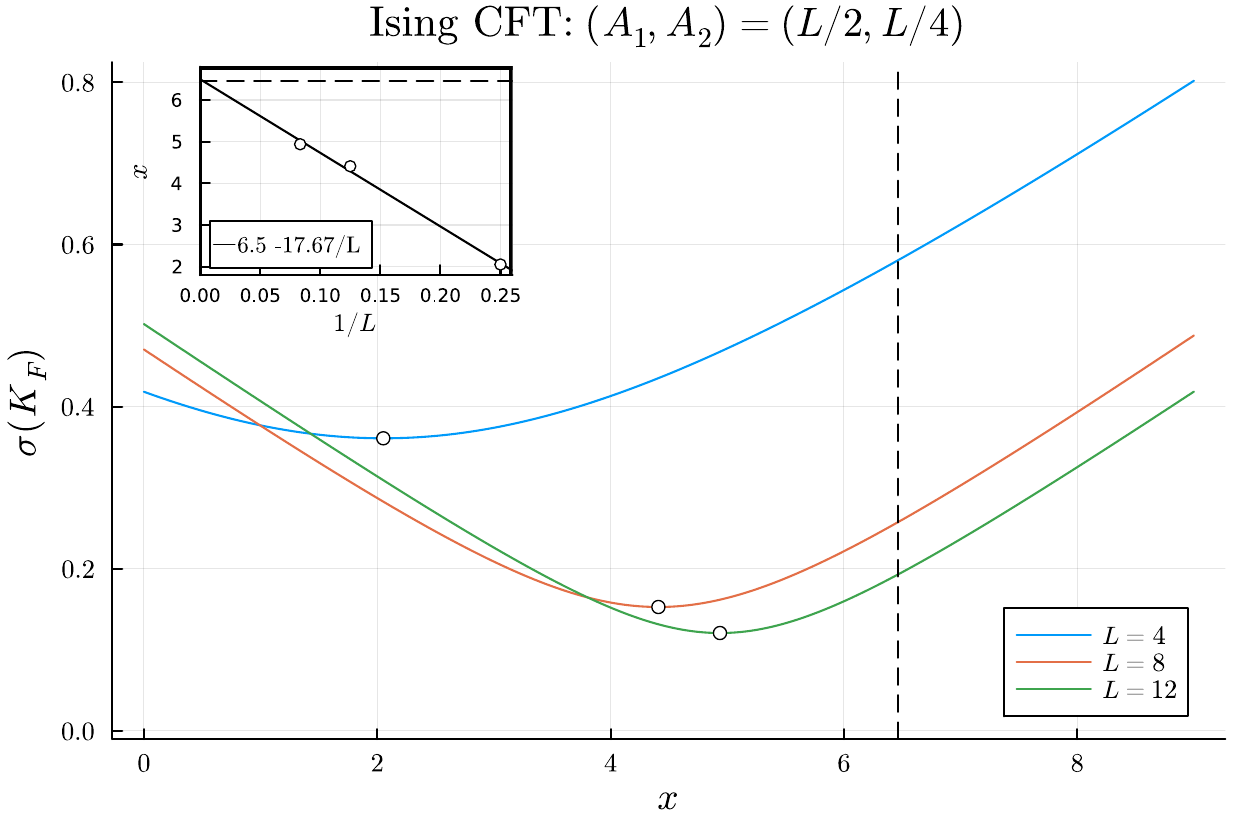}
    \caption{\justifying Error of the VFPE versus $x$ in $K_F \equiv x \overline{K_{A_2}} - (1+x) \overline{K_{A_1}}$ for the fuzzy sphere critical Ising model groundstate.
    We fix the ratio of subregion sizes $|A_2|/|A_1|$ and vary the total number of orbitals $L$.  
The ratio we choose is $|A_2|/|A_1| = 2$ with $A_1 = L/4$ and $A_2 = L/2$ for system sizes $L=\{4,8,12\}$. As we increase system size, the minima of the variance (white dots in the curves) approach the theoretical prediction (dashed vertical black line).
For the chosen ratio, the theoretical prediction is $x_\text{th} = 1/(\sqrt{4/3} -1) \simeq 6.46$ (dashed black line).
Note furthermore that the value of the minimal variance decreases as we increase the system size.
In the inset, we plot the $x$ value from the numerical minima of each of the curves as a function of $1/L$. We fit such points linearly and obtain $6.5 - 17.67/L$.  The extrapolated $y$-intercept is indeed close to the theoretical prediction $x_{th}$. 
}
    \label{fig:Ising-error-VFPE-parameters}
\end{figure}

In \Cref{fig:Ising-error-VFPE-parameters} we show the error of the VFPE for $k=2$ regions as a function of their relative coefficients. 
The group-theoretic method to do the averaging over angles is described in the Appendix \ref{app:Constructing-irreps}.  
As we increase the system size while fixing the subregion size ratio $|A_2|/|A_1|$, the minimum of the error indeed approaches the predicted value $x_\text{th}$.

Then we verify the spectrum of the dilatation operator $D_{\rec}$ in \eqref{eq:Hrec} for $k = 2$ with the coefficients fixed by \eqref{eq:cancel-rule} and \eqref{eq:overall}.
The reconstructed spectrum matches fairly well the conformal bootstrap values with a \emph{zero}-parameter fit. 
As we increase the system size, the agreement improves.  
Notice that the form of $\alpha\( \{ A_i \}\)$ does depend on the form of the coolness function $\beta_A$ in \eqref{eq:K-of-CFT}. 

\begin{figure}[!ht]
    \centering
    \includegraphics[width=\linewidth]{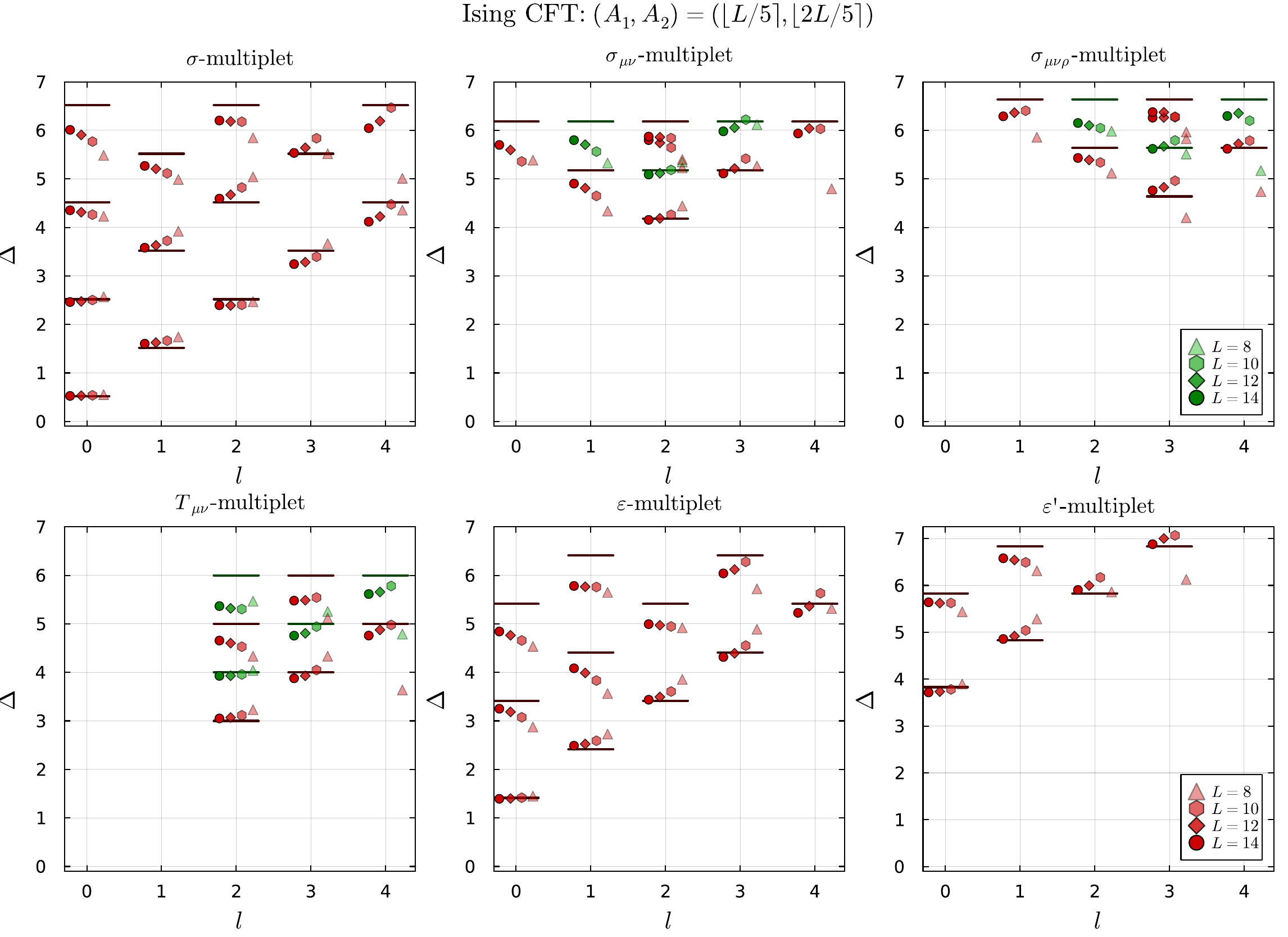}
    \caption{\justifying The spectrum of the reconstructed Hamiltonian from the fuzzy sphere critical Ising model groundstate versus angular momentum.  We fix the subregion sizes to scale with the system size as $(A_1,A_2) = (\left\lfloor L/5 \right\rceil, \left\lfloor 2L/5 \right\rceil )$, where $\left\lfloor \ldots \right\rceil$ denotes the nearest integer, and plot four system sizes $L= \{8,10,12,14\}$ with different markers and transparency.
    The horizontal lines indicate the conformal bootstrap values of the spectrum \cite{El-Showk:2012cjh, El-Showk:2014dwa, Kos:2016ysd}. 
    The two different colors (red and green) specify whether the state is particle-hole even or odd respectively. Notice that the state in the $\sigma_{\mu\nu}$-multiplet with $l=2$ and the state in the $\sigma_{\mu \nu \rho}$-multiplet with $l=3$ are doubly degenerate, which we indeed find. 
    }
    \label{fig:Ising-Hrec-k2}
\end{figure}

\begin{figure}
    \centering
    \includegraphics[width=0.95\linewidth]{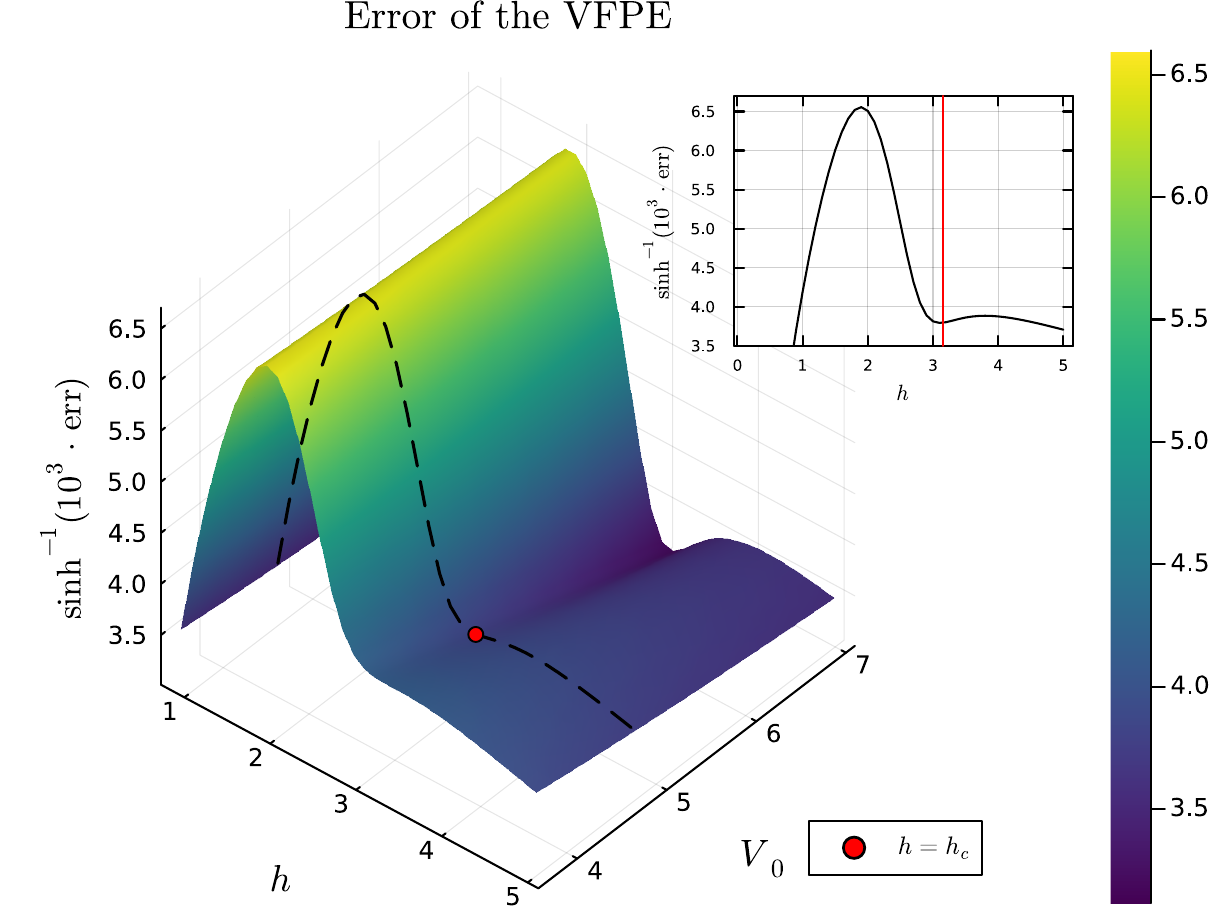}
    \caption{\justifying Heatmap of the error of the VFPE $\sigma(K_F)$ evaluated on the groundstate of the Ising Hamiltonian with system size $L = 12$, as the function of the couplings $(h,V_0)$. We fix the subsystem sizes to $(A_1, A_2)=(2,5)$, matching the choice used in the reconstructed spectrum. We used $\sinh^{-1}(10^3 \cdot \text{err})$ to make the shape more visible. The inserted subplot is a particular slice of $V_0 = 4.75$ shown by the dash line. A clear local minimum is visible near $h \simeq h_c$.}
    \label{fig:Ising-log-error-k2}
\end{figure}

Next (Fig.~\ref{fig:Ising-log-error-k2}), we explore the phase diagram of the transverse field Ising model Hamiltonian.  We vary both the transverse field and one parameter specifying the pseudopotentials, and study the error of the VFPE using the theoretically-predicted optimal value of the parameters $\lambda_A$.  We find that there are three regions of local minima: one in the ferromagnetic phase at small $h$, one in the paramagnetic phase at large $h$, and one in the neighborhood of the previously-identified Ising critical point.  

\section{Systematic improvement}

Suppose we are given the numerical groundstate of an arbitrary quantum critical Hamiltonian (perhaps the one people like best) as a finite-size approximation to a CFT groundstate. 
As we have seen in the examples above, the VFPE will have a finite error.  By doing gradient descent in the space of states (recall that this space is compact) on the error of the VFPE, we can 
reduce the error, which may correspond to a better finite-size realization of the CFT.

\begin{figure}[!ht]
    \centering
    \includegraphics[width=\linewidth]{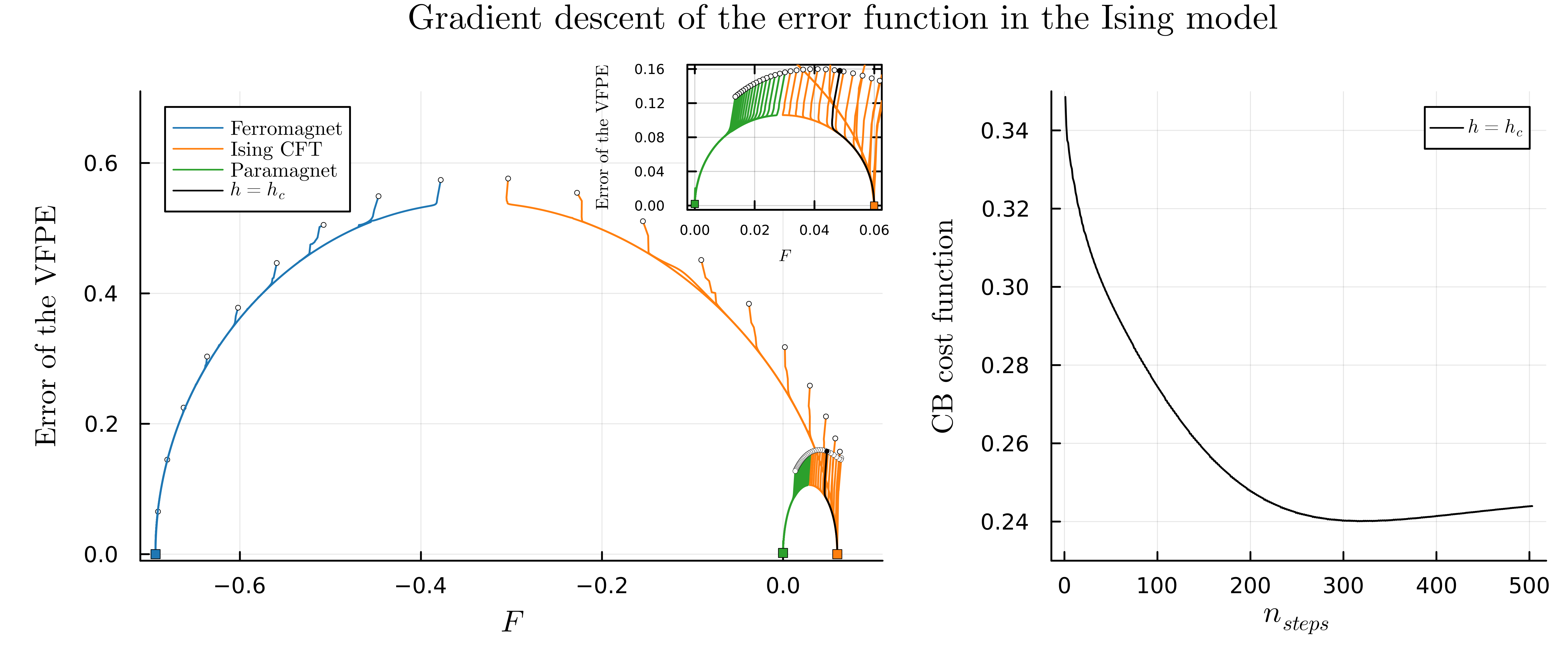}
    \caption{\justifying Left plot: Gradient descent trajectories of the error of the VFPE starting from the transverse-field Ising model groundstates, for $L=6$ orbitals. 
The abscissa is the value $F = \langle K_F \rangle$. 
We choose $k=2$ subregions $(A_1, A_2) = (1,2)$ and use the theoretically-predicted values of $\lambda_A$.  
    Initial states of the gradient descent are the ground states of the Ising Hamiltonian at $(V_0 = 4.75, h)$, 
    with 50 equally spaced values of $h \in [0.1, 6]$, represented by the white circles.  Final states are colored squares, colored according to their final $F$ value.     
    As $h$ increases from the ferromagnetic regime, the initial points move along the large arc from left to right.
    For small $h$ the gradient descent flows to the ferromagnetic minimum~(blue curves). 
    The groundstate here is a cat state, and this explains the value $F = - \log 2$; if we apply a training field to break the Ising $\IZ_2$ symmetry, we instead find $F=0$.  
    Approaching the critical field $h_c$, the initial points enter the basin of attraction of the Ising CFT and the trajectories converge to this fixed point~(orange curves). For larger $h$, the large arc folds in the plane, so that the ordering in $h$ continues along the small arc~(inset plot) from right to left. 
    As we continue to increase $h$, the trajectories ultimately flow to the paramagnetic minimum~(green curves).
    We also highlight the trajectory of the special critical value $h_c = 3.16$ (black curve). 
    Inset to the left plot: Zoom in to the $F$-function range between $F \in [0,0.06]$. We see that there exist forbidden regions when the error of the VFPE is small. 
    Right plot:  The cost function \eqref{eq:cost-function-def} (with $\Delta_\text{max} = 3$)
for the reconstructed spectrum for each point along the gradient descent, starting from the critical Ising groundstate.
    We observe that this cost function begins to increase slightly toward the end of the descent procedure.
    }
    \label{fig:Ising-GD-WS}
\end{figure}

One measure of improvement of the spectrum is a cost function measuring how close the low-lying spectrum is to a representation of the conformal algebra and to the known low-lying spectrum, as introduced in \cite{Fan:2024vcz,Fan:2025bhc}.  
In \Cref{fig:Ising-GD-WS}, we show the evolution of such a cost function (see App.~\ref{app:more-results-for-ising} for the precise definition) during gradient descent on the error of the VFPE. The initial states are taken to be the groundstates of the transverse-field Ising model on the fuzzy sphere, for various values of the transverse field, and optimized choices of pseudopotentials \cite{Zhu:2022gjc}.  
We then perform gradient descent within the space of $\gSO(3)$ singlets. Note that since the initial state is $\IZ_2$ even and has particle-hole symmetry, the resulting trajectory preserves these symmetries.

From the figure, we see that solutions of the VFPE correspond to fixed points of the RG. 
These fixed points match the ferromagnet and paramagnet, as well as Ising CFT.  
Also, as in $D=1+1$ \cite{Li:2025czz}, we observe that there are forbidden regions in the space of error versus RG monotone where no trajectories exist. 
Most practically, doing gradient descent on the error of the VFPE starting from the critical point of a human-constructed lattice model provides a systematic way to improve the low-lying spectrum at a fixed system size.  

\begin{figure}[!ht]
    \centering
    \includegraphics[width=0.95\linewidth]{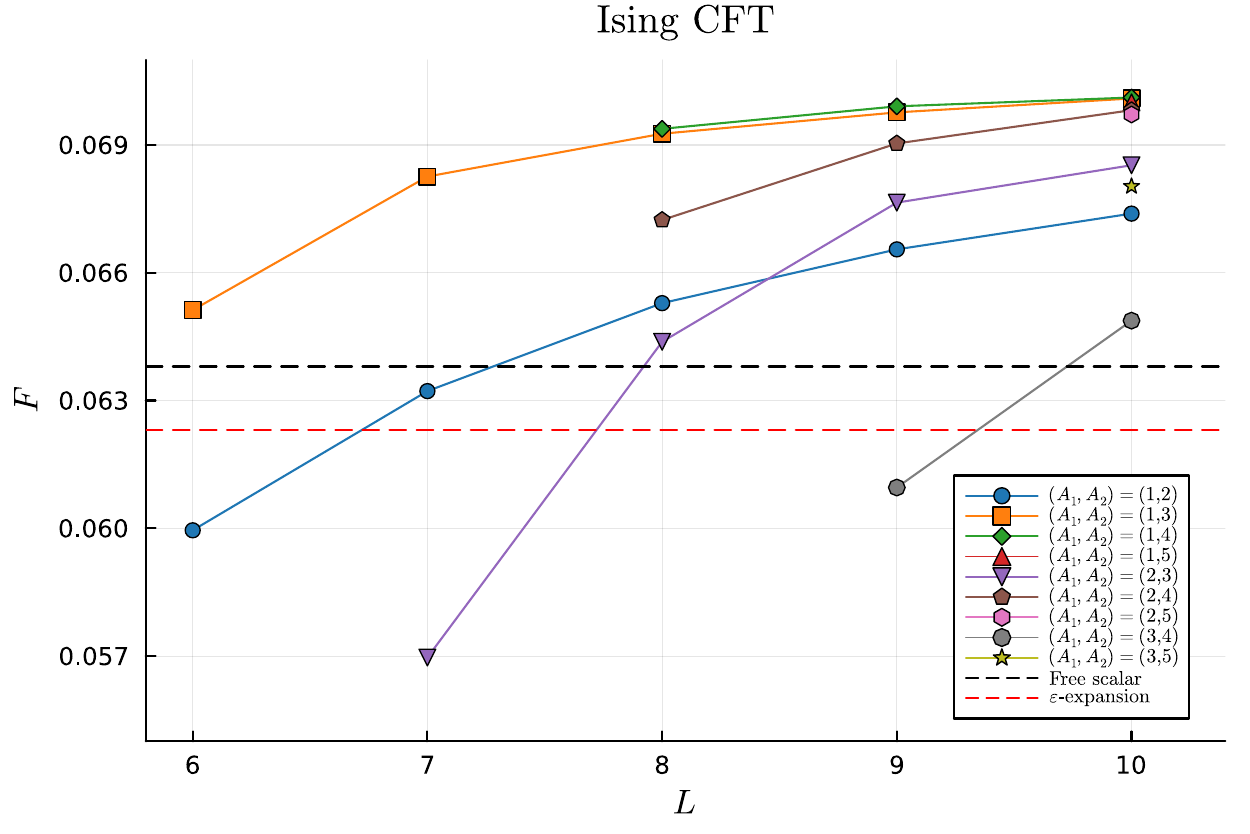}
    \caption{\justifying Converged values of the $F = \langle K_F \rangle$ function after minimizing the error of the VFPE. 
The initial state is the critical fuzzy sphere Ising model ground state at $(V_0, h_c) = (4.75, 3.16)$ for various system sizes $L$.
We run gradient descent on the error of the VFPE for $k=2$
(with various subregions $(A_1, A_2)$) until an error of 
$10^{-4}$, then switch to Newton's method and stop when the error of the VFPE is less than $10^{-8}$.
Horizontal dashed lines indicate the $F$-function for the free scalar~(black), and for the Ising CFT obtained via $\varepsilon$-expansion~(red).
}
    \label{fig:Ising-F-converged}
\end{figure}

Finally, we can consider the values of $F = \vev{K_F}$, and attempt to extrapolate them to large system sizes.  
This is the subject of \Cref{fig:Ising-F-converged}.  
The extrapolated values of $F$ depend on the choices of subregions used to define $K_F$.  As the system size grows, they approach a value closer to the approximate value computed using $\epsilon$-expansion.  However, they appear to be converging to a value that would violate the RG monotonicity of $F$ by a small amount, so we infer that some systematic error remains.  

In App.~\ref{app:ansatze}, we use these methods to study the analytical trial wavefunctions for the Ising and free scalar CFT groundstates of \cite{He:2025ong}.

The results of this systematic improvement of given CFT states using gradient descent on the error of the VFPE, and the similarity of Fig.~\ref{fig:Ising-GD-WS} with the results of the CFT search in $D=1+1$ \cite{Li:2025czz} make us optimistic that we can use these methods to identify new CFTs in $D=2+1$.

\begin{acknowledgments}

We thank Yin-Chen He and Zheng Zhou for helpful discussions.  Some of our numerics used the FuzzifiED package \cite{zhou2025fuzz}.  We used Julia. 
This work was supported in part by
funds provided by the U.S. Department of Energy
(D.O.E.) under cooperative research agreement
DE-SC0009919,
and by the Simons Collaboration on Ultra-Quantum Matter, which is a grant from the Simons Foundation (652264, JM).

\end{acknowledgments}

\clearpage
\let\addcontentsline\savedaddcontentsline 

\onecolumngrid 
\begin{center}
    \textbf{\large Table of Contents for Appendices}
\end{center}
\makeatletter
\print@toc{toc} 
\makeatother
\vspace{2.5em}
\twocolumngrid 

\appendix
\renewcommand{\theequation}
{\Alph{section}.\arabic{equation}}
 
\section{Details on the fuzzy sphere numerics}
\label{app:fuzzy-review}

In this appendix, we provide a brief review of the use of the fuzzy sphere as a regulator of 2+1d CFT (for more detail, see \cite{zhou2025fuzz}), and explain our numerical implementation of the entanglement bootstrap program in the fuzzy sphere.

{\bf Fuzzy sphere review.} 
Consider electrons of mass $M_e$ moving on the surface of a sphere of radius $R$ in the background of a monopole of charge $s$ placed at its center. 
The single-particle Hamiltonian reads
\be
    H_0 = \frac{1}{2M_eR^2} (\partial_\mu + i A_\mu)^2.
\ee
Its eigenstates are given by monopole harmonics with energies $E_n = [n(n+1) +(2n+1)s]/(2M_eR^2)$, where $n=0,1,2,\ldots$ is the Landau level index.

When the interaction scale is much smaller than the energy gap between Landau levels, we can project the system onto the lowest Landau level (LLL) $n=0$. This is spanned by the wavefunctions: 
\begin{equation}\label{eq:monopole-harmonics}
    \Phi_{m,s}(\Omega) = N_m e^{im\phi} \cos^{s+m}\left(\frac{\theta}{2}\right) \sin^{s-m}\left(\frac{\theta}{2}\right),
\end{equation}
with $m = -s, -s +1, \ldots, s$ and $N_m$ is a normalization factor. 
In second quantization, this is done by expressing the real-space operator in terms of the LLL operators via: 
\begin{equation}
    \psi_\sigma(\Omega) = \frac{1}{\sqrt{2s+1}} \sum_m \Phi_{m,s}^*(\Omega)~c_{m,\sigma}.
\end{equation}

This projection produces a finite-dimensional Hilbert space while preserving the full $\gSO(3)$ rotational symmetry of the sphere. This is the fuzzy sphere regularization. 
The method has been successfully applied to several critical models; see Refs.~\cite{Zhu:2022gjc,Hofmann:2023llr,Hu:2023xak,Zhou:2023qfi,Hu:2024pen,Voinea:2024ryq,Zhou:2024zud,He:2025ong,Zhou:2025rmv,Zhou:2025kng,Guo:2025odn,ArguelloCruz:2025zuq,Taylor:2025odf,Voinea:2025iun,Zhou:2025kng,Tang:2025wtj,dey2026:O3}. 

{\bf Entanglement in the orbital cut.}
We are interested in the reduced density matrix of a round ball, say centered at the North pole.  Because the orbitals \eqref{eq:monopole-harmonics} are localized around a fixed latitude determined by the $J^z$ quantum number $m$, we can approximate such a region as the support of the orbitals with the $l_A$ largest $m$.  The Hilbert space is indeed a tensor product in this basis: $ \CH = \CH_A \otimes \CH_{\bar A}$, where 
the north pole orbital space $\mathcal{H}_A$ is associated with the orbitals $A = \{s, s-1, \ldots, s-l_A+1\}$ and the south pole orbital space $\mathcal{H}_{\bar A}$ has orbitals $\bar A = \{s-l_A, \ldots, -s+1,-s\}$.
The Schmidt decomposition of a state in $\CH$ with respect to this bipartition is 
\be
    \ket{\psi} = \sum_{I,J} M_{IJ} \ket{I_{A}} \ket{J_{\bar A}},
\ee
where $\{\ket{I_A}, \ket{J_{\bar A}}\}$ spans the north/south pole spaces $\mathcal{H}_{A/\bar A}$, and we obtain the reduced density matrix by tracing out the $\bar A$ orbitals:
\be
    \rho_A = \Tr_{\hspace{-0.15cm}\bar A} \rho = {\bf M M^\dagger}. \label{eq:reduced-density-matrix-definition}
\ee

Because of symmetry, the matrix $M$ is block diagonal: each block carries quantum numbers corresponding to the Hilbert spaces $\mathcal{H}_A$ and $\mathcal{H}_{\bar A}$.
That is, 
\be
    M_{IJ} = \bigoplus_{\beta} M_{IJ}^{(\beta)},
\ee
where the $\beta$-sectors correspond to the quantum numbers of the subsystems. 
Which quantum numbers $\beta$ refers to depends on the model; we will discuss the example of the Ising critical point in detail below.
Using this fact, we can decompose the reduced density matrix into such blocks: 
\be
    \rho_A = {\bf M M^\dagger} = \bigoplus_{\beta} \rho_A^{(\beta)}. \label{eq:reduced-density-matrix-blocks}
\ee
Hence, we go sector-by-sector to do singular value decomposition and construct operators that depend on $\rho_A$ (such as the modular Hamiltonian $K_A = -\ln\rho_A$)~\footnote{Notice that, when the dimensions of region $A$ and $\bar A$ are not equal, the diagonal matrix in the SVD procedure will be padded with zeros. To avoid such non-physical values, we take the SVD values corresponding to the region with smallest dimension.}.

{\bf Averaging over $\gSO(3)$.}
The discussion so far is about the reduced density matrix of a specific ball $A_\Omega$ with center at a point $\Omega$. The next step is to average the entanglement Hamiltonian over $\Omega$.  To do this, we take advantage of the fact that the Hilbert space is a representation of $\gSO(3)$, and that the rotation of the center is accomplished by the action of $\gSO(3)$, so that 
averaging over the center is the same as averaging over the orbit of $\gSO(3)$:
\be
    \overline{O} = \int dg\, U(g)^\dagger O \, U(g), 
\ee
where $dg$ is the normalized Haar measure $\int dg = 1$, and $U(g)$ is the representation matrix corresponding to the group element $g$.

Rather than actually doing an integral over the group, we use a bit of representation theory. 
The averaged operator is a singlet under $\gSO(3)$: 
$\overline{O}$ commutes with the $\gSO(3)$ operators.
By Schur's Lemma, then, it cannot mix different irreps (spins $s$) or different magnetic quantum numbers ($m$), and it must act as a multiple of the identity within each irrep.
However, if there are multiple copies of the same irrep, the operator can still mix these copies. Using $i$ and $j$ to label this multiplicity space, we can express the averaged operator in a block-diagonal form:
\be \overline{O} = \sum_{s,m,i,j} \bar{O}^{(s)}_{ij} |s,m,i\rangle\langle s,m,j|
\label{eq:averaged-operator-irreps} \ee
Because the averaged operator is isotropic—meaning its action is entirely independent of the $J^z$ eigenvalue $m$—its effective matrix elements in the multiplicity space can be found by taking the partial trace over the $2s+1$ states of the irrep and dividing by its dimension:
\be
    \bar{O}^{(s)}_{ij} = \frac{1}{2s+1} \sum_{\tilde m = -s}^s \bra{s,\tilde m, i} O \ket{s,\tilde m, j}. \label{eq:averaged-operator-irreps-mel}
\ee

{\bf Selection rules.} 
Symmetries of the model play an important role in allowing us to study larger system sizes.  In particular, the reconstructed Hamiltonian $D_{rec}$ is block diagonal in the same symmetry sectors as the initial Hamiltonian.
Next we discuss the selection rules more in detail by taking as an example the Ising model~\cite{Zhu:2022gjc}. 

Let us remind ourselves of some important facts of the Ising model in the fuzzy sphere. The symmetries present are the full $SO(3)$~sphere rotation, Ising $\mathbb{Z}_2$, and particle-hole. Importantly, the two discrete symmetries are on-site symmetries and act on the spin degrees of freedom via ${\bf c}_m \to \sigma^x {\bf c}_m$ and ${\bf c}_m \to i \sigma^y {\bf c}_m^*, \, i \to -i$ respectively. Hence, eigenstates of the Hamiltonian are labeled by such quantum numbers $\ket{s,m,z_2,p,i}$ where $(s,m,i)$ labels an $\gSO(3)$~irrep, $z_2$ is the Ising $\mathbb{Z}_2$ quantum number, and $p$ is the particle-hole quantum number. In particular, the ground state is a $\gSO(3)$ singlet, $\mathbb{Z}_2$-even and particle-hole-symmetric state.

Consider now the operator of interest: the averaged modular Hamiltonian. 
And let us consider its matrix elements over the Ising model eigenstates. 
Since the ground state is $\mathbb{Z}_2$ even, the matrix elements $\langle s, m,z_2,p,i| K_A |s, m,z_2',p',j \rangle$ are only non-zero when $z_2 = z_2'$. That is, 
\be
    \langle s, m,z_2,p,i| K_A |s, m,z_2',p',j \rangle = 0 \text{ if } z_2 \neq z_2'.
\ee
Particle-hole symmetry is a little more interesting because it does not commute with the $\gSO(3)$ rotation group.   
The unitary operator corresponding to particle-hole symmetry acts as $U_{ph} \ket{s,m,z_2,p,i} = p \ket{s,-m,z_2,p,i}$. That is, it flips the angular momentum along the $z$-direction $m \to -m$. Hence, the sum~\eqref{eq:averaged-operator-irreps-mel} is only non-zero when $p=p'$. In other words, 
\be
    \sum_{ m = -s}^s \langle s, m, p,i | K_A | s, m, p', j\rangle = 0 \text{ if } p \neq p'.
\ee
Thus, the averaged modular Hamiltonian is block diagonal in the same symmetry sectors as the Hamiltonian. 
Similar statements can be made for the other critical models and their corresponding symmetries. 

{\bf Numerical implementation.} 
Here, we present the numerical details of our implementation of entanglement bootstrap on the fuzzy sphere. 
We use both exact diagonalization (ED) and density matrix renormalization group (DMRG) to obtain spectra. 
We use the package \emph{FuzzifiED} for the critical models on the fuzzy sphere \cite{zhou2025fuzz}.
We have also independently confirmed many of these results using our own code. 
A repository of our code can be found \href{https://github.com/xiangli-physics/EB-meets-fuzzy-sphere}{here}.

We are interested in computing the error of the VFPE and the reconstructed spectrum. 
In what follows, we discuss the construction of the modular Hamiltonian $K_A$ and the averaging over $\gSO(3)$ to find $\overline{K_A}$.  
These steps involve constructing operators of the form $K_A \otimes I_{\bar A}$ and computing matrix elements $\langle s, \tilde m, i | K_A \otimes I_{\bar A}| s,\tilde m,j\rangle$. 
Our task is then to implement these steps efficiently. We will explain this in more depth in the following paragraphs.

We start with the ground state of a local Hamiltonian $\ket{\psi} = \sum_I c_I \ket{I}$, expressed as a linear superposition of basis states $\ket{I}$ with definite quantum numbers. 
Here we discuss the orbital cut, where 
this state lives 
in a tensor product space $\mathcal{H} = \mathcal{H}_A \otimes \mathcal{H}_{\bar A}$,
\be \ket{I} = \ket{I_A} \ket{I_{\bar A}} \in \mathcal{H}_A \otimes \mathcal{H}_{\bar A}. \label{eq:orbital-cut-embedding}\ee
(For the real-space cut, we must explicitly double the Hilbert space, and embed our state $\ket{\psi}$ in this enlarged, tensor product Hilbert space.  We discuss this more-complicated calculation in Appendix \ref{app:real-space-cut}.)
Tracing over the subsystem $\bar A$, we obtain the reduced density matrix $\rho_A$. Using the block decomposition of the reduced density matrix in~\eqref{eq:reduced-density-matrix-blocks}, the modular Hamiltonian is constructed sector by sector: 
\be
    K_A = \bigoplus_{\beta} \sum_{I,I'} (K_{A}^{(\beta)})_{II'} |I^\beta_A\rangle \langle I_A'^\beta|,
\ee
where $\beta$ runs over the quantum numbers of subsystem $A$ and $ K_A^{(\beta)} = -  \ln \rho_A^{(\beta)}$.

To evaluate the matrix elements involved in the averaging, we need to bring $K_A$ back to the original Hilbert space $\mathcal{H}$. 
The matrix elements of $K_A \otimes I_{\bar A}$ to be used in \eqref{eq:averaged-operator-irreps-mel} are then 
\be
    \bra{ I_A^\beta, J_{\bar A}^{\bar \beta}} K_A \otimes I_{\bar A}
\ket{ I_A'^\beta, J_{\bar A}'^{\bar \beta}}
= \delta^{J_{\bar A} J_{\bar A}'}    
    (K_{A}^{(\beta)})_{II'}~.
\ee
Here $\bar \beta$ is the quantum number of subsystem $\bar A$ 
such that (using an additive notation for all symmetry groups) $ \beta + \bar \beta$ labels the symmetry sector in which $\ket{\psi}$ lives.  
Hence, we avoid ever constructing large matrices.

The remaining task is to  obtain the explicit states $\{\ket{s,m,i}\}$ of definite angular momentum and other discrete symmetry quantum numbers. 
For the error of the VFPE, only the $\gSO(3)$ singlets in the same symmetry sector as the ground state are needed. 
For the reconstructed spectrum, we construct only the low-angular momentum sectors to obtain the low-energy spectrum of the $D_{rec}$ operator.
We describe the procedure we implemented in the next appendix~\ref{app:Constructing-irreps}.

{\bf Gradient descent.} 
Lastly, we implement a gradient descent procedure to reduce the error of the VFPE. 
The idea is straightforward: Starting from a state~$\ket{\psi}$, we compute the gradient of the error of the VFPE and update the state along the direction given by steepest descent. 
We do this procedure iteratively until the objective function converges to an error less than $10^{-6}$.  For the last steps, it is convenient to use Newton's method instead.  

To implement the gradient descent, we first write the state $\ket{\psi}$ as:
\be
    \ket{\psi} = \sum_s c_s \ket{s}.
\ee
where $\{\ket{s}\}$ is the basis of states over which we will search. 
For the Ising model on the fuzzy sphere, the basis is the states that are $\gSO(3)$ singlets, $\mathbb{Z}_2$ even and particle-hole symmetric\footnote{Taking the basis over which we are doing gradient descent to be all the $\gSO(3)$ singlets would give the same result. Since the initial states in the Ising model are $\mathbb{Z}_2$ even and particle-hole symmetric, the selection rules exclude all but the $\mathbb{Z}_2$-even and particle-hole-symmetric $\gSO(3)$ singlets}. 

Next, we proceed to numerically get the gradient of the error of the VFPE. 
This is given by the equation $\nabla \sigma = \sum_s \partial_s \sigma(\ket{\psi}) \ket{s}$, where $\partial_s \sigma$ is: 
\be
    \partial_s \sigma = \frac{\sigma(\ket{\psi} + \epsilon \ket{s}) - \sigma(\ket{\psi} - \epsilon \ket{s})}{2\epsilon}.
\ee
In practice, we choose $\epsilon = 10^{-5}$.
Once we compute the gradient $\nabla \sigma$, we update the state via: 
\be
    \ket{\psi} \to \ket{\psi} - \gamma \nabla \sigma (\ket{\psi}),
\ee
where $\gamma$ is the step size. 
$\gamma$ is dynamically obtained using backtracking line search algorithm. 

Once the error hits a threshold~(around the order of $10^{-4}$), we can switch to Newton's method. 
The state is updated as:
\be
    \ket{\psi} \to \ket{\psi} - \beta \,  A^{-1} \cdot \nabla \sigma(\ket{\psi}),
\ee
where the matrix $A$ is the Hessian of the error of the VFPE and $\beta$ is the step size\footnote{Beware that the Hessian $A$ can contain zero eigenvalues, so its inverse is ill-defined. We regulate the inverse by the modification $A \to A + \delta I$ with $\delta \sim 10^{-6}$. }. We find that $\beta = 10^{-1}$ gives efficient convergence. We usually reach an error of less than $10^{-6}$ after just a few steps.

\section{Constructing $\gSO(3)$ irreps}
\label{app:Constructing-irreps}

In this appendix, we describe a method to explicitly decompose the many-fermion Hilbert space into irreps of $\gSO(3)$.  
While straightforward in principle, the matrices involved grow rapidly with system size, so we need to be efficient about it.  
The input to the algorithm is the Fock space made from fermionic creation operators with definite $L_z$, and the output is projectors onto definite angular momentum sectors.  

A straightforward way of finding all of the irreps would be to diagonalize the $L^2$ operator in the $L_z = 0$ sector, and select the eigenstates with angular momentum $l$. 
The rest of the multiplets can then be found using the raising and lowering operators $L_{\pm}$.
This method works for small system sizes (up to $N=12$ in the Ising model) where one can fully diagonalize the $L^2$ matrix.  
But as we increase the system size, this matrix becomes too large to be able to fully-diagonalize it. For example, for $N=14$ Ising model the $L^2$ matrix in the $\mathbb{Z}_2$ even, particle-hole symmetric sector is $184222 \times 184222$ using the FuzzifiED package.
This raises the question: Can we find an efficient way to construct the $\gSO(3)$ irreps without diagonalizing this big matrix?

\subsection{Counting irreps}
In this subsection we give a representation-theoretic algorithm to count the multiplicity of the irrep labelled $(\ell_\text{orbital}, s_\text{spin})$ in the decomposition of the many body Hilbert space.  
The full Hilbert space enjoys an action of $\gSU(2)_{orb} \times \gSU(2)_{spin}$, acting respectively by sphere rotations and flavor rotations.  (The fact that the latter is broken in the Ising model is not relevant here.)
The orbital label on the fermion operator transforms as the spin $j = (N-1)/2$ irrep of $\gSU(2)_{orb}$ while the flavor label is in the spin-$1/2$ irrep of $\gSU(2)_{spin}$.
The many-body Hilbert space for $N$ spinful fermions is therefore the antisymmetric product of $N$ copies of
$V = V_{j} \otimes V_{1/2}$, the one-particle Hilbert space: 
\be
    \mathcal{H}_N = \Lambda^N( V_{j} \otimes V_{1/2} ).
\ee
The goal is to decompose this space into irreps labelled by $(l,s)$ where $l,s \in \mathbb{Z}/2$.

We can temporarily regard $V_j$ as a representation of $\mathsf{GL}(N) \supset \gSU(2)_{orb}$. This is particularly useful as the decomposition of $\Lambda^N(V_j \otimes V_{1/2})$ into $\mathsf{GL}(N) \times \mathsf{GL}(2)$ irreps is given by: 
\be
    \Lambda^N(V_j \otimes V_{1/2}) = \bigoplus_{\lambda \vdash N} S_{\lambda}(V_j) \otimes S_{\lambda'}(V_{1/2}),
\ee
where the sum over $\lambda$ is over partitions of $N$, $\lambda'$ is its conjugate partition, and $S_{\lambda}$ is the Schur functor. 
This decomposition is a special case of Howe duality.

Since $V_{1/2}$ is the two-dimensional fundamental representation, the Schur functor $S_{\lambda'}(V_{1/2})$ is only non-zero if the Young diagram associated to the partition $\lambda'$ has at most two rows $\lambda' = (k_1,k_2)$ where $k_1+k_2 = N$ and $k_1 \geq k_2$. 
Under the $\gSU(2)_{spin}$ subgroup, the $\mathsf{GL}(2)$ irrep $S_{(k_1,k_2)}(\mathbb{C}_2)$ reduces to the $\gSU(2)$ irrep of spin $s = \frac{k_1-k_2}{2}$ with possible values $s\in \{\frac{N}{2}, \frac{N}{2}-1, \ldots \frac{1}{2} \text{ or } 0 \}$. 
Each of these spin-$s$ reps appears exactly once in the list of $\gSU(2)_{spin}$ factors. 

Now we decompose the $\mathsf{GL}(N)$ irreps into irreps of $\gSU(2)_{orb}$ subgroup. 
For a given spin-$s$, the $\gSU(2)_{spin}$ partition was $\lambda' = (k_1,k_2)$.
The corresponding $\gSU(2)_{orb}$ partition $\lambda$ is the conjugate of $\lambda'$: two columns of height $k_2$ and $k_1-k_2$ columns of height $1$. 
Let's denote this as $\lambda = (2^{k_2}, 1^{k_1-k_2})$. 
Then, we need to find the decomposition of the $\mathsf{GL}(N)$ irrep $\lambda$ when restricted to the $\gSU(2)_{orb}$ subgroup defined by $V_j$, the $N$-dimensional spin $j = \frac{N-1}{2}$ irrep.
This is a branching rule problem. 
The resulting $\gSU(2)_{orb}$ representation, $S_\lambda(V_j)$, will be a direct sum of irreps $(l)$.

We can solve this problem using characters.  
The character of the $\mathsf{GL}(N)$ representation $S_\lambda$ is a symmetric polynomial in $N$ variables $(x_1, \dots, x_N)$ called the Schur polynomial, $s_\lambda(x_1, \dots, x_N)$, a specific polynomial defined by the Young diagram $\lambda$.
We restrict the $\mathsf{GL}(N)$ rep to the $\gSU(2)_{orb}$ subgroup by replacing the $N$ variables $(x_1, \dots, x_N)$ with the $N$ weights of the $\gSU(2)_{orb}$ irrep $V_j$. 
The character of our $\gSU(2)_{orb}$ representation is the specialized polynomial $s_\lambda(q^{j}, \dots, q^{-j})$.
Since the character of any $\gSU(2)$ representation is a sum of $\gSU(2)$ irreducible characters, our specialized polynomial $s_{\lambda}(q^j, \ldots, q^{-j})$ must be such a sum.
The irreducible character for spin-$l$ is given by $\chi_l(q) = q^l + q^{l-1} + \ldots + q^{-l}$.
By peeling off the highest spin characters, we can find all the multiplicities $m_{l,s}$ of the spin-$l$ irrep corresponding to the spin-$s$ irrep.

Thus, the full decomposition is the sum over all $(l,s)$ pairs found this way: 
\be
    \mathcal{H}_N = \bigoplus_{l,s} \big((l)_{orb} \otimes (s)_{spin}\big)^{\oplus m_{l,s}}, \label{eq:full-decomp-irreps}
\ee
where $m_{l,s}$ is the multiplicity of the irrep $(l,s)$.

\subsection{Explicit decomposition}
The decomposition~\eqref{eq:full-decomp-irreps} is adapted to the full $\gSU(2)\times \gSU(2)$ symmetry. However, the numerical basis used by FuzzifiED implements the subgroup $\gU(1) \rtimes \mathbb{Z}_2 \subset \gSU(2)$ rather than the full $\gSU(2)$ spin rotation symmetry. 
We therefore need to rewrite the above decomposition in a form adapted to fixed $S_z$ sectors. 
In this basis, the Hilbert space decomposes as $\mathcal{H}_N = \bigoplus_{S_z} \mathcal{H}_{N,S_z}$.
Fixing $S_z$ is equivalent to fixing the numbers of up and down particles $N_{\uparrow} = \frac{N}{2}+S_z$, $N_{\downarrow} = \frac{N}{2} - S_z$. 
Hence, the same fixed-$S_z$ Hilbert space can be written as:
\be
    \mathcal{H}_{N,S_z} = \Lambda^{N_{\uparrow}} V_j \otimes \Lambda^{N_{\downarrow}} V_j.
\ee
Each single-species Hilbert space can be written as $\Lambda^{n} V_j = \bigoplus_{J,\alpha} V_{J}^{(\alpha)}$, where $J$ is the sphere angular momentum and $\alpha$ labels multiplicity. 
Hence, 
\be
    \mathcal{H}_{N,S_z} = \bigoplus_{(J_{\uparrow}, \alpha); (J_{\downarrow}, \beta)} V_{J_{\uparrow}}^{(\alpha)} \otimes V_{J_{\downarrow}}^{(\beta)}. 
\ee
Using the familiar sum-of-angular-momentum formulas $V_{J_\uparrow}\otimes V_{J_\downarrow} = \bigoplus_{l = |J_{\uparrow} - J_{\downarrow}|}^{J_{\uparrow} + J_{\downarrow}} V_l$, we can rewrite it as: 
\be
    \mathcal{H}_{N,S_z} = \bigoplus_{(J_{\uparrow}, \alpha); (J_{\downarrow}, \beta)\,}  \bigoplus_{\,l = |J_{\uparrow} - J_{\downarrow}|}^{J_{\uparrow} + J_{\downarrow}} V_l.
\ee

This decomposition is useful because it allows us to construct angular momentum eigenstates by coupling the single-species orbital multiplets:
\begin{widetext}
\be
    \ket{(J_{\uparrow}, \alpha), (J_{\downarrow}, \beta); l,m} = \hspace{-0.3cm}\sum_{  M_{\uparrow}+M_{\downarrow} = m} \hspace{-0.3cm} C^{l,m}_{J_{\uparrow}, M_{\uparrow}; J_\downarrow, M_{\downarrow}} \ket{J_{\uparrow}, M_{\uparrow}, \alpha} \ket{J_{\downarrow}, M_{\downarrow}, \beta},
\ee
\end{widetext}
where $C_{J_{\uparrow}, M_{\uparrow}; J_\downarrow,M_{\downarrow}}^{l,m}$ is the Clebsch-Gordan coefficient:
\be
    C^{l,m}_{J_{\uparrow},M_{\uparrow}; J_{\downarrow}, M_{\downarrow}} = \bra{J_{\uparrow}, M_{\uparrow}} \bra{J_{\downarrow}, M_{\downarrow}} J_{\uparrow}, J_{\downarrow}; l,m \rangle.
\ee
Therefore, we can construct the $\gSO(3)$ irreps with definite total angular momentum $l$ using the single-species multiplets. 
We can build such multiplets by obtaining the highest-weight state in the $M=J$ sector, and then constructing the rest of the multiplet using the lowering operator $J_-$.
More precisely, the highest-weight vectors of spin $J$ are the kernel of the raising operator $J_+$:
\be
    J_+ \ket{J,J,\alpha} = 0.
\ee 
The remaining states in the multiplet are generated by
\be
    \ket{J,M-1,\alpha}
    \propto J_- \ket{J,M,\alpha} .
\ee
Notice that the operators $J_{\pm}$ act only on the single-species Hilbert space and not on the full spinful Hilbert space. 
The essential point is that the raising and lowering operators used in
this construction act only on the single-species spaces $\Lambda^n V_j$,
whose dimensions are much smaller than the full spinful Hilbert space.
The large spinful problem is
replaced by the construction of multiplets in much smaller
single-species spaces, followed by Clebsch-Gordan coupling.
This is what makes the construction of larger-$N$ $\gSO(3)$ irreps
computationally feasible.


\section{Finite-size scaling}
\label{app:analysis-of-the-UV}

In this section,
  we explain why $k=2$ regions are sufficient to cancel the UV contribution to $H_\text{rec}$ in 3D CFTs.
Our strategy is to use the fuzzy sphere as a UV regulator
  and compute the UV contribution to $H_\text{rec}$ using perturbation theory.
We show that for $k=2$,
  the UV contribution vanishes in the limit $R \to \infty$,
In contrast, for $k=1$,
  some quantities remain UV independent and others do not,
  depending on the smallest nontrivial scaling dimension
  $\Delta_{\min}$ appearing in the UV contribution.

We organize the discussion of the UV contribution into three parts,
  each addressing a different aspect of $H_\text{rec}$.

The main takeaway is that
  although the entanglement Hamiltonian $K_A$ in QFTs
  is often considered ill-defined due to UV contributions,
  certain linear combinations of $K_A$ can in fact be UV independent.
This UV independence suggests that
  such combinations remain well-defined
  and can be meaningfully discussed in field theory.

Throughout, we use $O(\,\cdot\,)$ to denote scaling in $R$,
  though in practice the bounds are likely tight (i.e.\ $\Theta(\,\cdot\,)$),
  as the terms are generically nonzero.

  \subsection{The aspirational field theory quantities}

Given a set of angles $\{\theta_i\}$ and corresponding weights $\{\lambda_i\}$,
  we consider the following quantity in field theory.
\begin{equation}
  \ol{K} \text{ ``='' } \sum_{i=1}^{k} \lambda_i \ol{K}_{\theta_i},
\end{equation}
where $K_{\theta_i}$ denotes the entanglement Hamiltonian
  associated with the region defined by a polar cap of opening angle $\theta_i$,
  and $\ol{K}_{\theta_i}$ is the average of $K_{\theta_i}$
  over all configurations related by rotation symmetry.

We use quotation marks to emphasize that the expression is aspirational,
  since $\ol{K}_{\theta_i}$ is not well-defined in field theory due to UV divergences.
A precise definition only emerges after introducing a UV regulator and taking an appropriate limit.

To better understand the structure of $\ol{K}$,
  we decompose it into a scalar part and an operator part:
\begin{equation}
  F \text{ ``='' } \sum_{i=1}^{k} \lambda_i S_{\theta_i}
  \quad\text{and}\quad
  \ol{\wt K} \text{ ``='' } \sum_{i=1}^{k} \lambda_i \ol{\wt K}_{\theta_i},
\end{equation}
where each $K_{\theta_i}$ is decomposed into
  the scalar $S_{\theta_i}$
  and the operator $\wt{K}_{\theta_i}$,
  such that $\<\psi|\wt{K}_{\theta_i}|\psi\> = 0$,
  where $|\psi\>$ is the vacuum state.

\subsection{UV regulation via the fuzzy sphere}

To regulate the quantities introduced above,
  we use the fuzzy sphere as a UV regulator.
Concretely,
  we consider a family of systems realized on fuzzy spheres with varying radii $R$
  and study the limit $R \to \infty$.
(Recall that $R \sim \sqrt{L}$.)

Given a CFT,
  we pick a particular realization on the fuzzy sphere,
  which yields a family of Hamiltonians $\{H_R\}$
  parameterized by the radius $R$.
This family shares a common UV cutoff (analogous to lattice spacing $a=1$)
  while varying in system size.
Since the CFT is scale invariant, taking $R \to \infty$
  at fixed $a=1$ is equivalent to the more standard limit
  in which the sphere radius is held fixed while the lattice spacing is taken to zero.

For each system size $R$,
  we compute the quantities corresponding to the aspirational field theory quantities defined above.
\begin{equation}
  F_R = \sum_{i=1}^{k} \lambda_i\, S_{\theta_i, R}, \qquad
  \ol{\wt{K}}_R = \sum_{i=1}^{k} \lambda_i\, \ol{\wt{K}}_{\theta_i, R},
\end{equation}
where $S_{\theta_i, R}$ and $\wt{K}_{\theta_i, R}$
  are scalar and operator parts of the entanglement Hamiltonian $K_{\theta_i, R}$
  associated with the polar cap of opening angle $\theta_i$ on the fuzzy sphere of radius $R$.  In this Appendix, we do the analysis for the real-space cut (studied in Appendix \ref{app:real-space-cut}); we expect that the conclusions should apply as well to the orbital cut studied in the body of the paper.

The scaling analysis here applies to the real-space cut.  However, we observe below that the behavior of the UV contributions as a function of system size is quite similar between real-space and orbital cuts.  

\subsection{UV independence}

Based on this UV regularization scheme,
  we say a quantity is UV independent,
  if its value computed at each system size,
  converges as $R \to \infty$.
In particular,
  the quantities we study below are:
\begin{enumerate}
  \item the RG monotone $F_R$,
  \item whether $(\text{err}_R)^2 = \<0|\ol{\wt K}_R \ol{\wt K}_R|0\> \to 0$,
  \item the eigenvalues and eigenvectors of $\ol{\wt K}_R$.
\end{enumerate}

\subsection{Assumptions}

To carry out the analysis,
  we make the following assumptions about
  the entropy $S_{\theta_i, R}$ and
  the reduced entanglement Hamiltonian $\wt K_{\theta_i, R}$.
We first write down the ansatz for flat space,
  and then summarize the corresponding ansatz for the sphere.

On a flat space, where $A$ is a disk of radius $R$,
  we assume the entropy and reduced entanglement Hamiltonian take the form
\begin{equation}\label{eq:ansatz-S-flat}
  S_A = \mu_1 R - F + O(1/R)
\end{equation}
where $\mu_1$ is a UV dependent non-universal constant, and
\begin{equation}\label{eq:ansatz-K-flat}
  \begin{gathered}
    \wt K_A = \wt K_A^{CFT} + \wt K_A^{UV}, \\
    \wt K_A^{CFT} = \int_A d^2 x \beta_A(x) h(x), \\
    \wt K_A^{UV} = \int_{\partial A} d\ell \left( \cO^{\hat n}_{0,flat}(x) + \kappa \cO^{\hat n}_{1,flat}(x) + O(\kappa^2) \right).
  \end{gathered}
\end{equation}
Here $h(x)$ is the Hamiltonian density,
  $\hat n$ is the outward normal to $\partial A$ at $x$,
  $\kappa = 1/R$ is the boundary curvature,
  and each $\cO^{\hat n}_{i,flat}(x)$ is a bounded operator supported near $x$.
In particular, $\wt K_A^{UV}$ is supported near the boundary $\partial A$. 

\begin{remark}
  In a fully mathematical treatment of this problem, one would need to worry about additional error terms
    from operators with increasingly large support near the boundary, and increasingly small operator norm.  
  For simplicity we omit them from the ansatz;
    we expect their inclusion would not affect the conclusions.
\end{remark}

We claim that the operators $\cO^{\hat n}_{i,flat}(x)$ satisfy the parity relation
\begin{equation}\label{eq:parity}
  \cO^{-\hat n}_{i,flat}(x) = (-)^i\, \cO^{\hat n}_{i,flat}(x).
\end{equation}
This follows from purity of the ground state.
Purity implies $K_A|\psi\> = K_{A^c}|\psi\>$,
  and since $S_A = S_{A^c}$ we have
  $\wt K_A|\psi\> = \wt K_{A^c}|\psi\>$.
The CFT parts also match, because
  $\wt K_A^{CFT} - \wt K_{A^c}^{CFT}$ is a conformal generator that annihilates the vacuum,
  we have $\wt K_A^{CFT}|\psi\> = \wt K_{A^c}^{CFT}|\psi\>$.
Hence $\wt K_A^{UV}|\psi\> = \wt K_{A^c}^{UV}|\psi\>$.
Because replacing $A$ by $A^c$ flips the sign of $\kappa$ and of $\hat n$,
  \cref{eq:parity} follows.

\begin{remark}
  In the context of continuum QFT, where the groundstate is cyclic and separating, the vector relation $( K_A - K_{A^c}) \ket{\psi} =0$ implies the operator relation $K_A - K_{A^c} =0$.  We use this in the argument above.
\end{remark}

On the sphere, for each fixed $\theta_i$,
  we similarly expand $S_{\theta_i, R}$ and $\wt K_{\theta_i, R}$ in powers of $1/R$:
\begin{equation}\label{eq:ansatz-S}
  S_{\theta_i, R} = \mu_{1, \theta_i} R - F + O_{\theta_i}(1/R)
\end{equation}
where $\mu_{1, \theta_i}$ is a UV dependent non-universal constant, and
\begin{equation}\label{eq:ansatz-K}
  \begin{gathered}
    \wt K_{\theta_i, R} = \wt K_{\theta_i, R}^{CFT} + \wt K_{\theta_i, R}^{UV}, \\
    \wt K_{\theta_i, R}^{CFT} = \int_{A_{\theta_i}} d^2 x \beta_{A_{\theta_i}}(x) h(x), \\
    \wt K_{\theta_i, R}^{UV} = \int_{\partial A_{\theta_i}} d\ell \left( \cO^{\hat n}_0(x) + \frac{\alpha_{\theta_i}}{R} \cO^{\hat n}_{1}(x) + O_{\theta_i}\left(\frac{1}{R^2}\right) \right).
  \end{gathered}
\end{equation}
Here $A_{\theta_i}$ is the polar cap of opening angle $\theta_i$ on the sphere of radius $R$,
  and $\alpha_{\theta_i}$ is a constant depending on $\theta_i$.
Note that $\cO^{\hat n}_{1}(x)$
  may be different from the flat space counterpart,
  due to the additional curvature of the sphere.
Nevertheless,
  $\cO^{\hat n}_{1}(x)$ is again $R$ independent.
Additionally, 
  the parity relation \cref{eq:parity} is expected to hold
  in the spherical case by the same argument.

\subsection{The RG monotone $F$}

The goal is determine the number of angles $\theta_i$ required
  such that
\begin{equation}
  F_R \to F \quad\text{as}\quad R \to \infty.
\end{equation}
Since the leading term in $S_{\theta_i, R} = \mu_{1,\theta_i} R - F + O(1/R)$ is UV dependent,
  two radii are needed to extract the UV-independent quantity $F$,
  such that the UV dependent terms cancel out.

\widetext
\subsection{The error of the VFPE}

The goal is determine the number of angles $\theta_i$ required
  such that
\begin{equation}
  \<\psi|\ol{\wt K}_R \ol{\wt K}_R|\psi\> \to 0 \quad\text{as}\quad R \to \infty
\end{equation}

Let us see if one radius is sufficient.
Consider the case of a single angle $\theta$ with weight $\lambda = 1$,
  then $\ol{\wt K}_R = \ol{\wt K}_{\theta, R}$.
Applying the ansatz \cref{eq:ansatz-K},
\begin{align}
  \ol{\wt K}_R
  &= \ol{\wt K}{}_R^{CFT}
  + \sin(\theta)\,R \cdot
  \EE_{x \in S^2(R), \hat n} \left[ \cO^{\hat n}_0(x) + \frac{\alpha_{\theta}}{R} \cO^{\hat n}_1(x) + O(1/R^2) \right] \notag \\
  &= c_{\theta}\, R\, H^{CFT}_R
  + \sin(\theta)\,R \cdot
  \EE_{x \in S^2(R), \hat n} \left[ \cO^{\hat n}_0(x) + O(1/R^2) \right]
  \label{eq:KR-one-radius}
\end{align}
where $\sin(\theta)\,R$ is the boundary length
  and $c_{\theta}$ is a constant depending on $\theta$.
The $\cO^{\hat n}_1(x)$ term vanishes upon averaging over $\hat n$,
  by the parity relation \cref{eq:parity},
  $\cO^{-\hat n}_1(x) = -\cO^{\hat n}_1(x)$.

Because $H^{CFT}_R$ annihilates the vacuum,
  we have
\begin{align}
  \<\psi|\ol{\wt K}_R \ol{\wt K}_R|\psi\>
  &= \sin^2(\theta)\,R^2 \cdot
  \EE_{x,y \in S^2(R), \hat n, \hat m} \left[ \<\cO^{\hat n}_0(x) \cO^{\hat m}_0(y)\> 
\right. \nonumber\\  &+ \left.   O(\kappa^2) \right].
\end{align}
The correlator $\<\cO^{\hat n}_0(x)\,\cO^{\hat m}_0(y)\>$
  receives two contributions:
a contact term when $|x-y| = O(1)$, which contributes $O(1)$
  since the operators are bounded,
and a non-contact term when $x$ and $y$ are well-separated,
  which contributes $O(|x-y|^{-2\Delta_{\min}})$,
  where $\Delta_{\min}$ is the smallest nontrivial scaling dimension of the CFT.

Now, $\<\cO^{\hat n}_0(x) \cO^{\hat m}_0(y)\>$
  has two contributions,
  the contact term and the non-contact term.
The contact term happens when $x$ and $y$ are within $O(1)$ distance,
  and contribute $O(1)$ to the correlation function (because they are bounded operators).
The non-contact term happens when $x$ and $y$ are far apart,
  and contribute $O(1/|x-y|^{2\Delta_{\min}})$ to the correlation function,
  where $\Delta_{\min}$ is the smallest nontrivial scaling dimension of the CFT.
Thus, we have
\begin{align}
  \<\psi|\ol{\wt K}_R \ol{\wt K}_R|\psi\>
  &= O(R^2) \cdot
  \left[ O\left(\frac{1}{R^2}\right)
  + O\left(\frac{1}{R^4} \int_{x,y \in S^2(R), |x-y| > O(1)} \frac{dx dy}{|x-y|^{2\Delta_{\min}}} \right) \right] \\
 &= O(R^2) \cdot
  \left[ O\left(\frac{1}{R^2}\right)
  + O\left(\frac{1}{R^{2\Delta_{\min}}} \right) \right] \\
  &= \max(O(1), O(R^{2 - 2\Delta_{\min}})).
  \label{eq:uv-prediction}
\end{align}
Regardless of the value of $\Delta_{\min}$,
  we see that $\<\psi|\ol{\wt K}_R^2|\psi\>$ does not vanish as $R \to \infty$,
  so a single region is insufficient to cancel the UV contribution in $\ol{\wt K}_R$.

We now consider two radii at angles $\theta_1$ and $\theta_2$
  and choose weights $\lambda_1$ and $\lambda_2$
  such that the leading UV contribution from $\cO^{\hat n}_0(x)$ cancels,
  i.e. $\lambda_1 \sin(\theta_1) + \lambda_2 \sin(\theta_2) = 0$.
The leading UV term in $\ol{\wt K}_R$ then comes from $O(1/R^2)$:
\begin{align}
  \ol{\wt K}_R
  = c_{\theta}\, R\, H^{CFT}_R
  + \sin(\theta)\,R \cdot
  \EE_{x \in S^2(R), \hat n} \left[ O(1/R^2) \right].
  \label{eq:KR-two-radii}
\end{align}
Splitting into contact and the non-contact terms as before,
\begin{align}
  \<\psi|\ol{\wt K}_R \ol{\wt K}_R|\psi\>
  &= O(R^2) \cdot
  O(1/R^4)
  \left[ O(1/R^2) + O(1/R^{2\Delta_{\min}}) \right] \\
  &= \max(O(R^{-4}), O(R^{-2-2\Delta_{\min}}))
\end{align}
Since $\Delta_{\min} \geq 1/2$ for unitary CFTs,
  both terms vanish as $R \to \infty$.
Thus, two radii suffice for the error of the VFPE to vanish
  as $R \to \infty$.

\subsection{The eigenvalues and eigenvectors}

The goal is determine the number of angles $\theta_i$ required
  for the eigenvalues and eigenvectors of $\ol{\wt K}_R$ to converge as $R \to \infty$.

Let's see if one region is sufficient.
For a single angle $\theta$ with weight $\lambda = 1$,
  recall from \cref{eq:KR-one-radius} that
\begin{align*}
  \ol{\wt K}_R
  = c_{\theta}\, R\, H^{CFT}_R
  + \sin(\theta)\,R \cdot
  \EE_{x \in S^2(R), \hat n} \left[ \cO^{\hat n}_0(x) + O(1/R^2) \right]
\end{align*}
The question is
  whether $R \cdot \EE_{x,\hat n}\,\cO^{\hat n}_0(x)$
  is a small perturbation to $R\, H^{CFT}_R$ as $R \to \infty$.

To assess this, we rescale to the unit sphere.
Under this rescaling, $R\, H^{CFT}_R$ becomes $H^{CFT}_1$
  since the Hamiltonian density $h(x)$ has scaling dimension $3$,
  so the eigenvalues are $O(1)$.
Meanwhile, $R \cdot \EE_{x,\hat n}\,\cO^{\hat n}_0(x)$ scales as $O(R^{1-\Delta_{\min}})$,
  where $\Delta_{\min}$ is the smallest nontrivial scaling dimension of the CFT among operators allowed by the symmetry.
For $\Delta_{\min} > 1$ the perturbation vanishes as $R \to \infty$,
  and by second-order perturbation theory the eigenvalues converge at rate $O(R^{2-2\Delta_{\min}})$
  and the eigenvectors at rate $O(R^{1-\Delta_{\min}})$.
(We will discuss a subtlety in the next subsection.)

For $\Delta_{\min} \le 1$,
  we consider two radii at angles $\theta_1$ and $\theta_2$
  and choose weights $\lambda_1$ and $\lambda_2$
  such that the leading UV contribution from $\cO^{\hat n}_0(x)$ cancels.
The leading UV term in $\ol{\wt K}_R$ then comes from $O(1/R^2)$:
\begin{align}
  \ol{\wt K}_R
  = c_{\theta}\, R\, H^{CFT}_R
  + \sin(\theta)\,R \cdot
  \EE_{x \in S^2(R), \hat n} \left[ O(1/R^2) \right].
\end{align}
Now, $R \cdot \EE_{x,\hat n} O(1/R^2)$ scales as $O(R^{-1-\Delta_{\min}})$,
  which vanishes as $R \to \infty$ for any unitary CFTs.

Thus, two angles suffice for the eigenvalues and eigenvectors to converge as $R \to \infty$
  and a single angle already suffices when $\Delta_{\min} > 1$.

\subsection{The eigenvalues and eigenvectors: take 2}

The argument above suggests that for $\Delta_{\min} > 1$,
  a single angle suffices for the eigenvalues and eigenvectors to converge.
This is essentially correct,
  but there is a subtlety when $\Delta_{\min} \ge 2$.

Consider the perturbation $H^{CFT}_1 + \epsilon \int_{S^2(1)} d^2x\, \cO(x)$
  on the unit sphere in field theory.
If the second term were a bounded operator,
  convergence to zero as $\epsilon \to 0$ would be immediate.
However, since $\cO(x)$ is generally unbounded,
  for any fixed $\epsilon > 0$
  the perturbation may dominate the first term,
  making the limit $\epsilon \to 0$ potentially ill-defined.
This is indeed the case when $\Delta_{\min} \ge 2$.

The resolution is that the states on which the perturbation dominates
  are high-energy states of $H^{CFT}_1$.
Once a UV regulator is in place,
  these states are excluded from the spectrum,
  and the perturbation is genuinely small.
We describe this in detail below.

For simplicity, we focus on the case where $\Delta_{\min} > 1$.
The problematic high-energy state
$|\psi'\>$ that we consider corresponds to
  inserting $\cO$ into the ground state,
\begin{equation}
  |\psi'\> \;\text{``$\propto$''}\; \int_{S^2(1)} d^2x\, \cO(x) |\psi\>.
\end{equation}
Since this state is not normalizable in field theory,
  we instead insert $\cO$ slightly inside the unit sphere in the Euclidean path integral,
\begin{equation}
  |\psi'\> = \frac{1}{N} \int_{S^2(1-\delta)} d^2x\, \cO(x) |\psi\>
\end{equation}
  where $N$ is a normalization factor.

We will show that for $\Delta \ge 2$,
  the perturbation grows without bound
  as $\delta \to 0$ at fixed $\epsilon$.
However, with a UV regulator at scale $1/R$,
  we have $\delta \sim 1/R$,
  and combined with $\epsilon \sim R^{1-\Delta_{\min}}$
  the perturbation becomes controlled.

\begin{remark}
  Alternatively,
    one can work directly with the state on the fuzzy sphere of radius $R$,
  \begin{equation}
    |\psi'\> = \frac{1}{N} \EE_{x \in S^2(R),\, \hat n}\, \cO^{\hat n}_0(x)\, |\psi\>,
  \end{equation}
  in which case the perturbation is controlled directly from the fact that
    $\cO^{\hat n}_0(x)$ is a bounded operator,
    without invoking $\delta \sim 1/R$.
  The two approaches are equivalent:
    the boundedness of the operator on the fuzzy sphere
    plays the same role as the regulator $\delta \sim 1/R$.
\end{remark}

We first evaluate $N$.
Using the near and far decomposition of the correlator,
\begin{equation}
  N^2
  \approx \int_{S^2(1+\delta)} \int_{S^2(1-\delta)} d^2 x\, d^2 y\, \<\psi|\cO(x) \cO(y) |\psi\>
  = O\left(\frac{\delta^2}{\delta^{2 \Delta}}\right) + O(1)~.
\end{equation}
When $|x-y| = O(\delta)$,
  the correlator scales as $\<\psi|\cO(x) \cO(y) |\psi\> \sim 1/\delta^{2\Delta}$,
  and the integration measure contributes $O(\delta^2)$,
  giving $O(\delta^{2-2\Delta})$.
When $|x-y| \sim O(1)$,
  both factors are $O(1)$.
Since we assume $\Delta > 1$,
  the near term dominates and $N \sim \delta^{1-\Delta}$.

Next, we evaluate the matrix element of $H^{CFT}_1 + \epsilon V$,
  where $V = \int_{S^2(1)} d^2x\, \cO(x)$, in the basis $\{|\psi\>, |\psi'\>\}$:
\begin{align}
  &\=
  \begin{pmatrix}
    \<\psi|H^{CFT}_1|\psi\> & \<\psi|H^{CFT}_1|\psi'\> \\
    \<\psi'|H^{CFT}_1|\psi\> & \<\psi'|H^{CFT}_1|\psi'\>
  \end{pmatrix}
  + \epsilon \cdot
  \begin{pmatrix}
    \<\psi|V|\psi\> & \<\psi|V|\psi'\> \\
    \<\psi'|V|\psi\> & \<\psi'|V|\psi'\>
  \end{pmatrix} \\
  &=
  \begin{pmatrix}
    0 & 0 \\
    0 & \<\psi'|H^{CFT}_1|\psi'\>
  \end{pmatrix}
  + \epsilon \cdot
  \begin{pmatrix}
    0 & \<\psi|V|\psi'\> \\
    \<\psi'|V|\psi\> & \<\psi'|V|\psi'\>,
  \end{pmatrix}
\end{align}
using $H^{CFT}_1|\psi\> = 0$ and $\<\psi|\cO(x)|\psi\> = 0$.

We evaluate each matrix element in turn,
  keeping only the near term since it dominates for $\Delta > 1$.

\medskip\noindent
\textit{Off-diagonal perturbation:}
\begin{align}
  \<\psi|V|\psi'\>
  &= \frac{1}{N} \int_{S^2(1)} \int_{S^2(1-\delta)} d^2 x\, d^2 y\,\<\psi| \cO(x) \cO(y) |\psi\> \\
  &= \frac{1}{N} O(\delta^{2-2\Delta}) \\
  &= O(\delta^{1-\Delta}) \\
\end{align}

\medskip\noindent
\textit{Diagonal perturbation:}
\begin{align}
  \<\psi'|V|\psi'\>
  &= C \frac{1}{N^2} \int_{S^2(1+\delta)} \int_{S^2(1)} \int_{S^2(1-\delta)} d^2 x\, d^2 y\, d^2 z\, \<\psi| \cO(x) \cO(y) \cO(z) |\psi\> \\
  &= C \frac{1}{N^2} O(\delta^{4-3\Delta}) \\
  &= O(\delta^{2-\Delta})
\end{align}
where $C$ is the OPE coefficient of three $\cO$ operators,
The estimate follows from the near region $|x-y|, |y-z| = O(\delta)$,
  in which $\<\psi|\cO(x) \cO(y) \cO(z)|\psi\> \sim C/\delta^{3\Delta}$,
  with measure $O(\delta^4)$.

\medskip\noindent
\textit{Diagonal original Hamiltonian:}
Since $H^{CFT}_1$ acts as a radial derivative,
\begin{align}
  \<\psi'|H^{CFT}_1|\psi'\>
  &= \frac{1}{N^2} \int_{S^2(1+\delta)} \int_{S^2(1-\delta)} d^2 x\, d^2 y\, \<\psi| \cO(x) \partial_r \cO(y) |\psi\> \\
  &= \frac{1}{N^2} O(\delta^{1-2\Delta}) \\
  &= O(\delta^{-1})
\end{align}
where the near term has $\<\psi|\cO(x)\partial_r\cO(y)|\psi\> \sim O(1/\delta^{2\Delta+1})$
  with measure $O(\delta^2)$.

Assembling these estimates,
  the matrix in the basis $\{|\psi\>, |\psi'\>\}$ is
\begin{equation}
  \begin{pmatrix}
    0
    & O(\epsilon\,\delta^{1-\Delta}) \\
    O(\epsilon\,\delta^{1-\Delta})
    & O(\delta^{-1}) + O(\epsilon\,\delta^{2-\Delta})
  \end{pmatrix}
\end{equation}

We can now read off the behavior in each regime from the matrix above.
If we hold $\epsilon$ fixed and take $\delta \to 0$,
  then for $\Delta > 2$ the off-diagonal term $O(\delta^{1-\Delta})$ dominates the perturbation matrix
  ($1-\Delta < -1, 2-\Delta$),
  so the perturbation is large.
This means the eigenvalues and eigenvectors
  do not converge to those of $H^{CFT}_1$
  if the limits are taken in the order $\lim_{\epsilon \to 0} \lim_{\delta \to 0}$.

In our setup, however,
  the UV regulator implies $\delta \sim 1/R$ and $\epsilon \sim R^{1-\Delta}$,
  so the perturbation matrix becomes
\begin{equation}
  \begin{pmatrix}
    0
    & O(R^{1-\Delta} \cdot R^{\Delta-1}) \\
    O(R^{1-\Delta} \cdot R^{\Delta-1})
    & O(R) + O(R^{1-\Delta} \cdot R^{\Delta-2})
  \end{pmatrix}
  =
  \begin{pmatrix}
    0
    & O(1) \\
    O(1)
    & O(R)
  \end{pmatrix}
\end{equation}
The diagonal entry $O(R)$ reflects the energy of $|\psi'\>$.
Without a UV regulator,
  $\delta$ can be arbitrarily small and this energy is unbounded,
  but the UV regulator caps it at $O(R)$.
The off-diagonal entries are $O(1)$,
  so the mixing between $|\psi\>$ and $|\psi'\>$ is suppressed by $O(1/R)$
  relative to their energy separation.
Thus, the eigenvalues and eigenvectors converge to those of $H^{CFT}_1$ at rate $O(1/R)$.


To summarize,
  when there is a single angle, the behavior of the eigenvalues and eigenvectors of $\ol{\wt K}_R$ depends on $\Delta_{\min}$ as follows.
The transition at $\Delta_{\min} = 3/2$
  is based on whether the mixing with the low energy state leads to a lower energy or whether the mixing with the high energy state leads to a lower energy.
\begin{table}[h!]
\centering
\renewcommand{\arraystretch}{1.5}
\begin{tabular}{|c|c|c|c|}
\hline
Range of $\Delta_{\min}$ & Matrix & G.S. energy & G.S. eigenvector \\
\hline

$1 < \Delta_{\min} \le 3/2$
&
$\displaystyle
\begin{pmatrix}
0 & O(R^{1-\Delta_{\min}})\\
O(R^{1-\Delta_{\min}}) & O(1)
\end{pmatrix}
$
&
$\displaystyle O(R^{2(1-\Delta_{\min})})$
&
$\displaystyle \binom{1}{O(R^{1-\Delta_{\min}})}$
\\

\hline

$\Delta_{\min} > 3/2$
&
$\displaystyle
\begin{pmatrix}
0 & O(1)\\
O(1) & O(R)
\end{pmatrix}
$
&
$\displaystyle O(1/R)$
&
$\displaystyle \binom{1}{O(1/R)}$
\\

\hline
\end{tabular}
\caption{\justifying The first matrix uses the basis $\{|\psi\>, |\cO\>\}$
    where $|\cO\>$ is obtained via the state-operator correspondence,
    while the second matrix uses the basis $\{|\psi\>, |\psi'\>\}$.}
    \label{table-of-error-predictions}
\end{table}

\begin{figure}
    \centering
    \parfig{0.48}{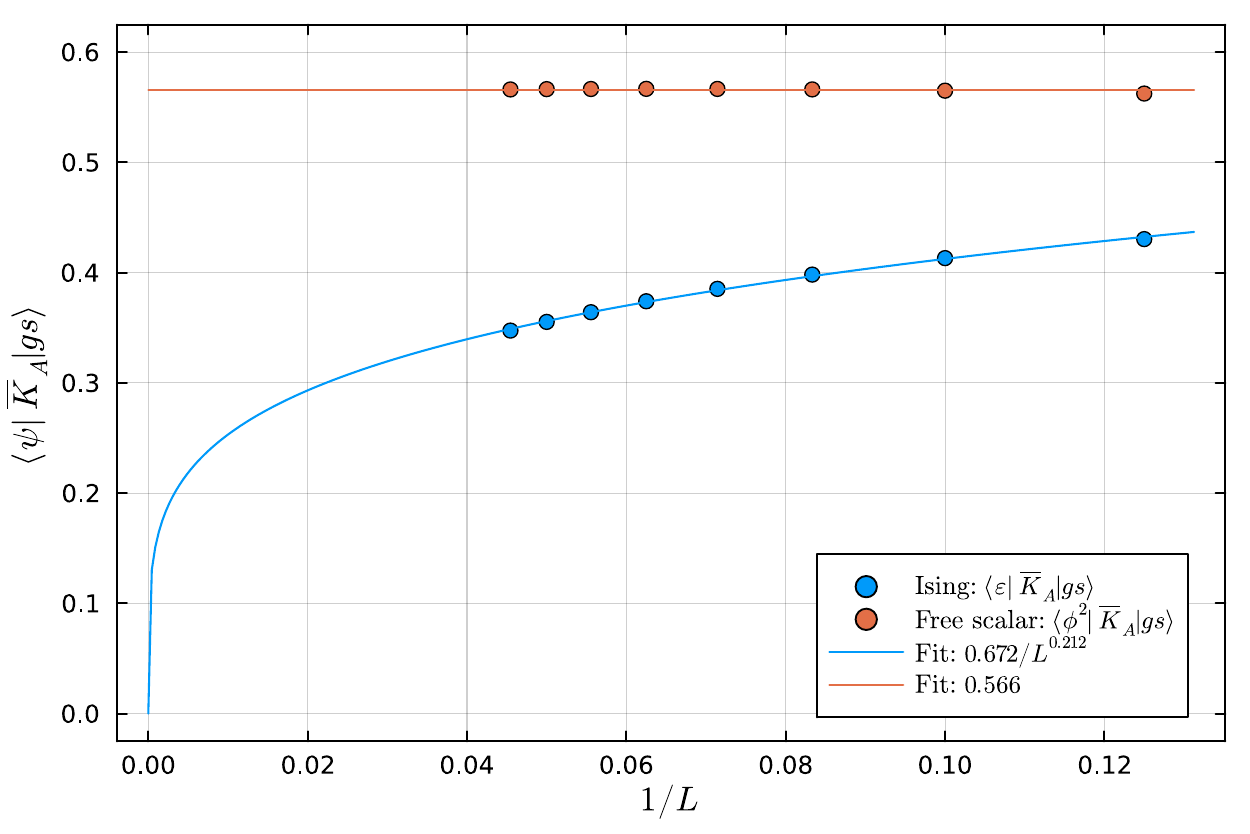}
    \caption{\justifying In this figure, we confirm two predictions of the description of the UV physics in this section.  We show $\langle \varepsilon | \overline{K_A} |gs \rangle$ where $\ket{gs}$ is the critical Ising model groundstate, $K_A$ is its entanglement Hamiltonian, and $\ket{\varepsilon}$ is the eigenstate corresponding to the energy operator $\varepsilon$.
    The plot also shows the corresponding power law fit. We see that this matrix element decays as a function of system size $L$ with a power law close to the expected $(\Delta_\varepsilon -1)/2 \simeq 0.206$ (using $\Delta_\varepsilon = 1.4126$) from Table \ref{table-of-error-predictions}.  
    In the free scalar CFT in $D=2+1$, $\phi^2$ has dimension $1$.  
    Therefore, we predict that $\langle \phi^2 | \overline{ K_A} | gs \rangle$ should be independent of $L$, which it is.}
    \label{fig:successful-test-for-primaries}
\end{figure}

Two successful tests of this picture are shown in 
Fig.~\ref{fig:successful-test-for-primaries}.

\begin{figure}
    \parfig{0.48}{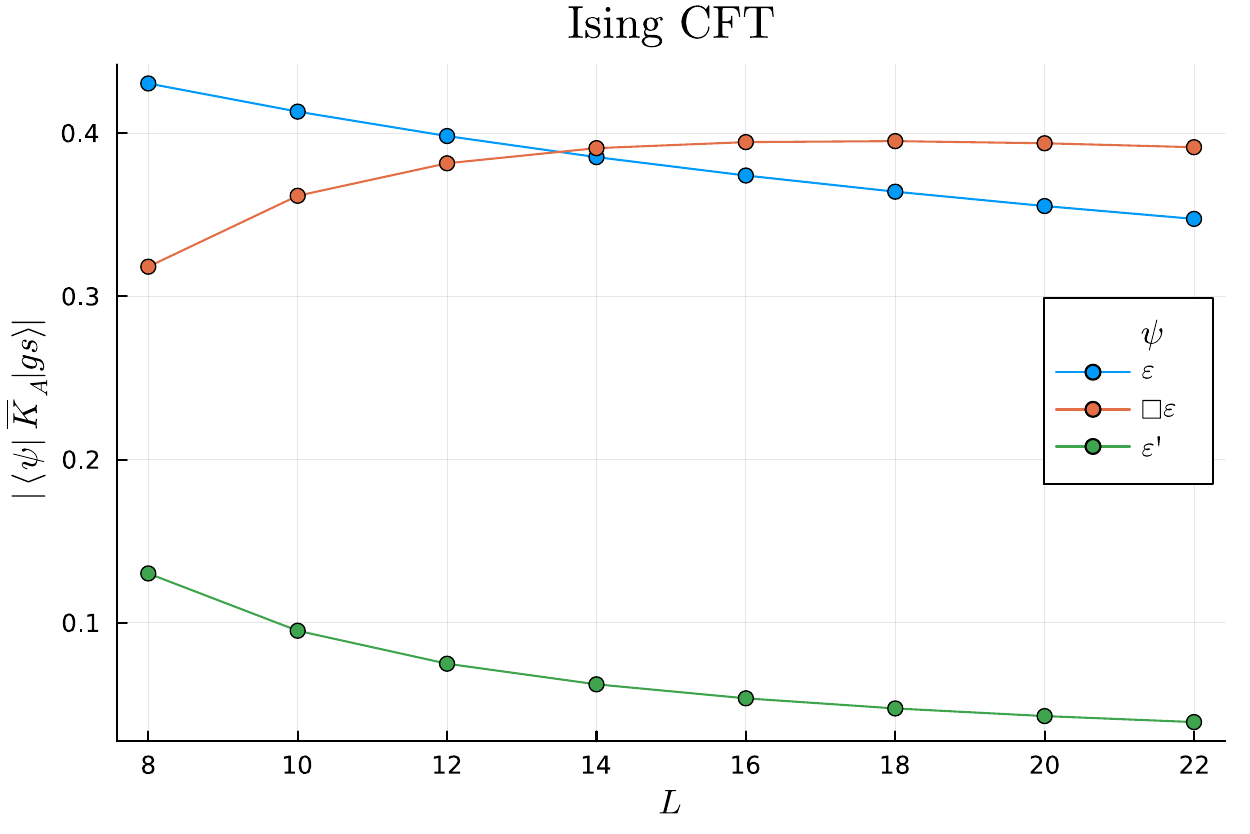}
\parfig{0.48}{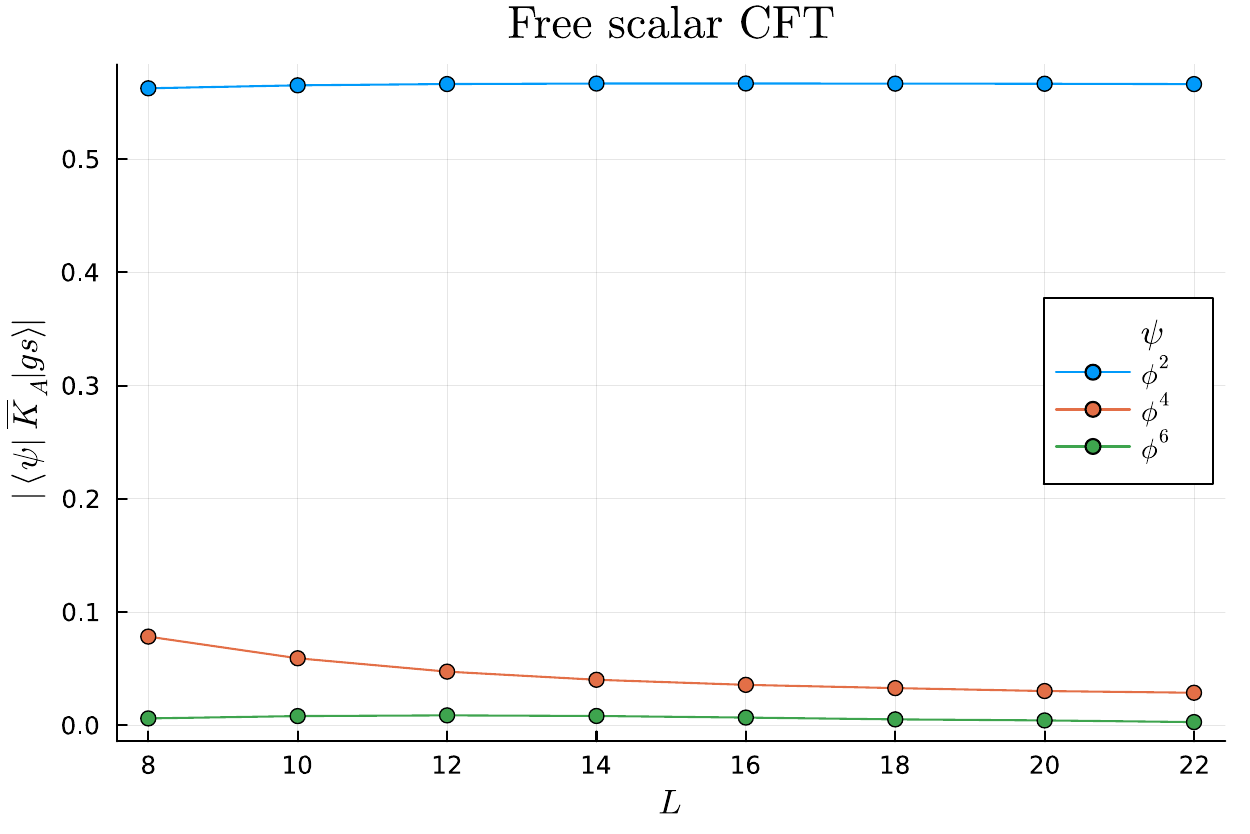}
    \caption{\justifying \label{fig:operator-error-Ising} Left: Matrix elements of $K_A$ in the Ising CFT between the groundstate and various scalar eigenstates of the original Ising Hamiltonian.  Data points up to $L=16$ were computed using both exact diagonalization and DMRG; the differences are of order $10^{-6}$.  Results for $L=18,20,22$ are computed using DMRG.  The growth of the matrix element with the descendant $\square \eps$ for small $L$ remains a mystery to us, though we do see it begin to decay at the largest sizes available.
    Analogous, similar results using the real-space cut are shown in Fig.~\ref{fig:KA-matrix-elements-Ising-RS}.  Right: The analogous matrix elements of $K_A$ in the free scalar CFT. }
\end{figure}

\endwidetext

\begin{figure}[!ht]
    \parfig{0.315}{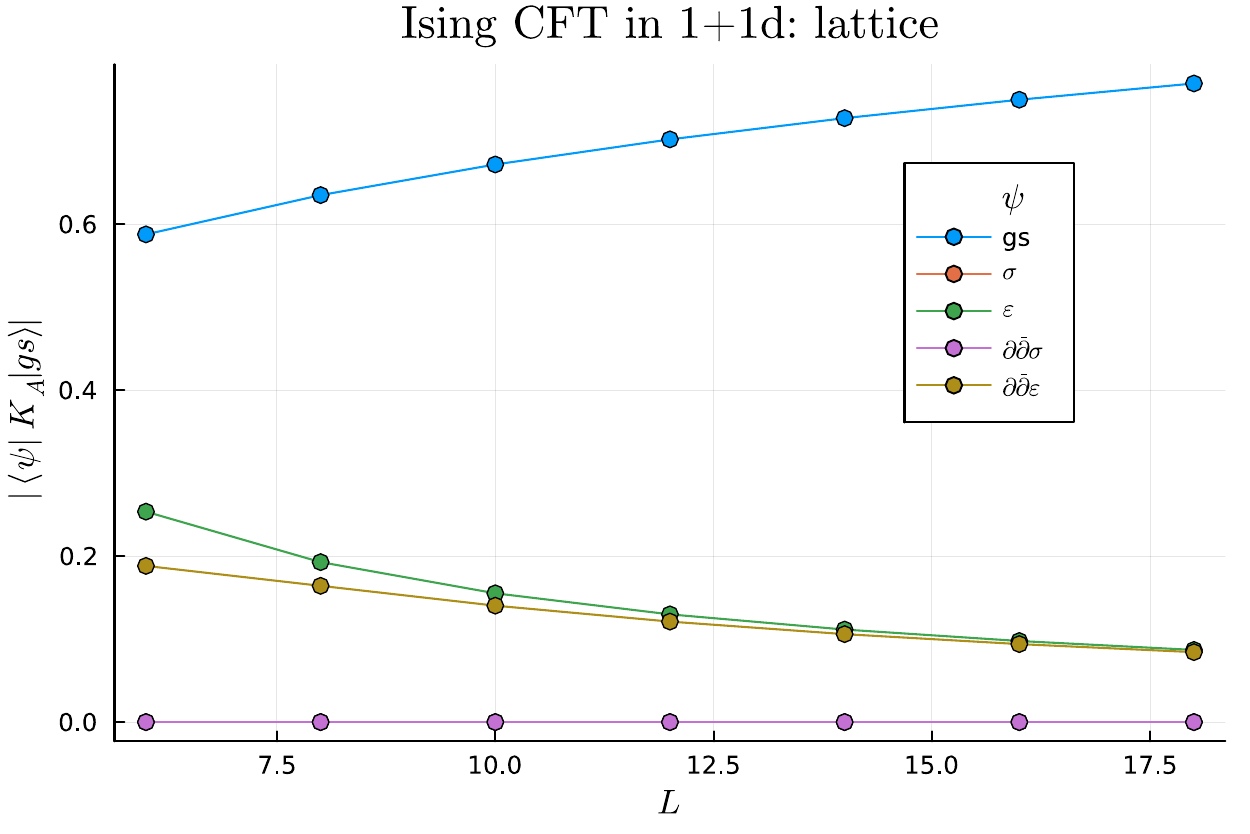}
    \parfig{0.31}{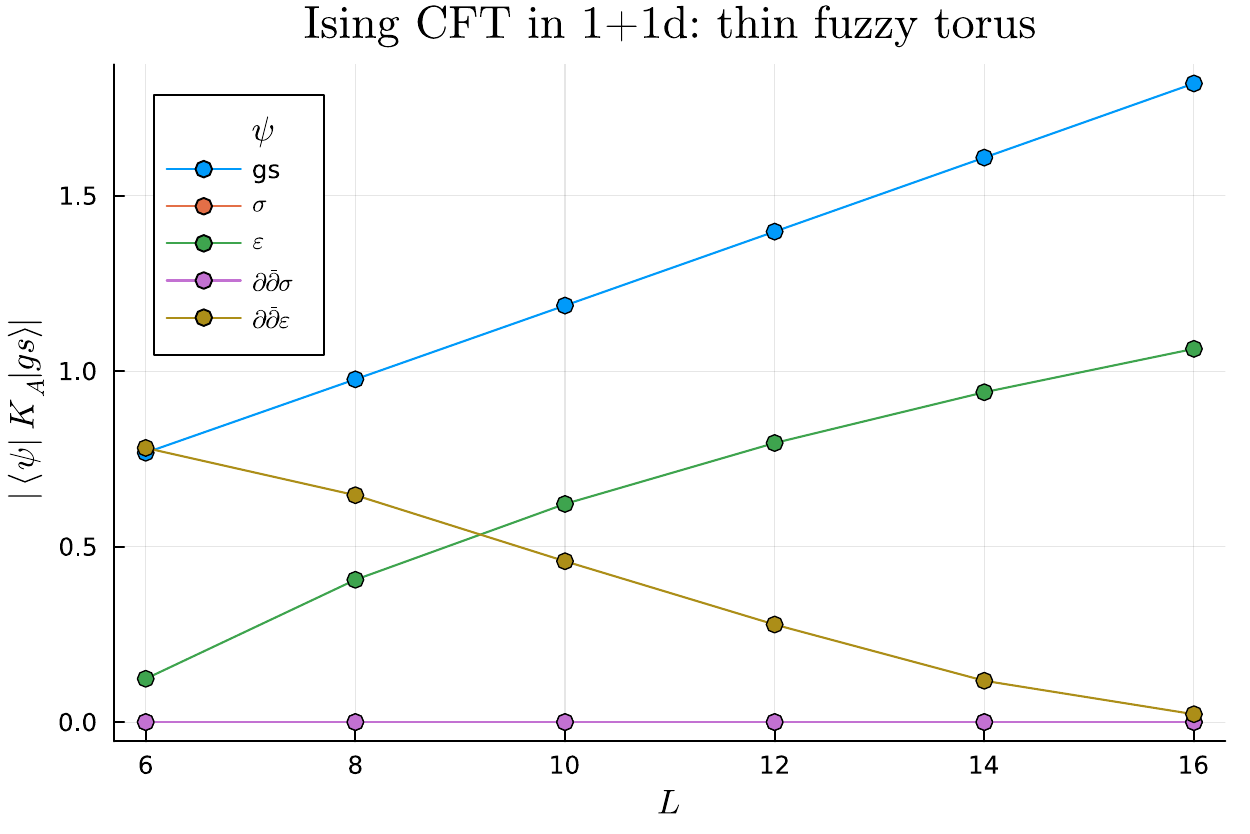}
    \parfig{0.31}{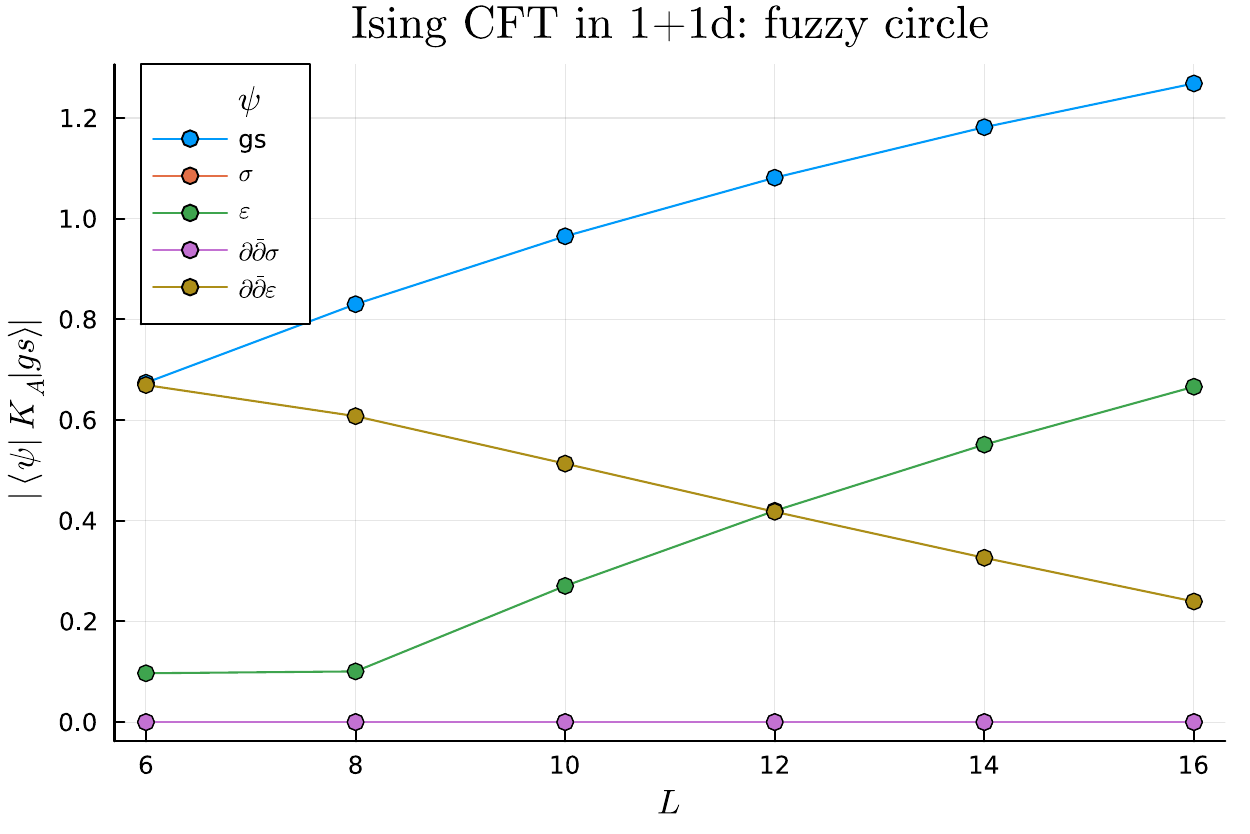}
    \caption{\justifying \label{fig:1+1d-Ising-regularizations} Matrix elements of $K_A$ in zero-momentum eigenstates of the critical transverse-field Ising model, as a function of system size for different regularizations: lattice (left), thin fuzzy torus (center), and fuzzy circle (right).
    We study the matrix elements $|\langle \psi|\overline{K}_A|gs \rangle|$ with respect to the ground state and the low-energy eigenstates, choosing the subregion size $A$ to be half the system size. 
    Because the groundstate is in the $\ell=0$ sector, these matrix elements of $K_A$ are the same as those of $\bar K_A$.  
    Matrix elements of $K_A$ with the $\sigma$-family are zero, the reason being that the ground state and $\sigma$ states live in different $\mathbb{Z}_2$ Ising sectors.
    The behavior of non-zero matrix elements depend on the regularization. 
    In the lattice regularization, all the matrix elements for states other than the identity decay with system size. 
    Conversely, in both fuzzy regularizations, one of the matrix elements with $\varepsilon$ primary seems to grow over available system sizes.  
    }
\end{figure}

\begin{figure}
    \centering
    \includegraphics[width=0.8\linewidth]{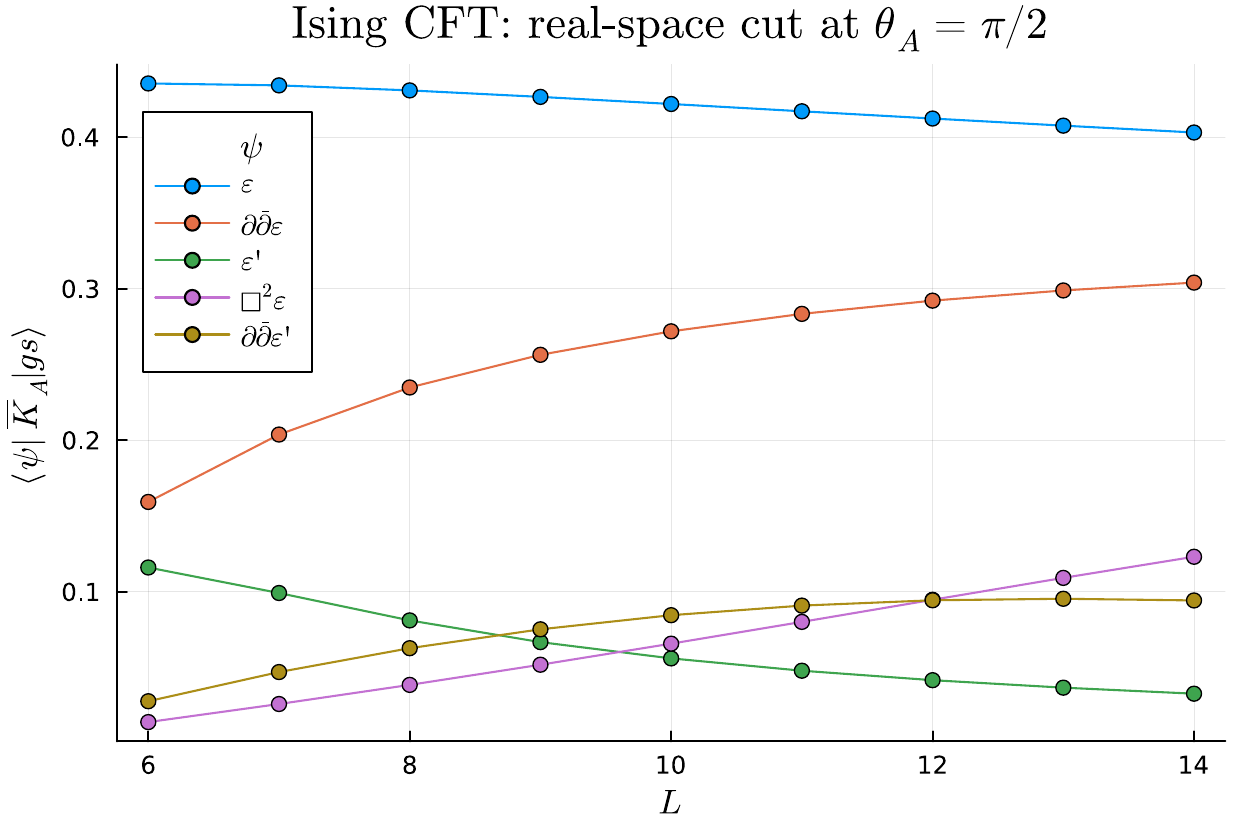}
    \caption{\justifying Matrix elements of $K_A$ for the 2+1d Ising critical groundstate of \cite{Zhu:2022gjc}, using the real-space cut.}
    \label{fig:KA-matrix-elements-Ising-RS}
\end{figure}

{\bf Fuzzy circle.} The behavior of the matrix element $\bra{\square \varepsilon} \overline{ K_A} \ket{\text{gs}}$ versus system size $L$ on the fuzzy sphere in Figs.~\ref{fig:operator-error-Ising} and Figs.~\ref{fig:KA-matrix-elements-Ising-RS} (for orbital and real-space cuts, respectively) is confusing from the point of view of the analysis described above.  We might ask whether it is a property special to the fuzzy regularization.  This is a question we can probe in $D=1+1$.  
First, we regularize the $D=1+1$ Ising CFT by the usual critical transverse-field Ising chain $H_{TFIM} = - \sum_{i=1}^L(Z_i Z_{i+1} + X_i )$.  The analogous matrix elements are shown in Fig.~\ref{fig:1+1d-Ising-regularizations}, and indeed they all decay with system size.  

To compare this to a fuzzy regularization, we need a notion of fuzzy circle. 
One such notion was studied in \cite{PhysRevB.111.085113}, which is just a thin fuzzy torus.  The results for the thin fuzzy torus are shown in Fig.~\ref{fig:1+1d-Ising-regularizations}.
A second, more intrinsic, notion of fuzzy circle is described next.  

The idea of fuzzy sphere is to replace the basis of continuum modes by a nice subset.  Nice can mean various things.  One meaning of nice is that the subset forms a representation of the desired symmetry group ($\gSO(3)$ in the case of fuzzy sphere).  Another is that we can define a family of subsets whose size grows slowly, so that one can get many data points for finite-size scaling.  A third meaning is that the modes are localized in spatial regions and have rapidly decaying overlaps.  The fuzzy sphere in 2d optimizes all these goals.  In other dimensions a solution that meets all these goals is not known to us\footnote{The construction of \cite{Gao:2025vho}, which puts a cutoff on angular momentum, meets the first goal, and to some extent, the third.}.

Let us consider the case of ${\cal M} = S^1$ with coordinate $\theta \in [0, 2\pi]$.  Then the full basis of functions is 
\be F_\infty \equiv \{ \phi_\ell(\theta), \ell \in \IZ \} \ee  
with
$ \phi_\ell(\theta) \equiv e^{ \ii \ell \theta}/\sqrt{2\pi}$.
The simplest fuzzy circle regulator is obtained by keeping only finitely many Fourier modes: 
\be F_L \equiv \{ \phi_\ell(\theta), \ell \in [-L, -L+1 ... L-1, L] \} ,\ee
and expanding
\be c(\theta) = \sum_{\ell=-L}^L c_\ell \phi_\ell(\theta), \ee
This is the analogue of the fuzzy sphere idea: the continuum Hilbert space of modes is replaced by a finite-dimensional subspace which preserves the spatial rotation symmetry of the target manifold.
The price of the sharp Fourier cutoff is that the position kernel is not sharply localized: 
\be
    \langle \theta | \theta' \rangle  =  \sum_{l = -L}^L \frac{e^{i l (\theta - \theta')}}{2\pi} = \frac{\sin\left( (L+\frac{1}{2})(\theta - \theta')\right)}{2 \pi \sin\left(\frac{1}{2}(\theta - \theta')\right)}.
\ee
Thus, unlike the fuzzy sphere, this regulator preserves the continuous spatial symmetry but does not produce a kernel with rapidly decaying spatial tails\footnote{A possible improvement is to replace the sharp cutoff by a smooth envelope, 
\be
    c(\theta) = \sum_{l=-L}^L \zeta_l \varphi_l(\theta),
\ee
with $\zeta_l$ decaying smoothly near the edge of the momentum window. This would replace the kernel by a smoother approximate delta function and be a basis function closer to the fuzzy sphere. We leave this smoother fuzzy circle regularization for future work. 
}.

Following the strategy of \cite{Zhu:2022gjc}, we can study the TFIM on this fuzzy circle. In the continuum, the Hamiltonian is given by: 
\be 
    H = \int d\theta d\theta' g(\theta - \theta') (\eta_0(\theta) \eta_0(\theta') - \eta_z(\theta) \eta_z(\theta')) - h \int d\theta \eta_x(\theta) 
\ee
where $\eta_\alpha(\theta) = c^\dagger (\theta) \sigma^\alpha c(\theta), \, \alpha = \{0,x,y,z\}$ and $g(\theta- \theta')$ is an ultra-local density-density interaction $g(\theta - \theta') = g_0 \delta(\theta - \theta') + g_2 \nabla^2 \delta(\theta- \theta')$. 
Notice that this Hamiltonian has the microscopic symmetries expected for the Ising transition: translation symmetry $c_l \to e^{-il\theta} c_l$, $\mathbb{Z}_2$ Ising acts on the spin degrees of freedom, $c_l \to \sigma^x c_l$, and particle-hole symmetry $c_l \to i \sigma^y c_l^\dagger$ at half-filling. 
The two limiting phases are also the expected ones: At small transverse field $h$, the ground state is an Ising ferromagnet, while at large $h$ it is a trivial paramagnet. 
Between these two regimes, we find the Ising CFT. For the choice of couplings used in our numerics and fixing the energy scale by setting $g_2=1$, the transition occurs near $(g_0,h)= (6 \pi, 7.29)$.

The computational effort involved in implementing this regularization of the 1+1d Ising CFT is comparable to that involved in lattice regularization.  One might have hoped for some benefit from the fact that this regularization preserves an exact translation symmetry, $c(\theta) \mapsto c(\theta + \alpha)$.  We did not observe any such benefit at the system sizes accessible to exact diagonalization.
For example, an analysis of the low-energy spectrum at criticality of the fuzzy circle shows the expected qualitative conformal towers, but the lattice regularization appears to converge faster.

The results for the matrix elements of $\overline{K_A}$ for the TFIM on the fuzzy circle are shown in Fig.~\ref{fig:1+1d-Ising-regularizations}.
In contrast to the lattice chain, one of the matrix elements with a primary seems to grow over the available system sizes. This is consistent with the result we found in the thin fuzzy torus and supports the conclusion that this phenomenon is a property of fuzzy regularization. 
\widetext 
{\,}
\endwidetext

\section{More numerical results for Ising CFT}
\label{app:more-results-for-ising}
\begin{figure}[!ht]
    \centering
    \includegraphics[width=\linewidth]{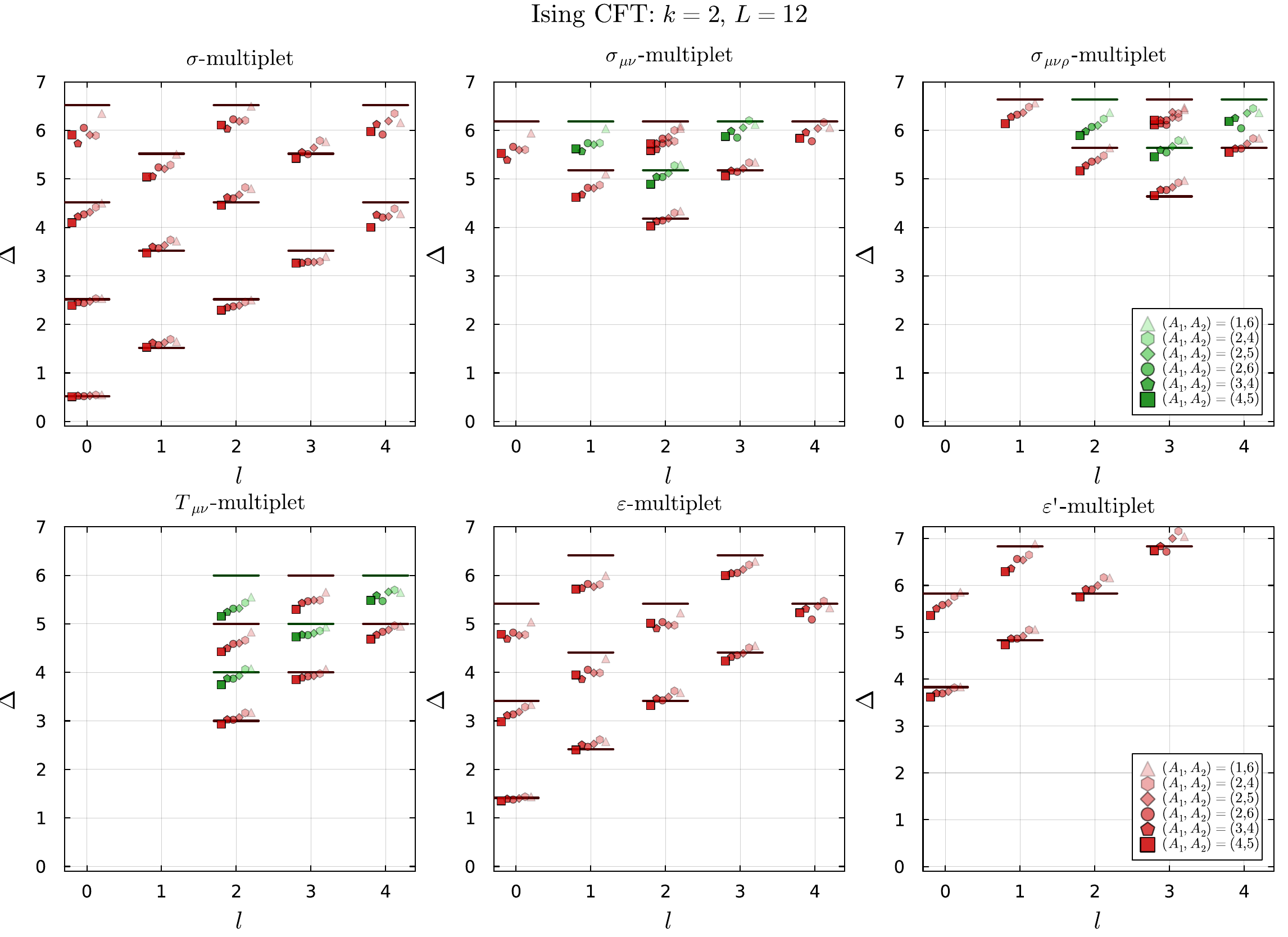}
    \caption{\justifying This plot is similar to \Cref{fig:Ising-Hrec-k2} but, instead of fixing the subregion size and increasing the system size, we fix the system size $L=12$ and plot the spectrum for various subregion sizes, using different markers and shades. 
    }
    \label{fig:Ising-Hrec-k2-L12}
\end{figure}

In \Cref{fig:Ising-Hrec-k2-L12} we show the spectrum of the reconstructed Hamiltonian for various choices of subregion sizes. We will use the phrase `reconstructed Hamiltonian' interchangeably with the phrase `reconstructed dilation operator'.  
This plot is important because it shows that the reconstructed spectrum looks qualitatively similar independent of the choice of subregion sizes.

In the body of the paper, we showed results for the reconstructed Hamiltonian with $k=2$.  A priori, based on the parametrization \eqref{eq:SRE-K_A}, one might have thought a priori that in $D=2+1$ one would need $k=3$ regions to cancel all the UV contributions.  Here we consider $k=3$ and demonstrate that $k=2$ is enough, consistent with our expectations.  
For $k=3$, the reconstructed Hamiltonian $H_{rec} = \sum_{i=1}^{k=3} \lambda^{th}_{A_i} \overline{K_{A_i}} -E_0$ where $\overline{K_{A_i}}$ is the averaged modular Hamiltonian defined on the region $A_i$, and the $\lambda_{A_i}^{th}$ are the theoretical parameters chosen to cancel the $UV$ contributions by:
\begin{enumerate}
\item $\sum_{i=1}^{k=3} \lambda_{A_i}^{th} = 0$, \quad \quad  \item $\sum_{i=1}^{k=3} \lambda_{A_i}^{th} |\partial A_i| = 0$, \quad \quad \item $\sum_{i=1}^{k=3} \lambda_{A_i}^{th} V_{A_i} = 1$, \end{enumerate}
    where $|\partial A_i| = \sin \theta_{A_i}$ and $V_{A_i} = 2\pi\sin(\theta_{A_i}/2)^4/\sin\theta_{A_i}$. As above, the angle $\theta_{A_i}$ is obtained by assuming that the ratio of the number of orbitals in region $A$ and the total number of orbitals is equal to the ratio of the area $A$ and the total area of the sphere: $\theta_{A_i} = \arccos(1-2|A_i|/L)$. 

\begin{figure}[!ht]
    \centering
    \includegraphics[width=\linewidth]{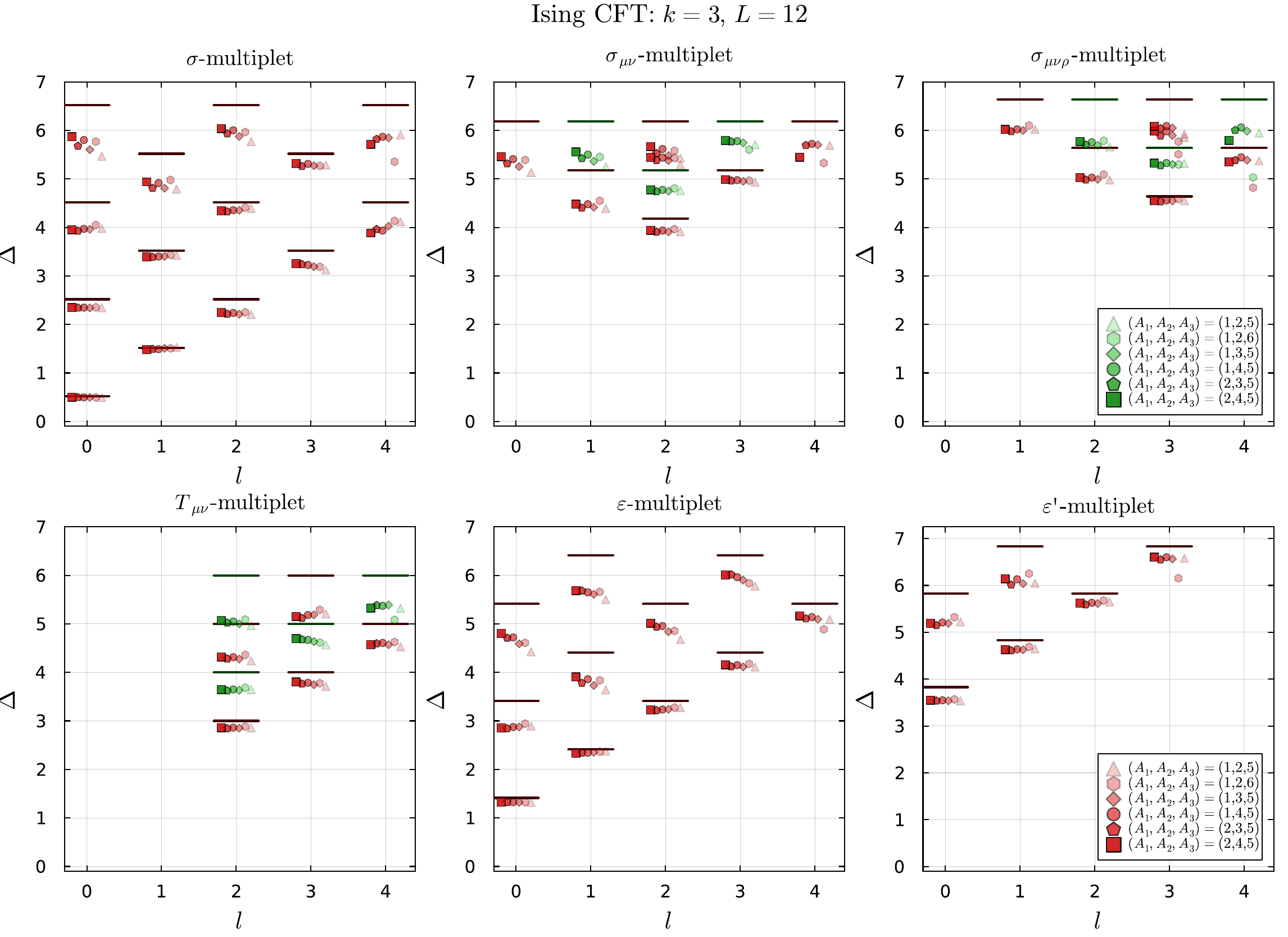}
    \caption{\justifying The spectrum of the reconstructed Hamiltonian for $k=3$.  
    We fix the system size $L=12$ and plot the spectrum for various subregion sizes, using different markers and shades. 
    Horizontal lines indicate the conformal bootstrap values of the spectrum. 
    Red or green indicates whether the state is particle-hole even or odd respectively. 
    }
    \label{fig:Ising-Hrec-k3}
\end{figure}

\begin{figure}[!ht]
    \centering
    \includegraphics[width=\linewidth]{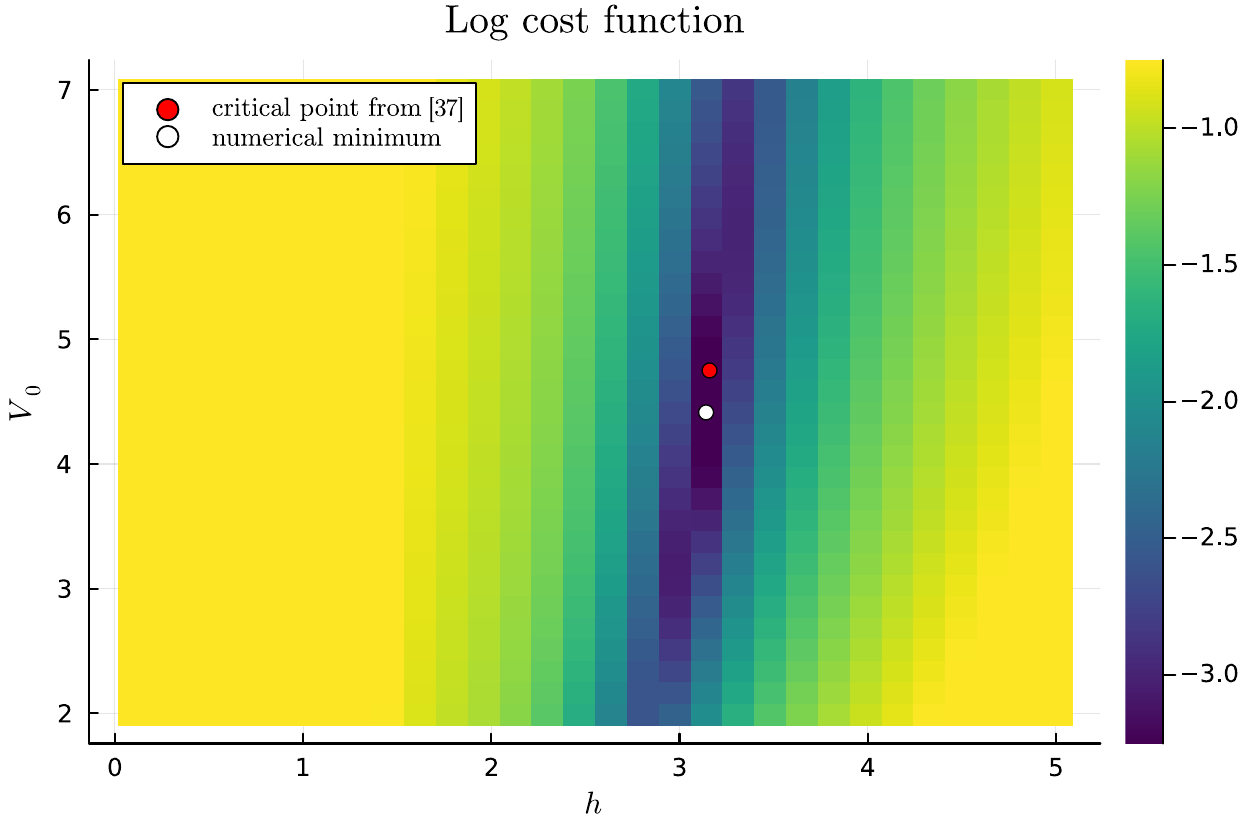}
    \caption{\justifying Heatmap of the logarithm of the cost function $f_c^{\Delta_\text{max} = 3}$ of the reconstructed Hamiltonian over the coupling space $(h,V_0)$ of the Ising model on the fuzzy sphere, for fixed system size $L=10$ and subregion size $(A_1,A_2)=(2,4)$ (matching the choice of subregions used in the reconstructed spectrum). 
    Indeed we find that the minimum of the cost function occurs in the neighborhood of the critical point identified by \cite{Zhu:2022gjc}.  Note the striking agreement with the location of the phase boundary of \cite{Zhu:2022gjc}.  
    }
    \label{fig:Ising-log-cost-function}
\end{figure}

In recent attempts to minimize finite-size effects on numerical approximations to CFT spectra, 
it has been useful to quantify the proximity of a given numerical spectrum to a partially-known CFT spectrum \cite{Fan:2024vcz,Fan:2025bhc}.  
We can apply this measure to our reconstructed spectrum, and this is what we do for the Ising phase diagram in \Cref{fig:Ising-log-cost-function}.  

    To define the cost function, let us denote by ${\bf \Delta} = (\Delta_\sigma, \Delta_{\partial \sigma}, \ldots)$ and ${\bf{E}} = (E_\sigma, E_{\partial \sigma},\ldots)$ as the vectors containing the expected scaling dimensions from CB and the energies obtained from reconstructing the spectrum respectively.  The cost function is defined by 
    \be \label{eq:cost-function-def} 
    f_c^{\Delta_\text{max}} = \sqrt{\tfrac{1}{N} \sum_{i=1}^N(\frac{{\bf E}_i - {\bf \Delta}_i}{ {\bf \Delta}_i} )^2},\ee 
    where $N$ is the smallest integer such that ${\bf \Delta}_i < \Delta_\text{max}$.
    Note that this differs somewhat from the cost function used in \cite{Fan:2024vcz,Fan:2025bhc},
$   \sqrt{ \Delta^2 - (\Delta \cdot E)^2/E^2 } $
which instead measures to what extent the vectors ${\bf \Delta}$ and ${\bf{E}}$ are parallel.  The difference is that our spectra have a preferred multiplicative normalization, and so we can use a more stringent cost function. 

\begin{figure}
    \centering
    \includegraphics[width=0.95\linewidth]{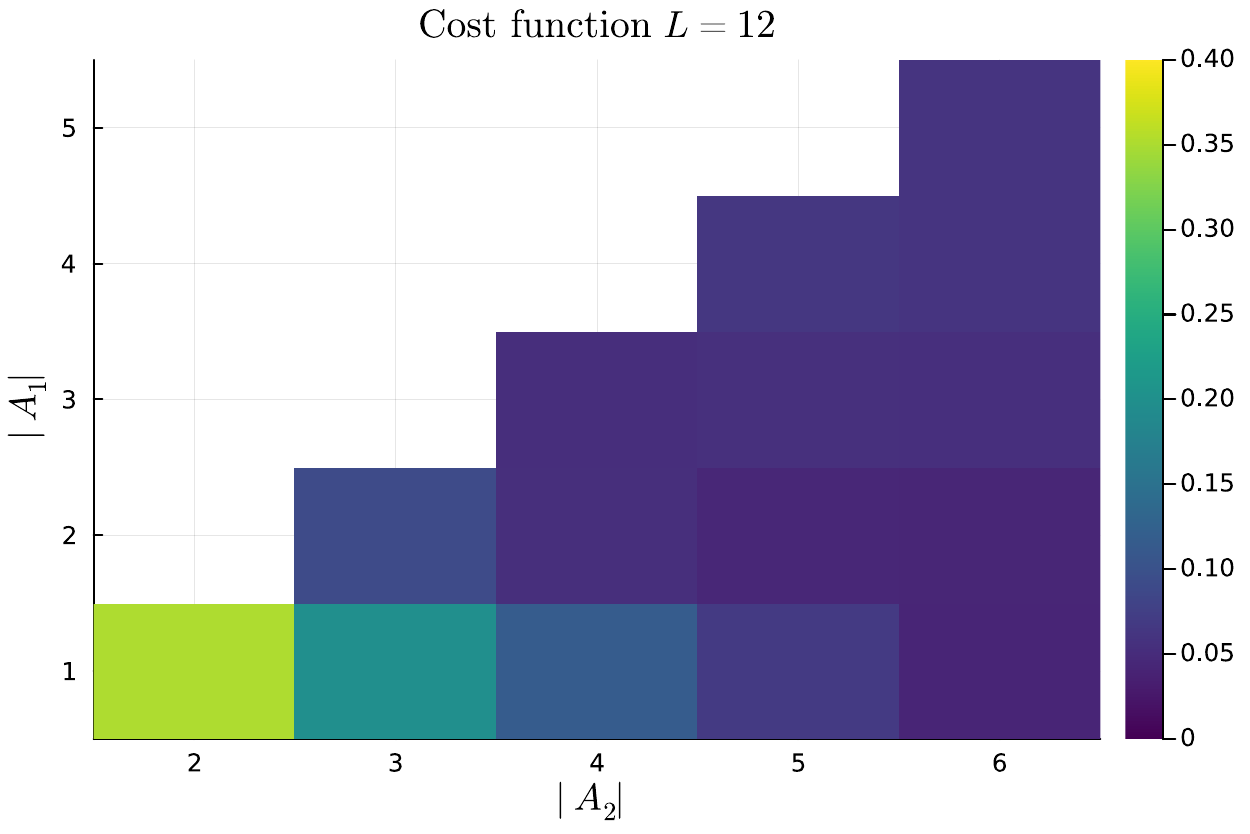}
    \caption{\justifying The cost function $f_c^{\Delta_\text{max} = 5}$ as a function of the choices of subsystem sizes $|A_{1,2}|$ for the Ising critical point of \cite{Zhu:2022gjc} at $L=12$.
    }
    \label{fig:error-vs-choice-of-region}
\end{figure}

In Fig.~\ref{fig:error-vs-choice-of-region}, we show the cost function of the reconstructed spectrum 
at the Ising critical point of \cite{Zhu:2022gjc}, versus the sizes of the subregions $A_{1,2}$.  The error decreases as the subregions grow.

\section{Results for other CFTs}
\label{app:other-CFTs} 

\begin{figure}[!ht]
    \centering
    \includegraphics[width=0.85\linewidth]{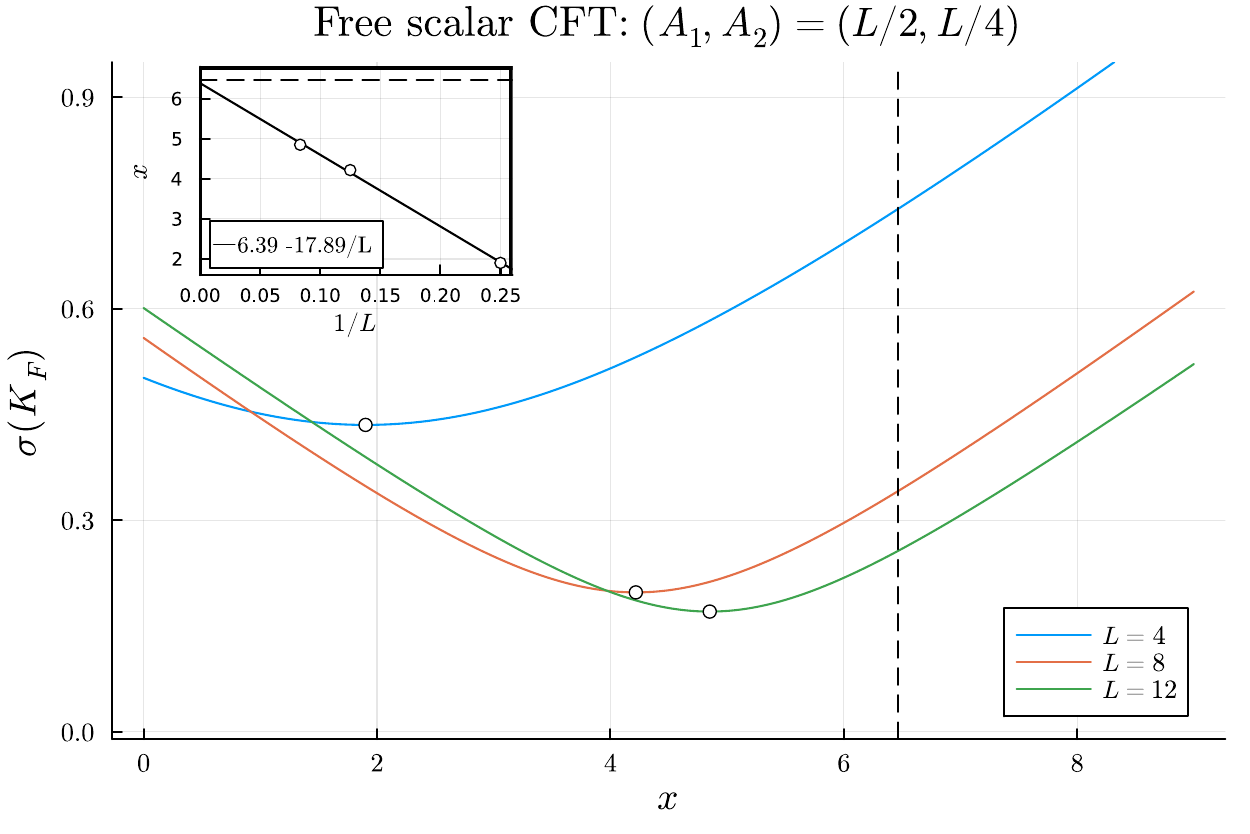}
    \caption{\justifying This plot is the same as Figure~\ref{fig:Ising-error-VFPE-parameters}, now for the free scalar CFT model~\cite{He:2025ong}.
    Inset: The linear fit obtained $6.39 - 17.89/L$ has similar values as the Ising CFT.}
    \label{fig:Free-scalar-error-VFPE-parameters}
\end{figure}

In \Cref{fig:Free-scalar-error-VFPE-parameters} we show 
the error of the VFPE versus the parameter $x$ in 
$K_F = x \overline{K_{A_2}} - (1+x) \overline{K_{A_1}}$ 
for the realization of the free scalar CFT described in \cite{He:2025ong}.  
Again the minimum error decreases with system size, and the extrapolation to $L\to \infty$ is close to the theoretical prediction $x_{th} = 6.46$ for the chosen ratio of subsystem sizes 
$|A_2|/|A_1| = 2$.  

\begin{figure}[!ht]
    \centering
    \includegraphics[width=0.95\linewidth]{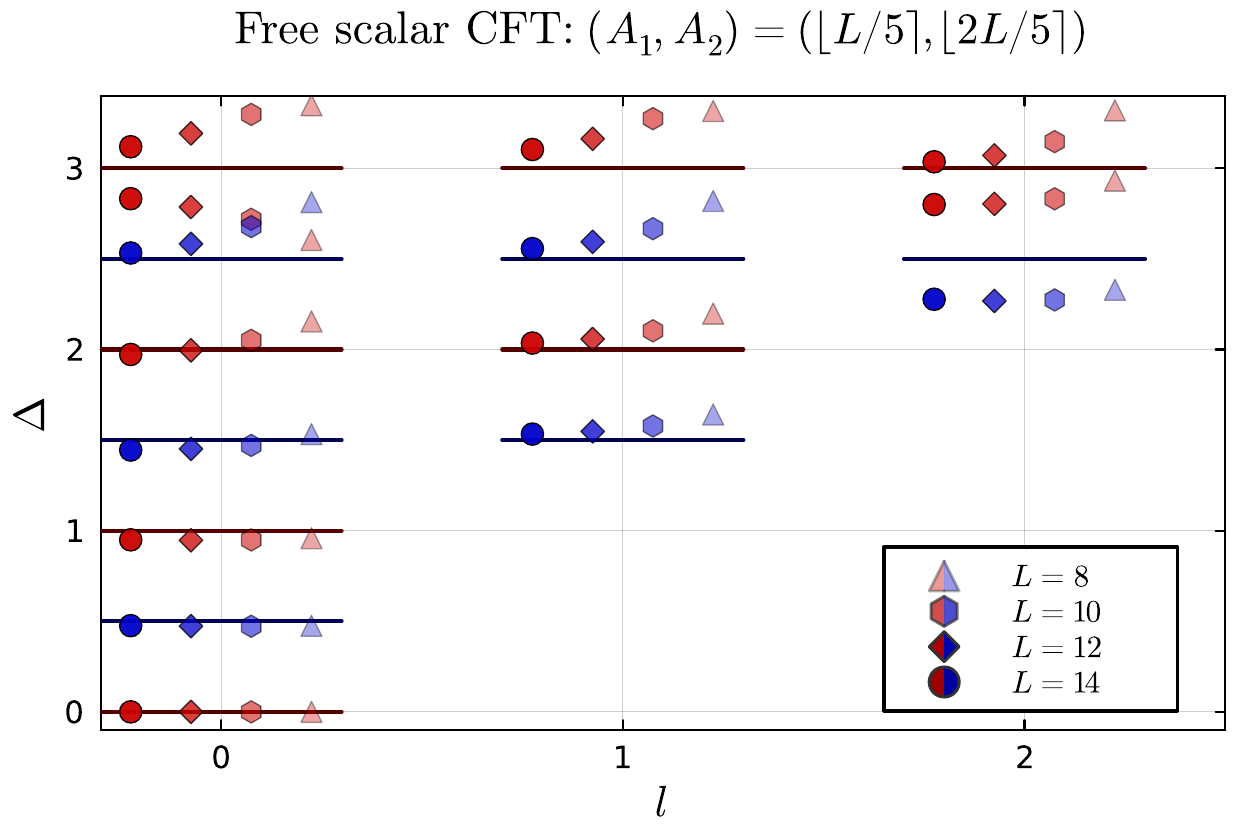}
    \caption{\justifying This plot is similar to Figure~\ref{fig:Ising-Hrec-k2}, now for the free scalar CFT model~\cite{He:2025ong}.
    The spectrum of the reconstructed Hamiltonian from the fuzzy sphere critical free scalar model groundstate versus angular momentum. We keep using the same subregion size scaling $(A_1, A_2) = (\left\lfloor L/5 \right\rceil, \left\lfloor 2L/5 \right\rceil )$ and plot four system sizes $L = \{ 8,10,12,14\}$ with different markers and transparency. 
    The horizontal lines indicate the theoretical values of the spectrum. 
    The two different colors~(red and blue) specify whether the state is $\mathbb{Z}_2$ even or odd respectively.}
    \label{fig:Free-scalar-Hrec}
\end{figure}

\begin{figure}[!ht]
\centering
{\centering
    \includegraphics[width=0.95\linewidth]{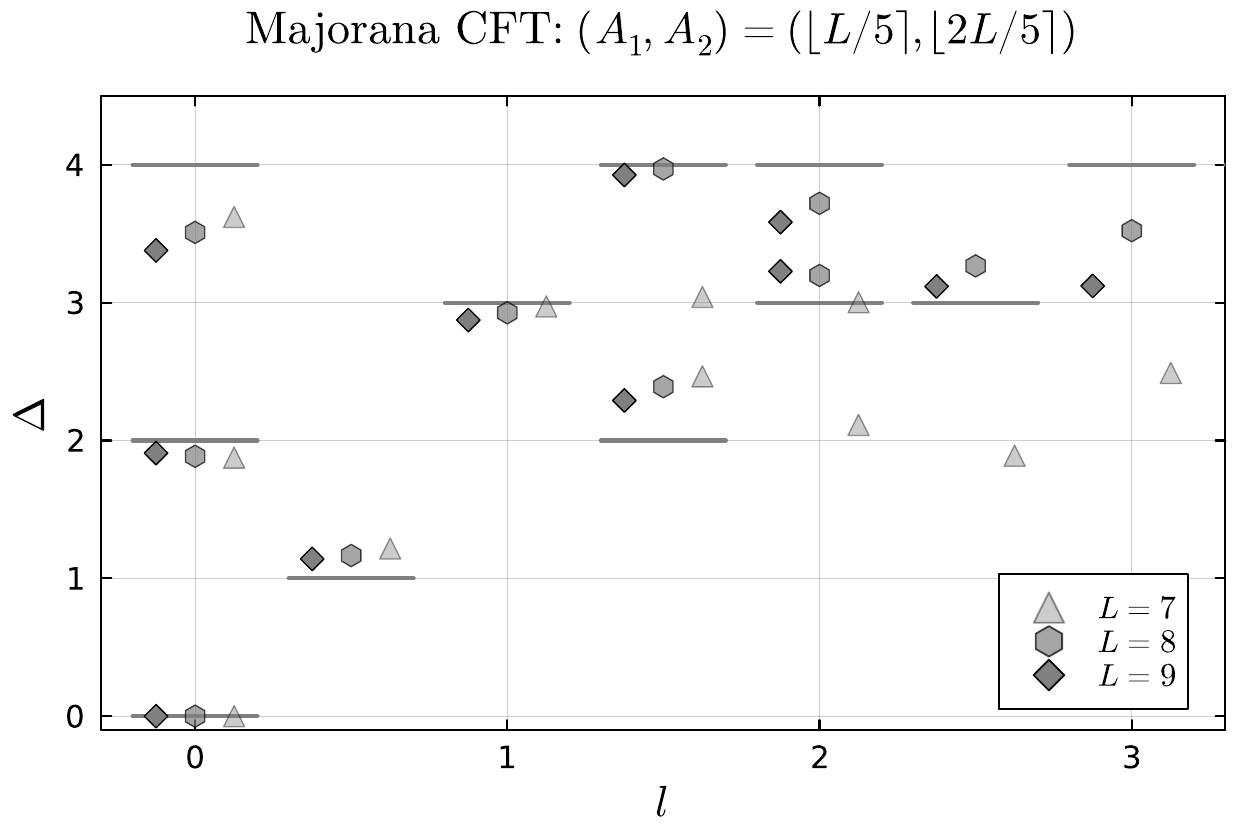}
    \par}
    \caption{\justifying 
    The spectrum of the reconstructed Hamiltonian from the fuzzy sphere critical Majorana model groundstate 
    \cite{Zhou:2025kng}
    versus angular momentum, analogous to Fig.~\ref{fig:Ising-Hrec-k2}. We use the same subregion size scaling $(A_1, A_2) = (\left\lfloor L/5 \right\rceil, \left\lfloor 2L/5 \right\rceil )$ and plot three system sizes $L = \{7,8,9\}$ with different markers and transparency. 
    The horizontal lines indicate the theoretical values of the spectrum. 
    Note that the number of orbitals in the LLL for the boson and the fermion differs by 1, and their approximate angular locations differ by a fraction of the spacing between orbitals.
    Therefore, an orbital cut that keeps the same number of orbitals for both fermions and bosons, defines slightly different polar angles $\theta_B$ and $\theta_F$ respectively.
    In specifying the size of a region $A$ in this system, we choose the angle which is the average of those of the regions associated with the boson and fermion orbitals.}
    \label{fig:Majorana-Hrec}
\end{figure}

\begin{figure}[!ht]
    \centering
    \includegraphics[width=0.95\linewidth]{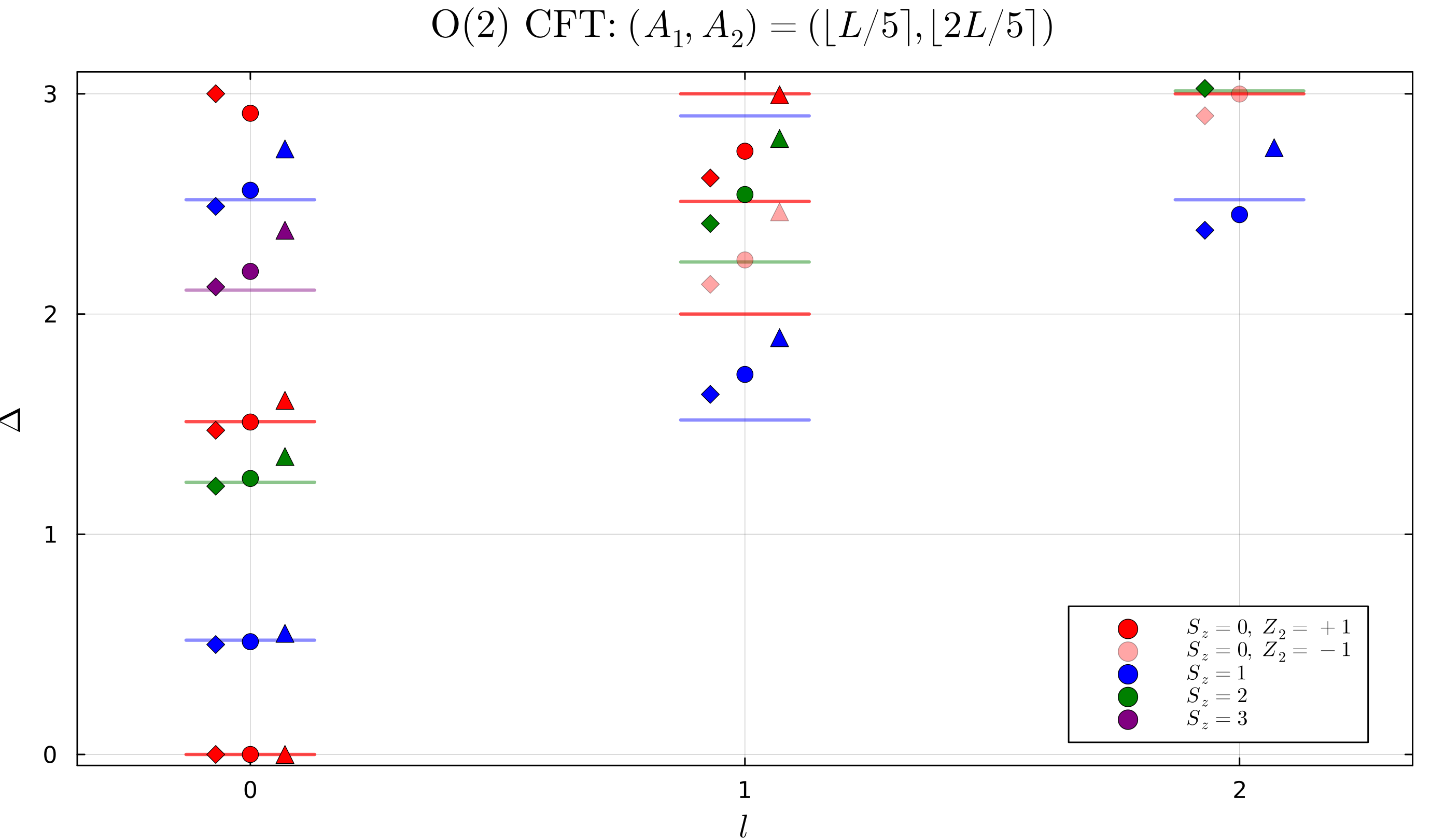}
    \caption{\justifying Reconstructed spectrum for the $\gO(2)$ CFT~\cite{Guo:2025odn}. We plot the sectors $(S_Z,Z_2) = (0,1), \, (0,-1)$ and $S_z = \{1,2,3\}$, for the low-energy states $\Delta \leq 3$. Colors identify the $S_z$ sector. 
$\gO(2) \cong U(1) \rtimes Z_2$ -- this improper $Z_2$ reverses the $U(1)$-charge: $S_z \to -S_z$, and so only commutes with the $U(1)$ in the $S_z = 0$ sector. Thus, the $S_z = 0$ sector can be further divided by $Z_2$ even or odd subsectors.
The two $Z_2$ subsectors of $S_z=0$
    are distinguished by the opacity: opaque red denotes $Z_2=1$ and transparent red $Z_2 = -1$. Marker shape distinguishes system size. We plot system sizes for $L=\{7,8,9\}$. All points are centered at their corresponding angular momentum l, with a small horizontal offset used only to separate different system sizes. The short horizontal segments indicate the corresponding conformal-bootstrap reference values.}
    \label{fig:O2-Hrec}
\end{figure}

\begin{figure}[!ht]
    \centering
    \includegraphics[width=0.95\linewidth]{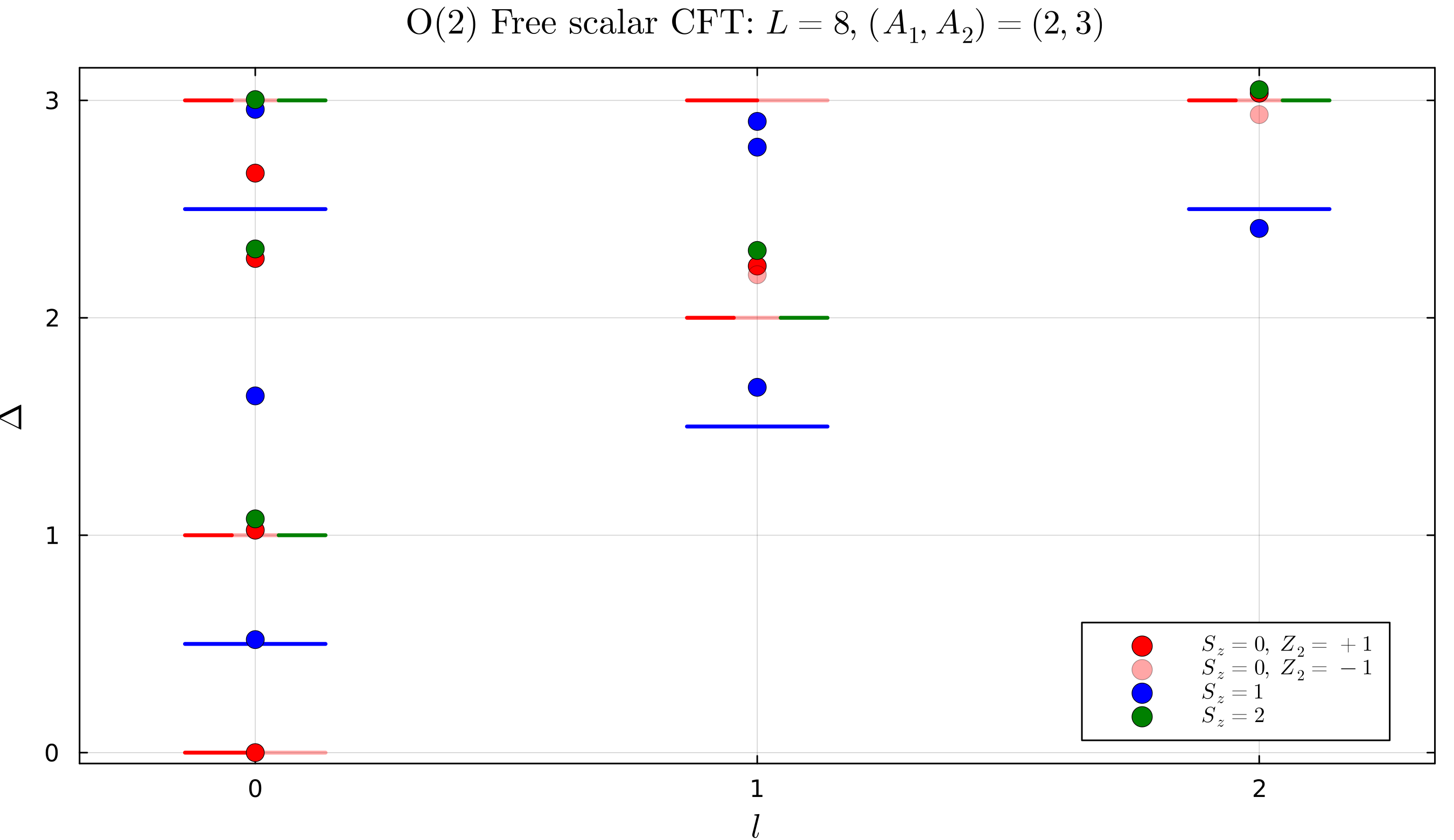}
    \caption{\justifying Reconstructed spectrum for the $O(2)$ Free scalar CFT~\cite{Guo:2025odn}. We plot the sectors $(S_Z,Z_2) = (0,1), \, (0,-1)$ and $S_z = \{1,2,3\}$, for the low-energy states $\Delta \leq 3$. Colors identify the $S_z$ sector, while within the $S_z=0$ sector the two $Z_2$ subsectors are distinguished by the opacity: opaque red denotes $Z_2=1$ and transparent red $Z_2 = -1$. We plot system size for $L=8$. The short horizontal segments indicate the corresponding conformal-bootstrap reference values.}
    \label{fig:O2-free-Hrec}
\end{figure}

\begin{figure}
    \centering
    \includegraphics[width=0.95\linewidth]{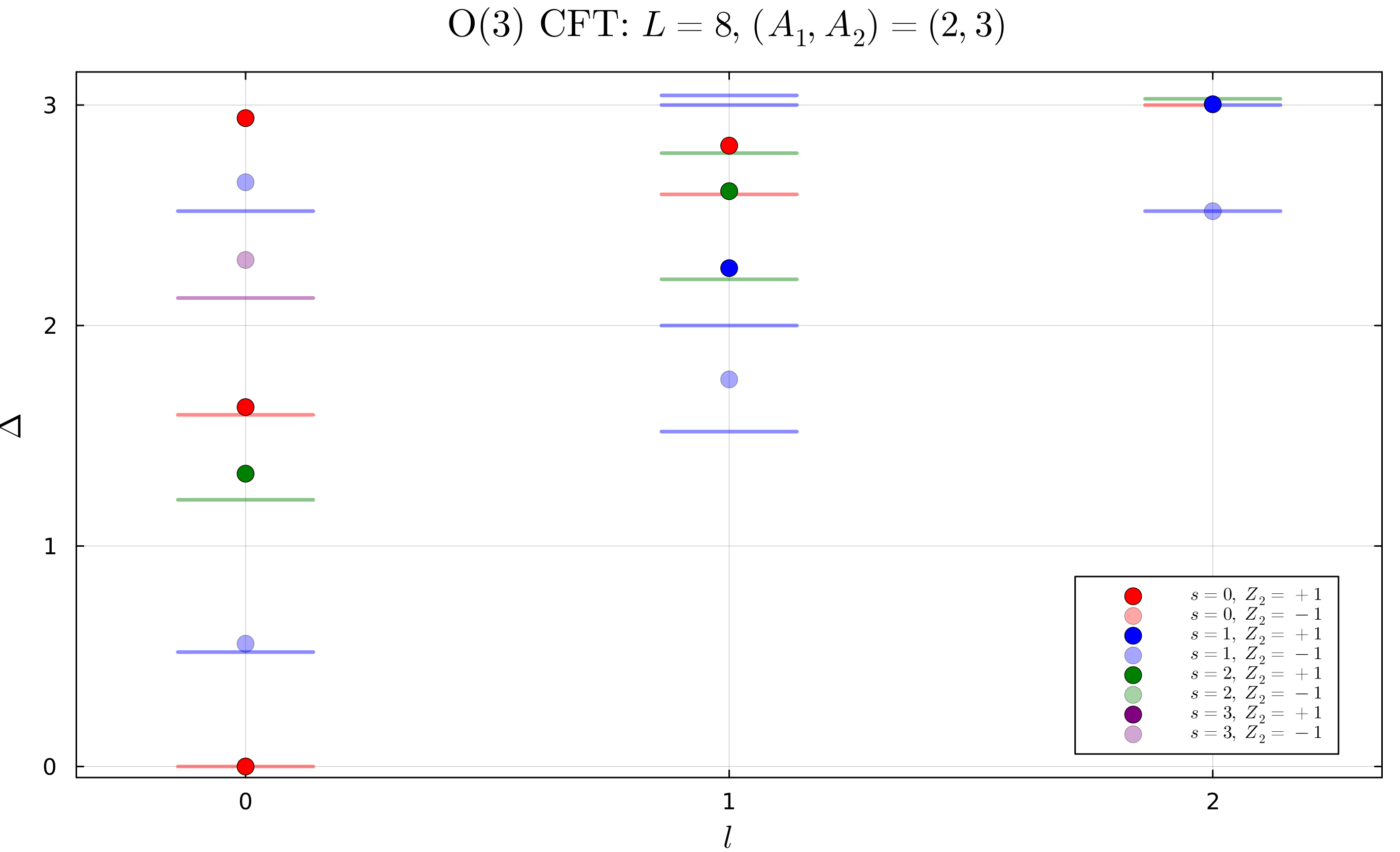}
    \caption{\justifying Reconstructed spectrum for the $O(3)$ CFT~\cite{dey2026:O3}.}
    \label{fig:O3-Hrec}
\end{figure}

In Figs.~\ref{fig:Free-scalar-Hrec},
\ref{fig:Majorana-Hrec},
\ref{fig:O2-Hrec},
\ref{fig:O2-free-Hrec},
\ref{fig:O3-Hrec}, we show the spectrum of the reconstructed Hamiltonian (dilation operator), analogous to Fig.~\ref{fig:Ising-Hrec-k2}, for other CFTs realized on the fuzzy sphere, respectively, the free scalar CFT \cite{He:2025ong}, the free Majorana CFT \cite{Zhou:2025kng}, the $\gO(2)$ Wilson-Fisher fixed point \cite{Guo:2025odn}, the free $\gO(2)$ model \cite{Guo:2025odn}, and the $\gO(3)$ Wilson-Fisher fixed point \cite{dey2026:O3}.  
Already for small system sizes, the results agree quite well with conformal bootstrap predictions.

\section{Entanglement bootstrap from trial CFT wavefunctions}
\label{app:ansatze}

Here we apply the entanglement bootstrap procedure directly to the trial wavefunctions $\ket{\psi_{\rm ansatz}(\Delta)}$ introduced in~\cite{He:2025ong}, corresponding to the free scalar CFT $(\Delta = \Delta_{\phi}= 0.5)$ and Ising CFT $(\Delta=\Delta_\sigma\simeq 0.518149)$. These proposed wavefunctions are motivated by an approximate harmonic oscillator algebra on the fuzzy sphere. They take the form of a squeezed state built from the density modes $n^\pm_{\ell m}$ acting on the product state $\ket{\varphi_0} = \prod\limits_{m} c^\dagger_{m,\downarrow}|0\rangle$. For more details, see~\cite{He:2025ong}. The motivation is the following. The construction of $D_\text{rec}$ uses only the groundstate wavefunction as input. Therefore, it gives a sharp test of whether a proposed wavefunction has the local entanglement structure of the desired CFT. 

\begin{figure}
    \centering
    \includegraphics[width=0.95\linewidth]{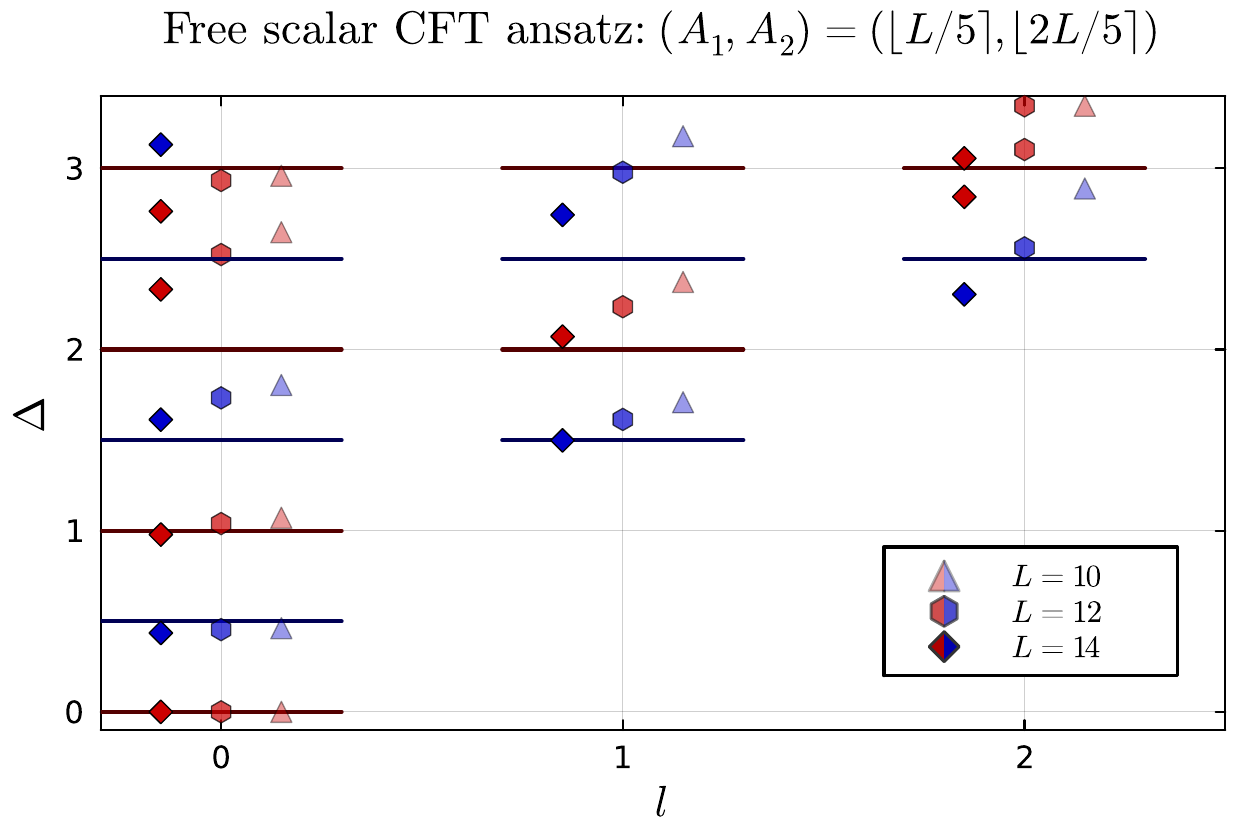}
    \caption{\justifying Reconstructed spectrum from the trial wavefunction for the 2+1d free scalar CFT. Horizontal lines correspond to the theoretical values of the free scalar spectrum.}
    \label{fig:trial-free-scalar-spectrum}
\end{figure}

Given $|\psi_{\rm ansatz}(\Delta)\rangle$, we repeat exactly the reconstruction procedure described in the main text.
We first consider the free scalar ansatz. In Fig.~\ref{fig:trial-free-scalar-spectrum}, we show the spectrum of $D_\text{rec}$ obtained from this wavefunction. The low-lying spectrum approaches the expected free scalar CFT values as the system size increases. 
This is consistent with the behavior of the correlation functions of the same wavefunction. Therefore, for the free scalar ansatz, the entanglement bootstrap and the correlation-function diagnostics give a consistent picture: the trial wavefunction appears to have the local entanglement structure of the free scalar CFT.

\begin{figure}
    \centering
    \includegraphics[width=0.95\linewidth]{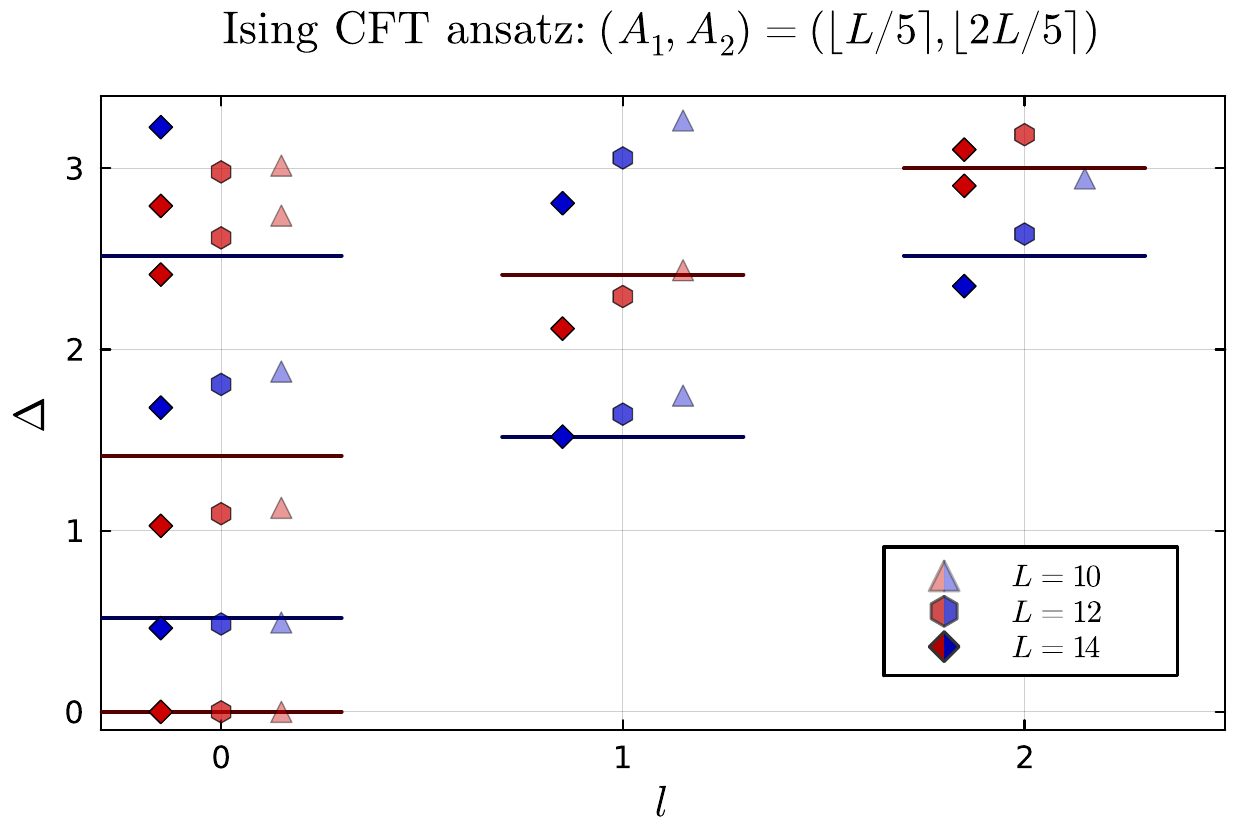}
    \caption{\justifying Reconstructed spectrum from the trial wavefunction for the 2+1d Ising CFT. Horizontal lines correspond to the CB values of the Ising CFT spectrum.}
    \label{fig:trial-ising-spectrum}
\end{figure}

The Ising ansatz gives a different result. In Fig.~\ref{fig:trial-ising-spectrum}, we show the corresponding reconstructed spectrum. Although the wavefunction has large overlap with the fuzzy-sphere Ising ground state, the reconstructed low-lying dimensions do not flow toward the Ising conformal bootstrap values. Instead, the first $\mathbb Z_2$-odd and $\mathbb Z_2$-even primaries extracted from $D_{\rm rec}$ approach values close to those obtained by fitting the antipodal correlators of the ansatz wavefunction \cite{He:2025ong}. Moreover, it looks as though the reconstructed spectrum is converging to the free scalar instead. The difference between the reconstructed spectrum from the free scalar and the Ising ansatz is quite small. 

\begin{figure*}[t]
    \centering
    \includegraphics[width=0.48\linewidth]{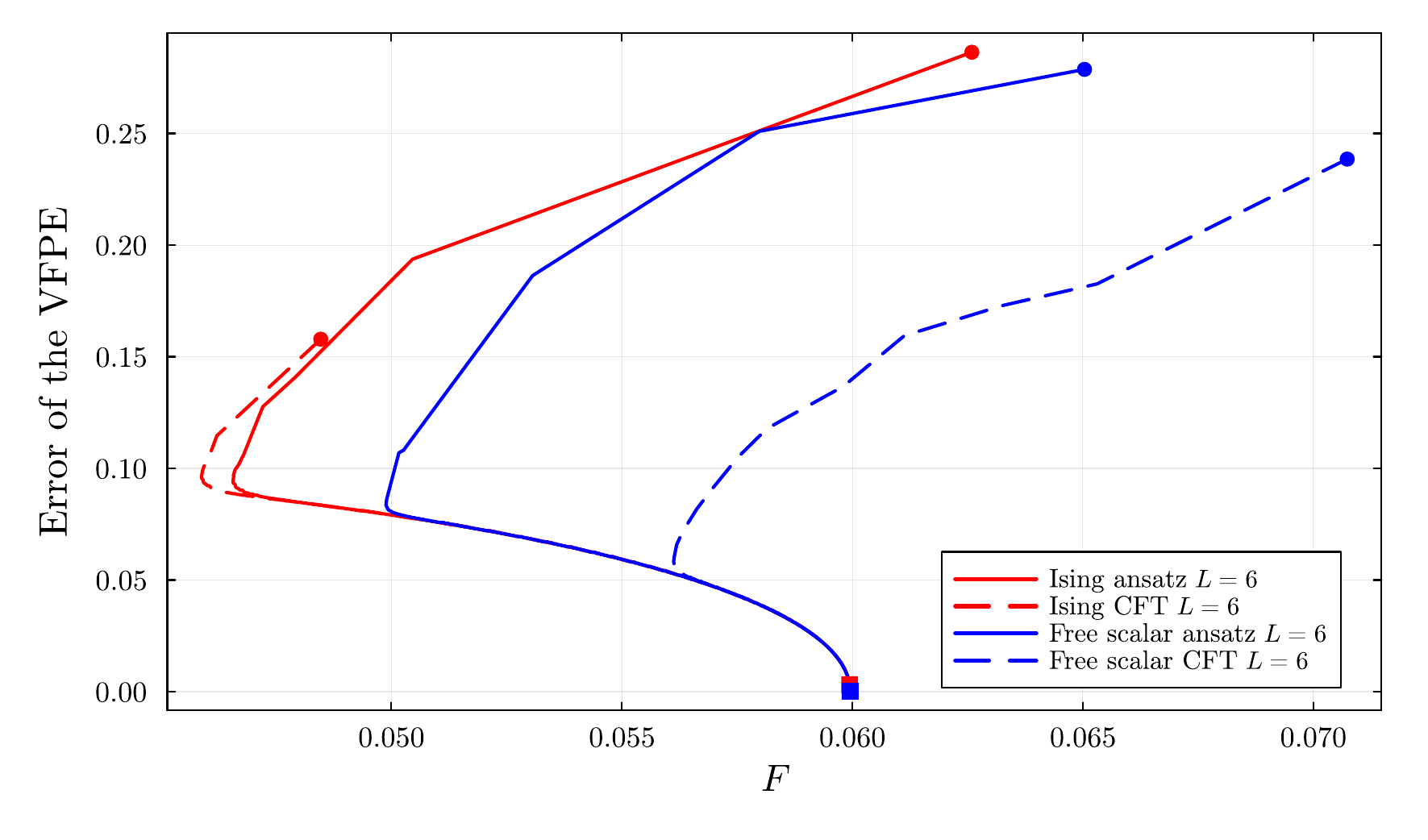}
    \includegraphics[width=0.48\linewidth]{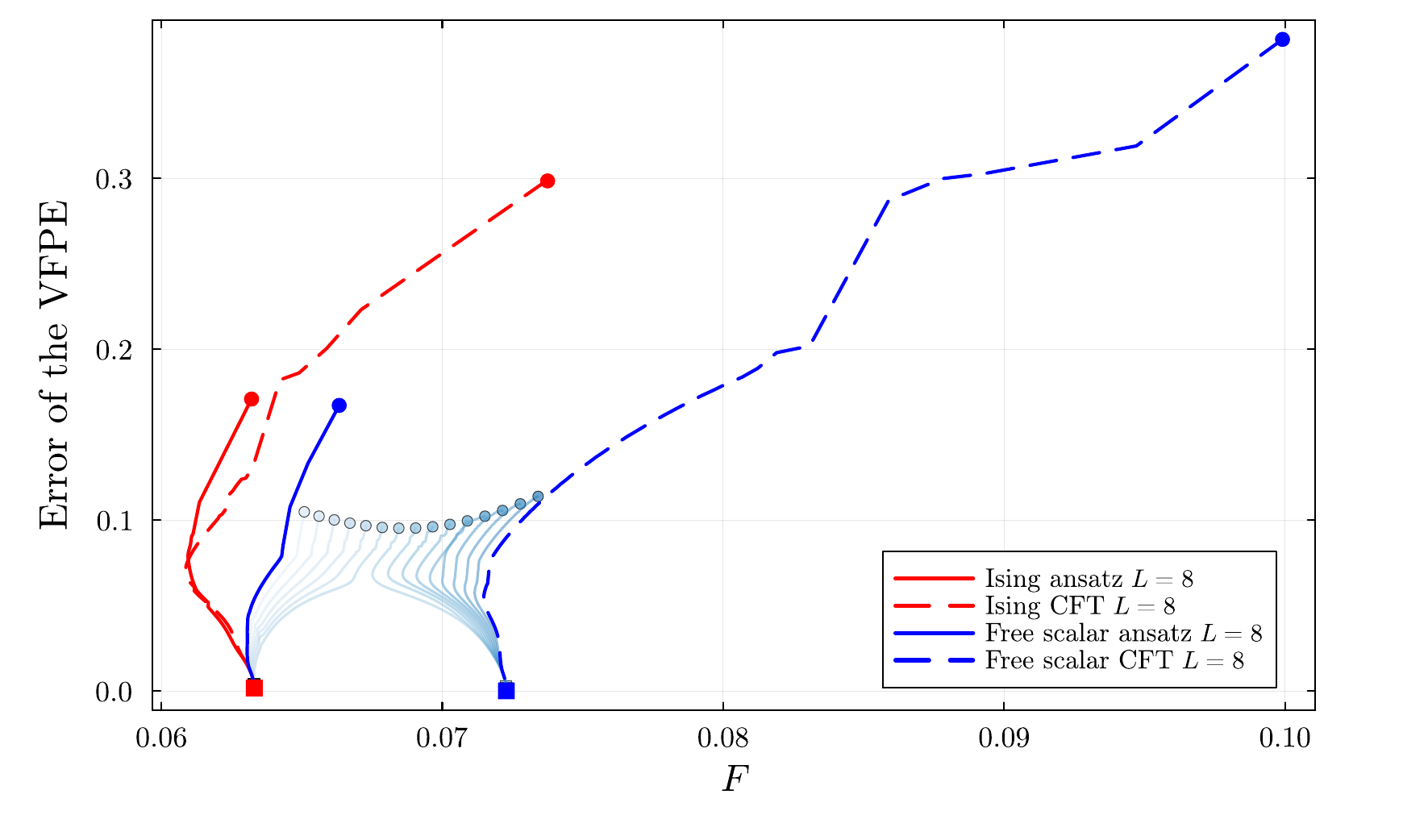}
    \caption{\justifying The trajectories of gradient descent on the error of the VFPE starting from the ansatze of \cite{He:2025ong} for the groundstates of free scalar and Ising CFTs, as well as the trajectories starting from the numerical groundstates of the corresponding critical Hamiltonians of \cite{Zhu:2022gjc} and \cite{He:2025ong} (denoted, perhaps optimistically, as `Ising CFT' and `Free scalar CFT', respectively).  The two plots show the results for $L=6$ and $L=8$ orbitals.  For $L=6$, all these initial states flow to the same Ising fixed point.  For $L=8$, the groundstate of the Hamiltonian for the free scalar CFT identified in \cite{He:2025ong} flows to a different state, but the analytical ansatz flows to the Ising state.  
    In the $L=8$ plot (right), we also show the results of GD starting from an interpolation between states along the flow from the free scalar ansatz and the free scalar critical groundstate.
    We observe the presence of a void in the space of low-error states in between the Ising and free scalar fixed points (similar to the void between trivial and Ising fixed points outlined in Fig.~\ref{fig:Ising-GD-WS}, and to the voids observed in \cite{Li:2025czz}).  For larger system sizes, we expect that the free scalar ansatz would find its way around this void and flow to the correct fixed point. }
    \label{fig:GD-on-ansatze}
\end{figure*}

However, there is a sense in which the ansatz of \cite{He:2025ong} for the Ising CFT groundstate is morally correct.  Starting from this state and doing gradient descent on the error of the VFPE, the end result is the same state as we found starting from the groundstate of the critical Ising Hamiltonian (see Fig.~\ref{fig:GD-on-ansatze}).  For the free scalar ansatz, the gradient descent trajectory (at the system sizes we studied) {\it also} ends at the Ising fixed point.

\section{Real space cut}

\label{app:real-space-cut}

In this appendix, we redo our analysis using the real-space cut.  The advantage is that the subsystem really is a local region of space, and its size can be varied continuously.  The main disadvantage, as we review next, is that the calculation requires doubling the size of the Hilbert space.  A second disadvantage, as observed in \cite{Hu:2024pen}, is that the real-space cut also cuts through the integer quantum Hall (IQH) degrees of freedom, and many of the features of the spectrum of $K_A$ in the real-space cut can be traced to IQH physics.  With our current level of understanding of how to separate these contributions, the real space cut results (\eg~for the reconstructed spectrum) are only slightly better than the orbital cut results, but are only available for small system sizes.  

Let $ \CH_1 \equiv \text{span} \{ c^\dagger_{m_1\sigma_1} \cdots c^\dagger_{m_n\sigma_n} \ket{0} \} $
be the physical Hilbert space of fermions on the fuzzy sphere.
More generally, given a collection of canonical fermion mode operators $ \{ c^\dagger_\alpha\} \equiv \CA$, 
let's define 
$$\CH_\CA \equiv \text{span} \{ c^\dagger_{\alpha_1} \cdots c^\dagger_{\alpha_n} \ket{0} \} .$$
So in this notation $\CH_1 \equiv \CH_{\{ c_{m, \sigma = \up,\down}\}}$.

In order to define real-space entanglement (between $A$ and $\bar A$) of a state $\ket{\psi}$ in $\CH_1$ it is useful to embed it in a larger bipartite Hilbert space 
\be \CH_2 \equiv \CH_A \otimes \CH_{\bar A} \ee
where $\CH_A$ is defined as follows. 
The mode operator 
\bea 
c_{m\sigma}  &=& \int d\Omega \Phi_m^\star(\Omega) \psi_\sigma(\Omega) 
\nonumber
\\ &=& \int d\Omega \Phi_m^\star(\Omega)\Theta_A(\Omega) \psi_\sigma(\Omega) 
+ \int d\Omega \Phi_m^\star(\Omega)\Theta_{\bar A}(\Omega) \psi_\sigma(\Omega)  \nonumber
\\ \nonumber
&\equiv& p_{mA} c_{m\sigma A} + p_{m\bar A} c_{m\sigma \bar A} . \eea
If $\{ {1\over p_{mA}} \Phi_m^\star(\Omega)\Theta_A(\Omega) \} $ is an ON basis on $A$ 
(as is the case when $A$ is a polar region of the fuzzy sphere) then the $c_{mA}$ and $c_{m\bar A}$ are a collection of (twice as many) canonical fermion modes.
So in the notation above $ \CH_A = \CH_{\{ c_{m\sigma A} \} }$, 
$ \CH_{\bar A} = \CH_{\{ c_{m\sigma \bar A} \} }$. 

Now for each such region we define an isometry 
\be \varphi_A: \CH_1 \to \CH_2 \ee
by the map 
\be \varphi_A \( c_{m_1}^\dagger \cdots  \ket{0} \) 
=  \( p_{m_1A} c_{m_1\sigma A}^\dagger + p_{m_1\bar A} c_{m_1\sigma \bar A}^\dagger \) \cdots \ket{0} . \ee
Let $\CH_3 \equiv \text{Im} \varphi_A(\CH_1) $ be the image of $\CH_1$ in $\CH_2$.  

Notice that $\CH_2$ is a scary place with many more states than $\CH_1$.
In particular, consider the operator $L_z = \sum_m m c_m^\dagger c_m$ on $\CH_1$
(ignore spin which is irrelevant for this discussion).  
A naive lift of this operator to $\CH_2$ gives  
\bea \tilde L_z &=& \sum_m m \( p_{mA}^2 c_{mA}^\dagger c_{mA}^\nd + p_{m\bar A}^2 c_{m \bar A}^\dagger c_{m \bar A}^\nd 
\right. \\  \nonumber && \left.
+ ~~ p_{mA} p_{m\bar A} \( c_{mA}^\dagger c_{m\bar A}^\nd + c_{m\bar A}^\dagger c_{mA}^\nd \) \) 
\\ & = & 
\( c_{mA}^\dagger, c_{m \bar A}^\dagger \) \begin{pmatrix} p_{mA}^2 & p_{m \bar A} p_{mA} \\  \nonumber
p_{mA} p_{m\bar A} & p_{m\bar A}^2 \end{pmatrix} \begin{pmatrix} c_{mA}^\nd \\ c_{m\bar A}^\nd \end{pmatrix}
\\ \nonumber & = & 
\( c_{mA}^\dagger, c_{m \bar A}^\dagger \) 
\begin{pmatrix} p_{mA} \\  p_{m \bar A}  \end{pmatrix} 
\(  p_{mA},  p_{m\bar A}\) \begin{pmatrix} c_{mA}^\nd \\ c_{m\bar A}^\nd \end{pmatrix}. \eea
Thus the eigenvalues of this matrix are $1$ and $0$, and we learn that the modes 
$ p_{mA} c_{mA} + p_{m\bar A} c_{m \bar A} $ carry angular momentum $m$ 
and the (new) modes 
$ p_{m\bar A} c_{mA} - p_{m A} c_{m \bar A} $ carry angular momentum {\it zero}. 

Now we come to some an important question:
Is $ K_A \varphi_A \ket{\psi} \in \CH_3 $?  
Based on what we just learned about the angular momentum generators, we cannot do the averaging on $\CH_2$, but the map $\varphi^{-1}$ is only defined on $\CH_3$.
In fact, the answer is `no'.   
To explain how we deal with this, we must make a brief digression.

The doubled Hilbert space $\CH_2^A$ for the region $A$ is a subspace of a much larger Hilbert space $ \CH_\infty$, which is the Hilbert space of free fermions in the continuum on the sphere.  $\CH_\infty$ does have a nice action of $\gSO(3)$, and so the averaging makes sense there, in principle.  
To compute the error of the vector fixed point equation,
  we would hope to compute $\overline{K_A |\psi\>}$ in $\CH_\infty$.
  
  There are some problems with this goal.  
First, this space is not numerically tractable.
Moreover, $K_A |\psi\>$ is generally not a vector in the LLL, $\cH_1$;
  in fact it is not even in the doubled Hilbert space $\cH_2$.
Instead, we would like to find the projection to the lowest Landau level of the result of this averaging.  
The best approximation is obtained by the average
\be \overline{\varphi^{-1} P_{\CH_3^A}K_A\ket{\psi}}\ee
which, as we show, can be computed using finite resources.
So, in fact, instead of computing the full error and requiring
  $|\psi\> \propto \overline{K_A |\psi\>}$,
  we impose a weaker condition,
  namely,
  $|\psi\> \propto P_{LLL}\overline{K_A |\psi\>}$,
  where $P_{LLL}$ is the projector to the lowest Landau level $\cH_1$.
This means that error of the VFPE that we quote should really be regarded as a lower bound on the error in the ideal Hilbert space $\CH_\infty$.

Explicitly, choose a basis $\{ \ket{i} \}$ of $\CH_1$ 
(or some subspace of $\CH_1$ of states with definite particle number etc). 
Then let $ \varphi_A \ket{i} = \ket{\varphi_i} $ be their images, which form a basis of $\CH_3 \subset \CH_2$.  
The inverse of $\varphi$ on $\CH_3$ is well-defined and given by 
\be \varphi^{-1} \( \ket{\alpha } = \sum_i \alpha_i \ket{\varphi_i} \in\CH_3 \) = \sum_i \alpha_i \ket{i} . \ee
Now, when we want to integrate some entanglement Hamiltonian over regions we must first act with $\varphi^{-1}$ to get back a state on the actual Hilbert space.  Then we can just use the ordinary rotation generators on the fuzzy sphere.  
Similarly, when we want to make $K_B$ for some other region, we can do so in $\varphi_B(\CH_1)$ just as above and then map it back by $\varphi_B^{-1}$.  So, importantly, we never need to make a tripartition of the Hilbert space.

\begin{figure}
    \centering
    \includegraphics[width=0.8\linewidth]{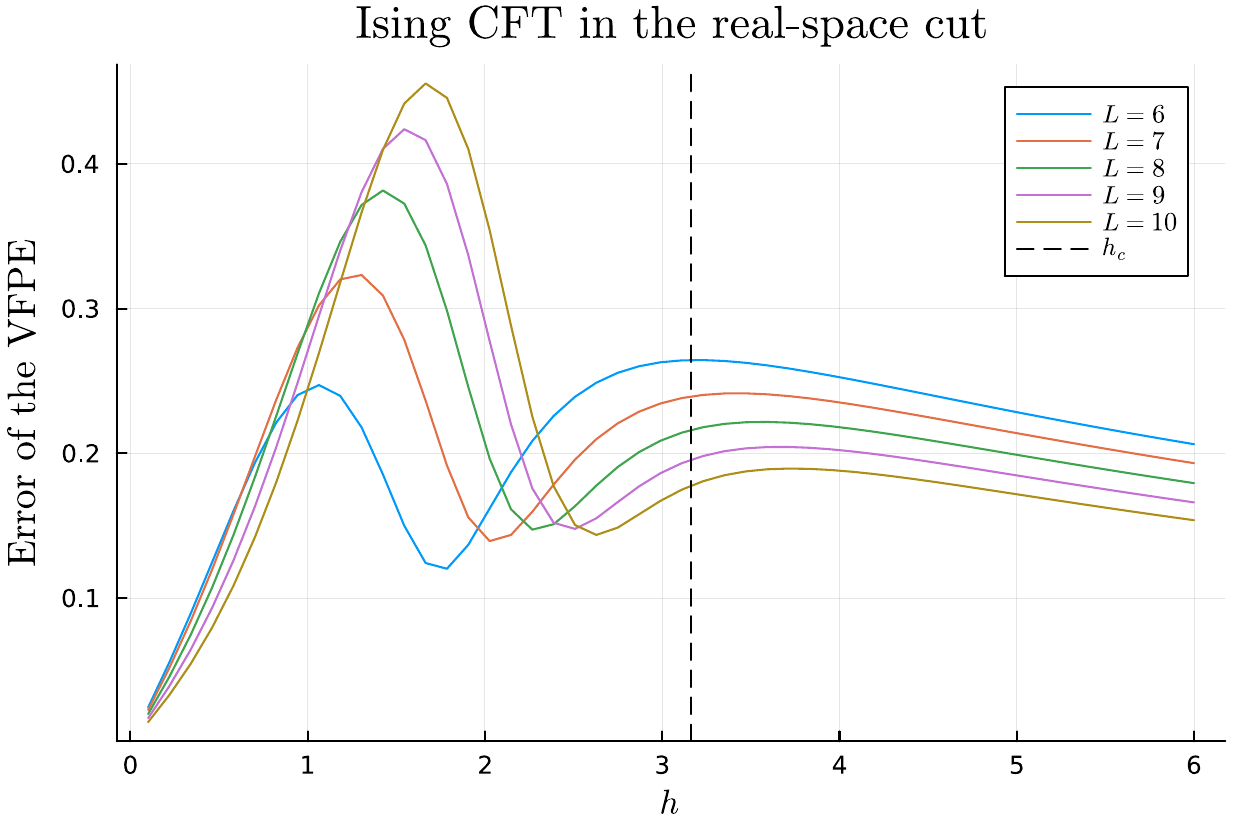}
    \caption{\justifying Error of the VFPE in the real-space cut evaluated on the ground state of the 2+1d Ising model Hamiltonian, as a function of the transverse field $h$ and fixed pseudopotential coupling $V_0 = 4.75$. 
    Here, we fixed the subregion sizes to $(\theta_{A_1}, \theta_{A_2}) = (\pi/4, \pi/2)$ for different system sizes. 
    The dashed line indicates the critical value $h_c = 3.16$, identified in~\cite{Zhu:2022gjc}.
    As we increase system size, we note that the minima of the error approach the critical value $h_c$. 
    }
    \label{fig:Ising-phase-diagram-RS}
\end{figure}

\begin{figure}
    \centering
    \includegraphics[width=0.8\linewidth]{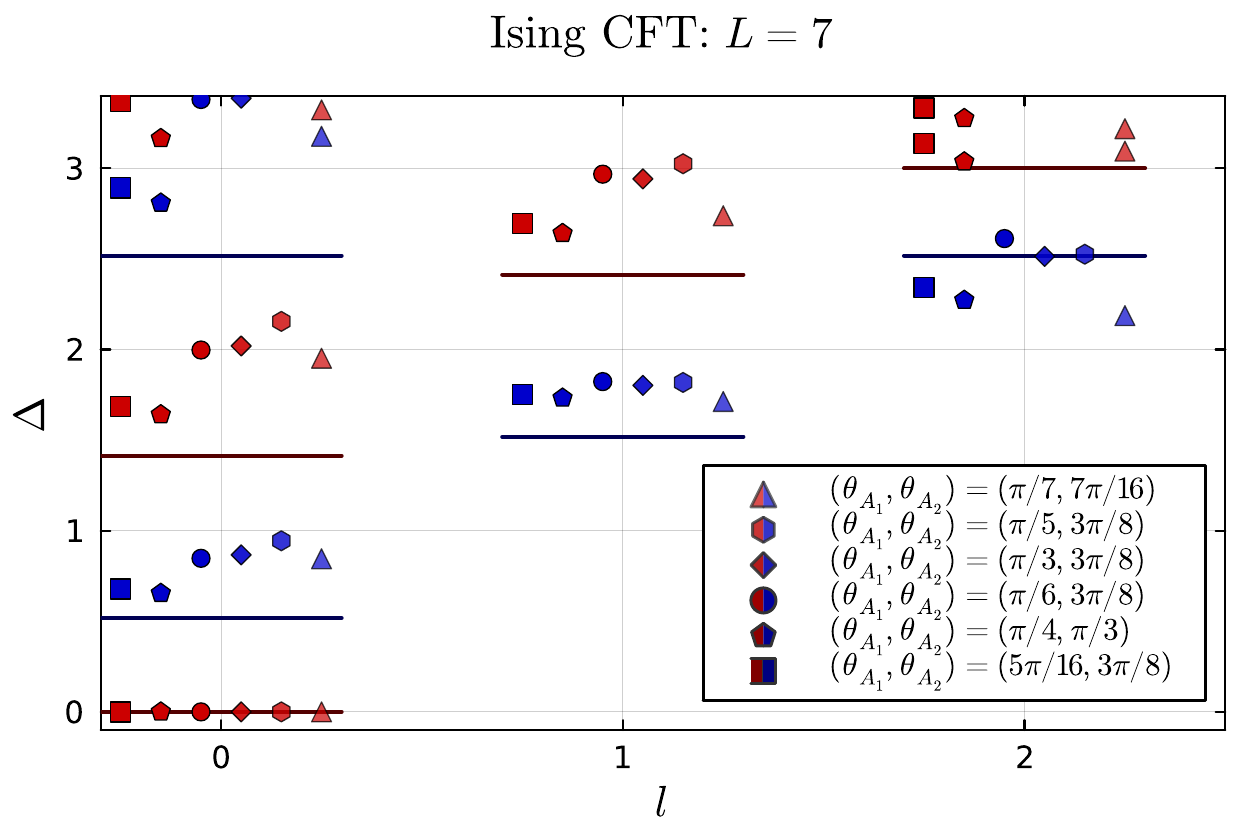}
    \caption{\justifying Reconstructed spectrum for $k=2$ 
    for the $L=7$ Ising critical groundstate of \cite{Zhu:2022gjc},
    using the real-space cut for different subregions.}
    \label{fig:Hrec-RS-Ising}
\end{figure}

\begin{figure}
    \centering
    \includegraphics[width=0.8\linewidth]{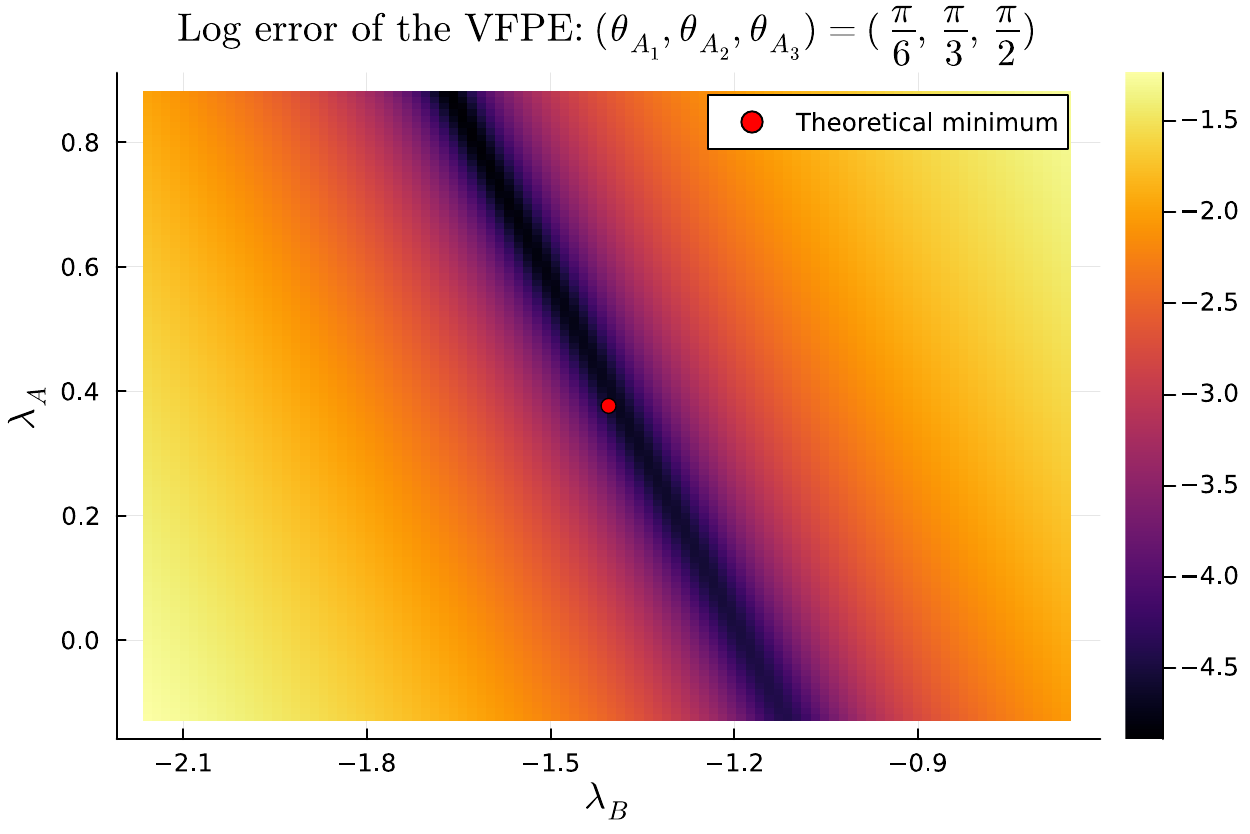}
    \caption{\justifying Log error of the VFPE 
    for the $L=10$ Ising critical groundstate of \cite{Zhu:2022gjc}, using 
    the real-space cut, with $k=3$ regions.  There is a narrow, flat valley of minima containing the theoretical prediction.}
    \label{fig:Log-error-k3-RS-Ising}
\end{figure}

\clearpage

\bibliographystyle{ucsd}
\bibliography{collection}
\end{document}

%% file: jm-tex-macros-public.tex
\usepackage{amssymb}
\usepackage{amsmath,bm}
\usepackage{amssymb}
\usepackage{graphicx}
\usepackage{amsfonts}         
\usepackage{fancybox}

\usepackage[usenames,dvipsnames]{xcolor}%

\usepackage[pagebackref,  %
bookmarks={false}, pdfauthor={John McGreevy}, pdftitle={Yay, physics!}]{hyperref}
\hypersetup{colorlinks=true, 
linkcolor=BrickRed, 
citecolor=Violet, 
filecolor=OliveGreen, 
urlcolor=RoyalBlue, 
filebordercolor={.8 .8 1}, 
urlbordercolor={.8 .8 0}}%
\usepackage{soul}%
\setstcolor{Red}

\definecolor{darkgreen}{rgb}{0,0.4,0}
\definecolor{darkred}{rgb}{0.4,0,0}
\definecolor{darkblue}{rgb}{0,0,0.4}
\definecolor{lightblue}{rgb}{.6,.6,0.9}

\definecolor{uglybrown}{rgb}{0.8,  0.7,  0.5}

\definecolor{palatinatepurple}{rgb}{0.41, 0.16, 0.38}
\definecolor{celebrationcolor}{rgb}{0.75,  0.0,  0.9}

\definecolor{Mathematica1}{rgb}{0.368417, 0.506779, 0.709798}
\definecolor{Mathematica2}{rgb}{0.880722, 0.611041, 0.142051}
\definecolor{Mathematica3}{rgb}{0.560181, 0.691569, 0.194885}

\definecolor{shadecolor}{rgb}{0.90,0.90,0.90}
\definecolor{DVcolor}{rgb}{0.95,  0.5,  0.2}
\definecolor{lightbluemuons}{rgb}{0.0,.65,1.0}

\definecolor{chartreuse}{rgb}{0.70, 1.00, 0.00}

\usepackage{tikz}
\usetikzlibrary{shapes}
\usetikzlibrary{trees}
\usetikzlibrary{snakes}
\usetikzlibrary{matrix,arrows} 				%
\usetikzlibrary{positioning}				%
\usetikzlibrary{calc,through}				%
\usetikzlibrary{decorations.pathreplacing}  %
\usepackage[tikz]{bclogo} 					%
\usepackage{pgffor}							%

\usetikzlibrary{decorations.markings}
\usetikzlibrary{intersections}				%

\usetikzlibrary{arrows,decorations.pathmorphing,backgrounds,positioning,fit,petri,automata,shadows,calendar,mindmap, graphs}
\usetikzlibrary{arrows.meta,bending}

\tikzset{
    vector/.style={decorate, decoration={snake}, draw},
    fermion/.style={postaction={decorate},
        decoration={markings,mark=at position .55 with {\arrow{>}}}},
    fermionbar/.style={draw, postaction={decorate},
        decoration={markings,mark=at position .55 with {\arrow{<}}}},
    fermionnoarrow/.style={},
    gluon/.style={decorate,
        decoration={coil,amplitude=4pt, segment length=5pt}},
    scalar/.style={dashed, postaction={decorate},
        decoration={markings,mark=at position .55 with {\arrow{>}}}},
    scalarbar/.style={dashed, postaction={decorate},
        decoration={markings,mark=at position .55 with {\arrow{<}}}},
    scalarnoarrow/.style={dashed,draw},
	vectorscalar/.style={loosely dotted,draw=black, postaction={decorate}},
}

\def\centerarc[#1](#2)(#3:#4:#5)%
    { \draw[#1] ($(#2)+({#5*cos(#3)},{#5*sin(#3)})$) arc (#3:#4:#5); }

\usepackage{enumitem}

\usepackage{slashed}

\usepackage[font=small,labelfont=bf]{caption}
\DeclareCaptionFont{tiny}{\tiny}
\usepackage{epstopdf}
\usepackage{wasysym}

\usepackage[yyyymmdd,hhmmss]{datetime}   
\usepackage{framed}  %
 \usepackage{mdframed}
 \usepackage{wrapfig}
 \usepackage{yfonts}  

\def\ketbra#1#2{ | #1 \rangle\hskip-2pt\langle #2|}

\newmdenv[%
        backgroundcolor=lightgray,
    linecolor=black,
    outerlinewidth=2pt,
]{boxedandshaded}

\def\parfig#1#2{
\parbox{#1\textwidth}
{\includegraphics[width=#1\textwidth]{#2}}
}

\numberwithin{equation}{section}

\renewcommand{\theequation}{\arabic{section}.\arabic{equation}}

\def\nd{{ \vphantom{\dagger}}}

\newcommand{\vev}[1]{\langle #1 \rangle}

\newlength{\extraspace}
\newlength{\extraspaces}
\def\be{\begin{equation}}
\def\ee{\end{equation}}

\newcommand{\bea}{\begin{eqnarray}}
\newcommand{\eea}{\end{eqnarray}}

\newtheorem{theorem}{Theorem}

\def\eps{\epsilon}

\def\Tr{{{\rm Tr~ }}}

\def\bra#1{\left\langle#1\right|}
\def\ket#1{\left|#1\right\rangle}

\def\vev#1{\left\langle{#1}\right\rangle}

\def\CA{{\cal A}}

\def\CH{{\cal H}}

\def\CO{{\cal O}}%

\def\II{\relax{I\kern-.10em I}}

\def\IZ{\mathbb{Z}}
\def\IB{\relax{\rm I\kern-.18em B}}

\def\ID{\relax{\rm I\kern-.18em D}}
\def\IE{\relax{\rm I\kern-.18em E}}
\def\IF{\relax{\rm I\kern-.18em F}}
\def\IG{\relax\hbox{$\inbar\kern-.3em{\rm G}$}}
\def\IGa{\relax\hbox{${\rm I}\kern-.18em\Gamma$}}
\def\IH{\relax{\rm I\kern-.18em H}}
\def\II{\relax{\rm I\kern-.18em I}}
\def\IK{\relax{\rm I\kern-.18em K}}

\def\inbar{\,\vrule height1.5ex width.4pt depth0pt}

\def\IR{\mathbb{R}}

\def\simgt{\hskip0.05in\relax{ 
\raise3.0pt\hbox{ $>$
{\lower5.0pt\hbox{\kern-1.05em $\sim$}} }} \hskip0.05in}

\def\lp10{\ell_p^{10}}
\def\lp11{\ell_p^{11}}
\def\R11{R_{11}}

\def\frac#1#2{{#1 \over #2}}

\def\up{\uparrow}
\def\down{\downarrow}

\def\Ione{\hbox{$1\hskip -1.2pt\vrule depth 0pt height 1.53ex width 0.7pt
                  \vrule depth 0pt height 0.3pt width 0.12em$}}

\newdimen\tableauside\tableauside=1.0ex
\newdimen\tableaurule\tableaurule=0.4pt
\newdimen\tableaustep
\def\phantomhrule#1{\hbox{\vbox to0pt{\hrule height\tableaurule width#1\vss}}}
\def\phantomvrule#1{\vbox{\hbox to0pt{\vrule width\tableaurule height#1\hss}}}
\def\sqr{\vbox{%
  \phantomhrule\tableaustep
  \hbox{\phantomvrule\tableaustep\kern\tableaustep\phantomvrule\tableaustep}%
  \hbox{\vbox{\phantomhrule\tableauside}\kern-\tableaurule}}}
\def\squares#1{\hbox{\count0=#1\noindent\loop\sqr
  \advance\count0 by-1 \ifnum\count0>0\repeat}}
\def\tableau#1{\vcenter{\offinterlineskip
  \tableaustep=\tableauside\advance\tableaustep by-\tableaurule
  \kern\normallineskip\hbox
    {\kern\normallineskip\vbox
      {\gettableau#1 0 }%
     \kern\normallineskip\kern\tableaurule}%
  \kern\normallineskip\kern\tableaurule}}
\def\gettableau#1 {\ifnum#1=0\let\next=\null\else
  \squares{#1}\let\next=\gettableau\fi\next}

\def\({\left(}
\def\){\right)}

\def\ii{{\bf i}}

\def\EE{{\bf E}}

\def\lsim{\mathrel{\mathstrut\smash{\ooalign{\raise2.5pt\hbox{$<$}\cr\lower2.5pt\hbox{$\sim$}}}}}
\def\gsim{\mathrel{\mathstrut\smash{\ooalign{\raise2.5pt\hbox{$>$}\cr\lower2.5pt\hbox{$\sim$}}}}}

\def\overleftrightarrow#1{\vbox{\ialign{##\crcr
     $\leftrightarrow$\crcr\noalign{\kern-0pt\nointerlineskip}
     $\hfil\displaystyle{#1}\hfil$\crcr}}}
     
     \def\overleftarrow#1{\vbox{\ialign{##\crcr
     $\leftarrow$\crcr\noalign{\kern-0pt\nointerlineskip}
     $\hfil\displaystyle{#1}\hfil$\crcr}}}

\def\eg{{\it e.g.}}

\def\gSO{\textsf{SO}}

\def\gO{\textsf{O}}
\def\gSU{\textsf{SU}}
\def\gU{\textsf{U}}

\newif{\ifeq}           %
\eqtrue                 %

\newcounter{lecturecounter}

%% file: macros.tex
\newtheorem{remark}[theorem]{Remark}

\newcommand{\cH}{\mathcal{H}}

\newcommand{\cO}{\mathcal{O}}

\newcommand{\wt}{\widetilde}

\renewcommand{\ol}{\overline}
\newcommand{\interior}[1]{%
  {\kern0pt#1}^{\mathrm{o}}%
}

\def\={\phantom{{}={}}}